# Toward Polyoxometalate Nanoelectronics.

Dominique Vuillaume,[1,*] and Anna Proust.[2,*]

*1) Institute for Electronics Microelectronics and Nanotechnology (IEMN), CNRS, Av. Poincaré, F-59650 Villeneuve d'Ascq, France.*

*2) Institut Parisien de Chimie Moléculaire (IPCM), CNRS, Sorbonne Université, 4 Place Jussieu, F-75005 Paris, France.*

** Corresponding authors: dominique.vuillaume@iemn.fr ; anna.proust@sorbonne-universite.fr*

**Abstract.**

Polyoxometalates form a large family of molecular oxide clusters of the early transition metals with unique and tunable properties (multi-redox, thermal and chemical robustness, magnetic). We review more than 30 years of experimental research on the electron transport properties of polyoxometalates devices, from thin films and self-assembled monolayers down to single-molecule junctions. We focus on the relationship between the polyoxometalate structures (structural type, nature of metals and heteroatoms, role of the counterions, redox states, electrode linkers and functional ligands) and the electronic structures of the polyoxometalate-based devices (energy positions of the molecular orbitals, energy offset at the interfaces). Then, we critically discuss the performances of polyoxometalates in nanoelectronics devices: capacitance and resistive switching memories, spintronics, quantum bits and neuromorphic devices. We conclude with a discussion about pending issues and perspectives.

## Contents.

## 1. INTRODUCTION.

The ever-increasing demand for information and communication technologies, combined with the rapid expansion of deep learning and artificial intelligence, places considerable pressure on computational performance and energy efficiency. With the channel length of the transistor approaching the size of about 5 atoms in the silicon crystal (i.e. ≈ 2 nm), silicon-based technologies face intrinsic scaling limitations and naturally come to the end of the so-called "Moore law".[1, 2] Further innovations come by adding functionalities, reducing power consumption, taking advantage or gaming with device variability, for example. This has driven the exploration of alternative materials and devices in a "More-than-Moore" approach. Molecules and hybrid molecular systems present a compelling opportunity to meet the growing need for tailored, multifunctional devices suited to mobile and diverse computing environments.[3-9] Molecules offer unique advantages: atomic-level structural precision, reproducibility, monodispersity. They are inherently compatible with nanometer-scale integration and bottom-up fabrication methods. Chemical synthesis allows fine-tuning of their electronic behavior, enabling the design of switches/memories[10-15] and quantum components such as qubits.[16-18] Additionally, solution processing facilitates low-cost manufacturing, on various substrates, also suited for flexible electronics and internet of things.[15, 19] Molecular engineering also enables the integration of multifunctionality, yielding multi-addressable systems responsive to chemical, optical, electrical, and/or magnetic stimuli.[20, 21]

Among molecules, polyoxometalates (POMs) are standing out.[22, 23] They form an ever-expanding family of molecular oxide clusters of the early transition metals,[24] that can be used as building blocks for functional nanoscale systems[25] or integrated into composite materials for their unique added-value properties.[26] Several classes of molecular devices with resistive and memory switching capabilities have been developed based on various physicochemical properties, *e.g.* photochromic molecules (diarylethene, spiropyran, azobenzene),[27-36] redox molecules (ferrocene, viologen, porphyrin, and many other molecules, see recent reviews),[37-39] mechanically interlocked molecular machines (rotaxanes, catenanes),[40-45] spin crossover molecules.[36, 46-50] Compared to these approaches, POMs combine the redox properties and the robustness of extended metal oxides with the diversity and tunability of molecules. They are redox-active with discrete redox states that can be separately and reversibly addressed.[51, 52] They have thus found many redox-driven applications in photo/electrocatalysis,[53, 54] photo/electrochromic materials[55-57] and energy conversion/storage systems.[58] However, their potential in the field of nanoelectronics is still under-explored.

POM-based materials are of interest for the three types of technologies and devices on which we focus this review. A first type concerns resistive switching devices that are extensively explored to address requirements for high-density data storage, low-power computing, encryption, data security, data analysis, recognition and classification tasks.[59] To address the environmental and energy cost issues, one approach relies on a paradigm shift from conventional von Neumann architectures, where memory and processing units are physically separated, to brain-inspired in-memory computing.[60-64] In an another approach, huge efforts are directed toward the development of quantum computing hardware,[65-67] especially by chemical engineering and manipulating molecular spin qubits, networks and interfaces.[68-70]

For example, the stabilization of multiple reduction states of polyoxovanadates immobilized onto a gold electrode has been demonstrated and related to conductance changes by scanning tunneling spectroscopy.[71] The possibility of addressing multiple charge states corresponding to logic ON/OFF states opens a wide range of opportunities for digital data processing and memory technologies. In 2007, the incorporation of POMs into the floating gate of a field-effect transistor for flash-type memories was patented[72] and in 2014 three different types of electronic devices integrating POMs were published, exemplifying multi-level capacitive flash memories,[73] hybrid molecular/semiconductor capacitor[74] and resistive memories,[75] respectively. These early examples showing that the charge trapping and redox-switching ability of POMs can be harnessed either in capacitive or memristive devices, depending on the environment,[76] have driven further investigations. These have been partially reviewed, mostly as a section in more general papers devoted to the charge carrier behavior of POMs,[77] substrate functionalization[78] and mainly oriented towards energy-related applications,[79] or limited to non-volatile memories.[80-82] None of them focus on comparative electron transport (ET) analysis and performance metrics so that a clear picture of the POM-based electronic devices is still to draw. An updated and tutorial review dedicated to ET of POM based electronic devices, including perspectives for the design of the next generation of POM-based devices for in-memory, neuromorphic and quantum computing, is thus meaningful.

Polyoxometalates are readily accessible by condensation of oxometalates under pH control. Figure 1 presents some POM archetypes. They obey the general formula $[M_mO_y]^{q-}$, for isopolyanions and $[X_xM_pO_z]^{n-}$ for heteropolyanions (M = Mo(VI), W(VI), V(V), Nb(V); X = S, P, As, Si, Al …). These include the Lindqvist $[M_6O_{19}]^{2-}$, Anderson-Evans $[M'M_6O_{24}]^{n-}$ (M' = Al(III), Mn(III), Fe(III), Co(III), Ni(II), Zn(II), Te(VI) …), Keggin $[XM_{12}O_{40}]^{n-}$ (M = Mo, W, X = P with n = 3; X= Si with n =4 …) and Wells-Dawson $[X_2M_{18}O_{62}]^{p-}$ (M = Mo, W; X = P with p = 6; X = S with p = 4 …) types. Formal loss of one or several $\{M^{VI}=O\}^{4+}$ unit(s) gives lacunary species such as $[XW_{11}O_{39}]^{(n+4)-}$, with vacant sites ready to be functionalized, either by the introduction of extra transition metal cations or organic extensions. The introduction of organic tethers can also rely on the reactivity of the metal-oxo bonds to yield organic-inorganic hybrids,[83-86] among which some of them have been investigated in the context of nanoelectronics.[81] Polyoxometalates are polyanions and thus carry counter-cations, be they alkali or organic, such as tetraalkylammonium, which role is much more subtle than a simple charge balance.[87]

The POM shaping process onto electrodes is a preliminary and integral step, which determines the properties of their molecular devices and their reproducibility. Unlike organic molecules and most organometallic complexes, POMs are polyanions and their ionic character precludes their sublimation under (ultra)high vacuum. Furthermore, they are prone to aggregate and crystallize on surfaces, which is a limitation to get a uniform distribution. However, various solution processing techniques have proved efficient: the simpler ones involve the exchange of the POM counter cations to form Langmuir-Blodgett films,[88, 89] Layer-by-Layer assemblies,[56, 90-93] or to deposit them on pre-assembled positively charged self-assembled monolayers by dip-coating.[94-97] Encapsulation in polymers has also been widely used.[98] Recent examples pertinent to the topic of this review will be presented in the following sections. A more elaborated immobilization route includes the use of organic-inorganic POM hybrids featuring remote and reactive organic functions to graft them covalently on a substrate. This can be achieved either in one step (direct grafting on the electrode)[99-104] or in two steps (cross-coupling reactions between a POM hybrid and a substrate functionalized with a complementary chemical function).[105-107] Note that the use of organically augmented POMs is not restricted to the assembly of a large number of POM units, since single-molecule devices have also been implemented through this route.[108] Pyrene derivatives of POMs have also been immobilized onto carbon nanotubes through π-π interactions.[109, 110] The tools of supramolecular chemistry on surfaces[111, 112] are still under-exploited in this context of the POM-shaping process onto electrodes. Examples of H-bonding to steer the formation of POM networks directly on surfaces[113] or the use of preformed molecular templates to periodically arrange individual POMs[114, 115] are still very scarce. This requires further effort towards POM functionalization and multi-functionalization. Finally, the development of soft-landing of mass-selected ions from the gas phase[116, 117] offers new opportunities to handle individual POMs.[118, 119] A more thorough description of substrate functionalization by POMs is out of the scope of this review but more details can be found in a recent review and references therein.[78]

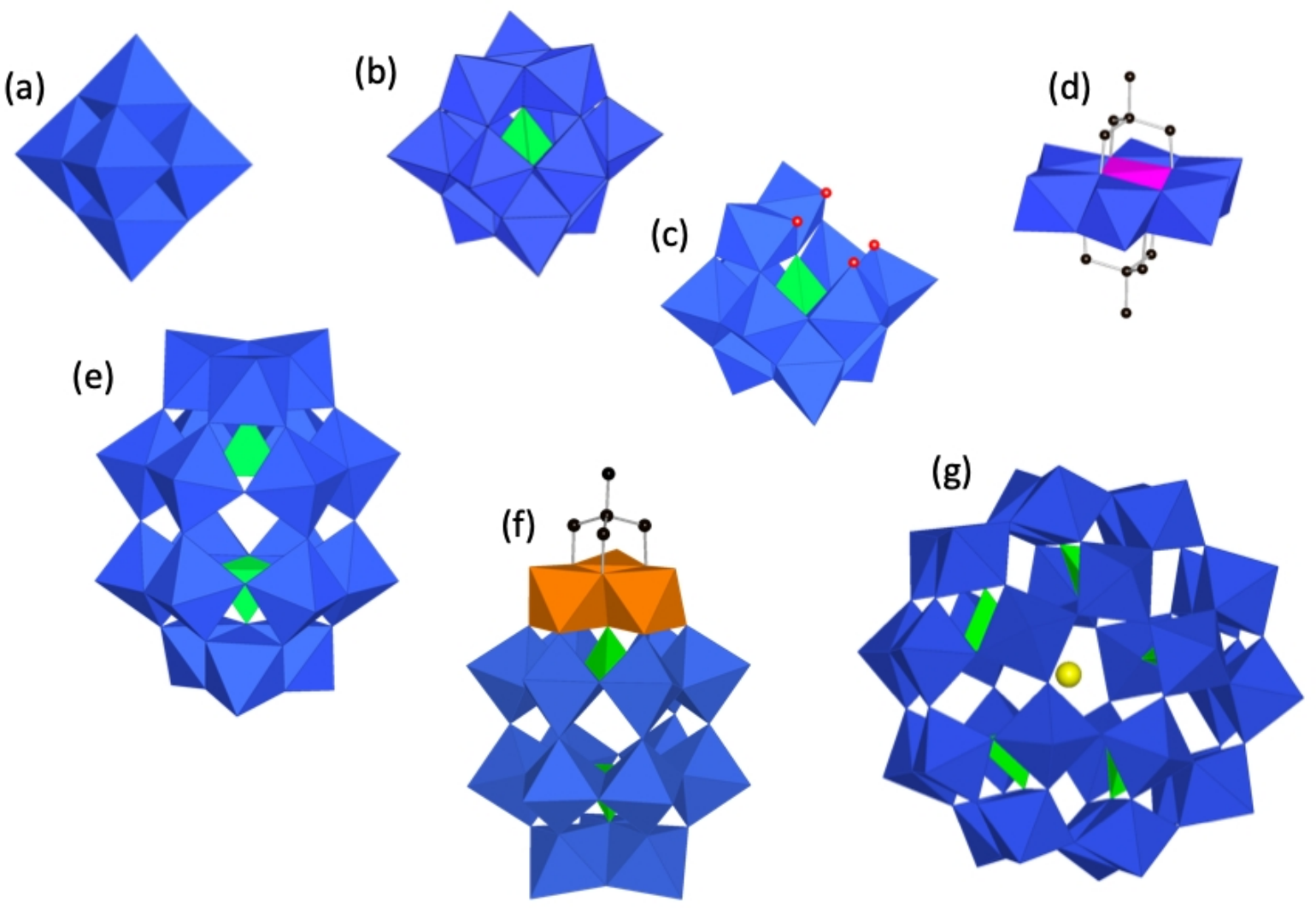

***Figure 1**. Selected examples of POMs and POM organic-inorganic hybrids:* ***(a)*** *Lindqvist* $[M_6O_{19}]^{2-}$, ***(b)*** *Keggin* $[XM_{12}O_{40}]^{n-}$, ***(c)*** *lacunary Keggin* $[XM_{11}O_{39}]^{(n+4)-}$, ***(d)*** *Anderson-Evans* $[MnMo_6O_{18}\{(OCH_2)_3CCH_3\}_2]^{3-}$, ***(e)*** *Wells-Dawson* $[X_2M_{18}O_{62}]^{p-}$ *and* ***(f)*** $[P_2W^{VI}_{15}V^{V}_3\{(OCH_2)_3CCH_3\}]^{9-}$,

***(g)*** *Preyssler $[NaP_5W_{30}O_{110}]^{14-}$ (X = P, As, Si, Al…. and M = Mo or W). Color code: blue octahedra, $\{WO_6\}$ or $\{MoO_6\}$; pink octahedra, $\{MnO_6\}$; orange octahedra, $\{VO_6\}$; green tetrahedra, $\{PO_4\}$.*

Compared to the general class of transition metal complexes used in molecular electronics,[39, 120-125] the electronic structure of POMs is akin to that of oxides: the highest occupied molecular orbitals (HOMO) are composed of non-bonding lone pairs on oxygens, while the lowest unoccupied orbitals (LUMO) are essentially non-bonding and are mainly composed of metallic d orbitals, like in the d band of oxides.[126, 127] These account for the redox properties of POMs that display successive and reversible reduction processes, which number generally exceeds that presented by conventional metal coordination complexes and organometallic molecules. The low structural reorganization due to the non-bonding character of the LUMO is favorable for long-term stability/endurance in programming/erasing cycles. Delocalization of the added electrons over part or whole of the POM skeleton (depending on the nature of the POM scaffold and the metallic centers) by ground state delocalization, as well as thermally- or photo-chemically activated electron hopping, has been substantiated by various spectroscopic techniques and is supported by theoretical calculations.[128-132] The unique set of POM properties, endless chemical diversity (structural type, metal addenda, heteroatoms, counter cations...) at the nanometric size, air stability and thermal robustness, compatible with CMOS technology, fine engineering of LUMO energy levels, existence of discrete, readily accessible and reversible multi-level quantum states for high density storage, and low-cost processability from solutions,[78] is driving a substantial interest in their implementation in nanoelectronic devices, even if some issues remain to be clearly understood (*e.g.*, such as the precise role of the unavoidable presence of counter-cations to the electron transport properties,[87, 133-135] *vide infra*). However, the full potential of POMs for in-memory, spintronics and quantum computing has not yet been realized.[136]

This review is organized as follows. Section 2 is devoted to discussing the basic electron transport mechanisms through POM-based systems, from thin films, monolayer-based molecular junctions and down to single molecule junctions. In section 3, we review the application of POMs in memory devices, while Section 4 is devoted to spintronics and quantum computing, and Section 5 to neuromorphic devices.

## 2. BASIC ELECTRON TRANSPORT PROPERTIES OF POLYOXOMETALATES.

In this section we review the basic electronic properties of POMs, from thin films, down to monolayers and single molecules. The device applications will be discussed in Sections 3 - 5.

**2.1. Thin films and bulk materials.**

In 2001, N. Glezos and coworkers presented a study on the electronic properties of POMs embedded in resists (polymers) for e-beam lithography, with the objective of developing materials with dual properties: a material suitable for active electronic devices (by virtue of the electronic properties of the POMs) that can be directly patterned without the need for additional lithographic steps.[137] The POM of choice was a Keggin $H_3(PW_{12}O_{40})$ embedded in the standard positive-tone polymer resist PMMA (poly(methyl methacrylate)) and a negative-tone copolymer PHECIMA (poly(2-hydroxyethylmethacrylate-*co*-cyclohexylmethacrylate-*co*-isobornylmethacrylate-*co*-acrylic acid).[138] Other polymers, like poly(vinyl alcohol), were also tested with less success. In a series of works[139-143] (mainly focusing on the $H_3(PW_{12}O_{40})$/PMMA system that is more stable than with the PHECIMA copolymer, which deteriorates with time), the same group further investigated the role of several parameters on the electron transport: the density of POMs in the polymer host (*i.e.*, the mean inter-POM distance), the device size and geometry (vertical *vs.* planar devices). The $H_3(PW_{12}O_{40})$/PMMA ratios were varied from 1:4 to 5:1 w/w, leading to a mean nearest neighboring distance of 2.8 nm to 0.5 nm between the POMs. However this was estimated from geometrical considerations based on the POM diameter of ≈ 1 nm and the POM density, since the exact organization of the POMs in the polymer matrix is not known (no TEM cross-section image, for example). To fabricate the planar devices, thick films (50-300 nm thick) were deposited between two electrodes (mainly Al) separated by a gap length L = 10 nm to 40 µm lithographed on an insulating substrate ($SiO_2$ 200 nm thick). For the vertical devices, thinner films (10-60 nm thick) were sandwiched between a bottom Al (on glass) or n-type doped silicon electrode and a top Al electrode of millimeter lateral size (evaporated through a stencil mask). The main results are summarized as follows. For planar devices, the current-voltage (I-V) curves are strongly non-linear with "steps" and "plateaus" (Fig. 2-a).[139] For $H_3(PW_{12}O_{40})$/PMMA concentration ratios ≥ 1:1 and electrode gaps L < 50-100 nm, 3 electron transport mechanisms were proposed:[139] (i) at low voltages (typically < 2-3 V), the electrons are sequentially tunneling from POM to POM via a variable range hopping (VRH) mechanism, considering the likely disorder of the POMs organization in the PMMA matrix. Depending on the electrode gap and the POM density, ≈ 5 to 10 POMs and tunneling events are involved in the conducting channel. Despite this, a standard "single" tunnel model (Simmons model[144, 145]) was used to fit the data, giving a mean tunnel energy barrier (combining the POM/PMMA and film/ metal electrode energy barriers) of ≈ 0.2-0.65 eV. Consequently, it is difficult to obtain a clear picture of the energetics of the POMs in these hybrid materials, *i.e.*, to precisely determine the energy position of the LUMO of the POM with respect to the electrode Fermi energy. (ii) The "step" in the I-V corresponds to a peak of the conductance (Fig. 2-a), which was ascribed to field-assisted Fowler-Nordheim tunneling between the adjacent POMs, since the electric field starts to be in the MV/cm range. Again the same fuzziness as above holds for the determination of the electronic structure of the POMs. (iii) At higher voltages (Fig. 2-a), the I-V curve follows a square voltage behavior typical of a space charge limited current (SCLC).[146-148] Increasing the voltage, more and more electrons can be trapped in the easily reduced POMs (POMs are electron sponges with many reduction states), creating a negative space charge in the material, which opposes electron transport (Coulomb repulsion). The electron mobility in this POM/PMMA material is low ($4x10^{-5}$ - $4x10^{-3}$ $cm^2V^{-1}s^{-1}$), well below the charge mobility of the best organic semiconductors (a few $cm^2V^{-1}s^{-1}$)[149] and this feature precludes their use in field-effect transistors. In all these studies, the device conductance remains below 1 nS. For L > 100 nm, only the SCLC behavior was observed. The same conclusions hold for $H_3(PW_{12}O_{40})$/PHECIMA [141] and $H_3(SiW_{12}O_{40})$/PMMA.[143] In the vertical devices, for films of $H_3(SiW_{12}O_{40})$ thinner than 100 nm, only the two tunneling regimes as for the planar devices (*vide supra*) were observed[140] (Fig. 2-b). A negative differential resistance behavior (*i.e.*, a decrease in the current when increasing the voltage, Fig. 2-c) was also observed for $H_3(PW_{12}O_{40})$/PMMA film with a thickness of 10 nm, but its physical origin remained elusive[140] (a cascade resonant tunneling through the POM molecular orbital was suggested but not firmly proven). We also note that all the films studied by these authors have been annealed after deposition (typically at 120 °C for 2 min) and it was observed that annealing can induce gap states at the POM/electrode interface that contribute (enhance) the conductance of the device (*vide infra*).[150] Albeit this latter result was observed for another POM and at 140 °C for 10 min and that the validity of a generalization remains to be explored, cautions are to be taken to determine the intrinsic electrical behavior of the POMs in that case. The use of Al electrodes, unavoidably covered by a thin native oxide, also complicates the interpretation of the I-V curves. A comparison of the same POM/PMMA films with Au electrodes (planar configuration)[143] showed that the currents are higher in the latter case (no oxide). The non-linear behavior of the I-V is still observed even if the characteristic features (I-V "steps", conductance peaks) are slightly voltage-shifted due to the difference in the energy alignment between the electrode Fermi energy and the POM molecular orbitals.

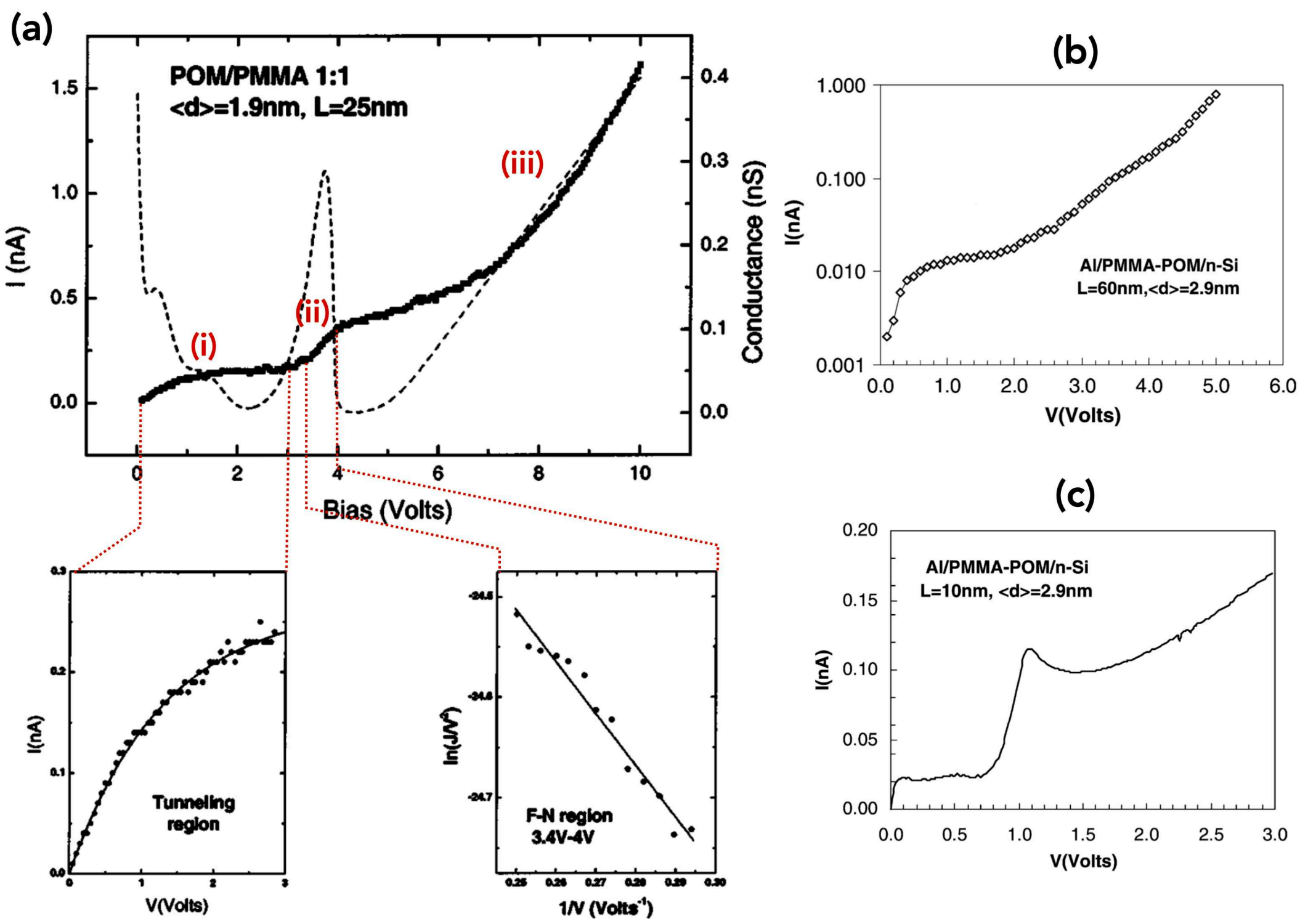

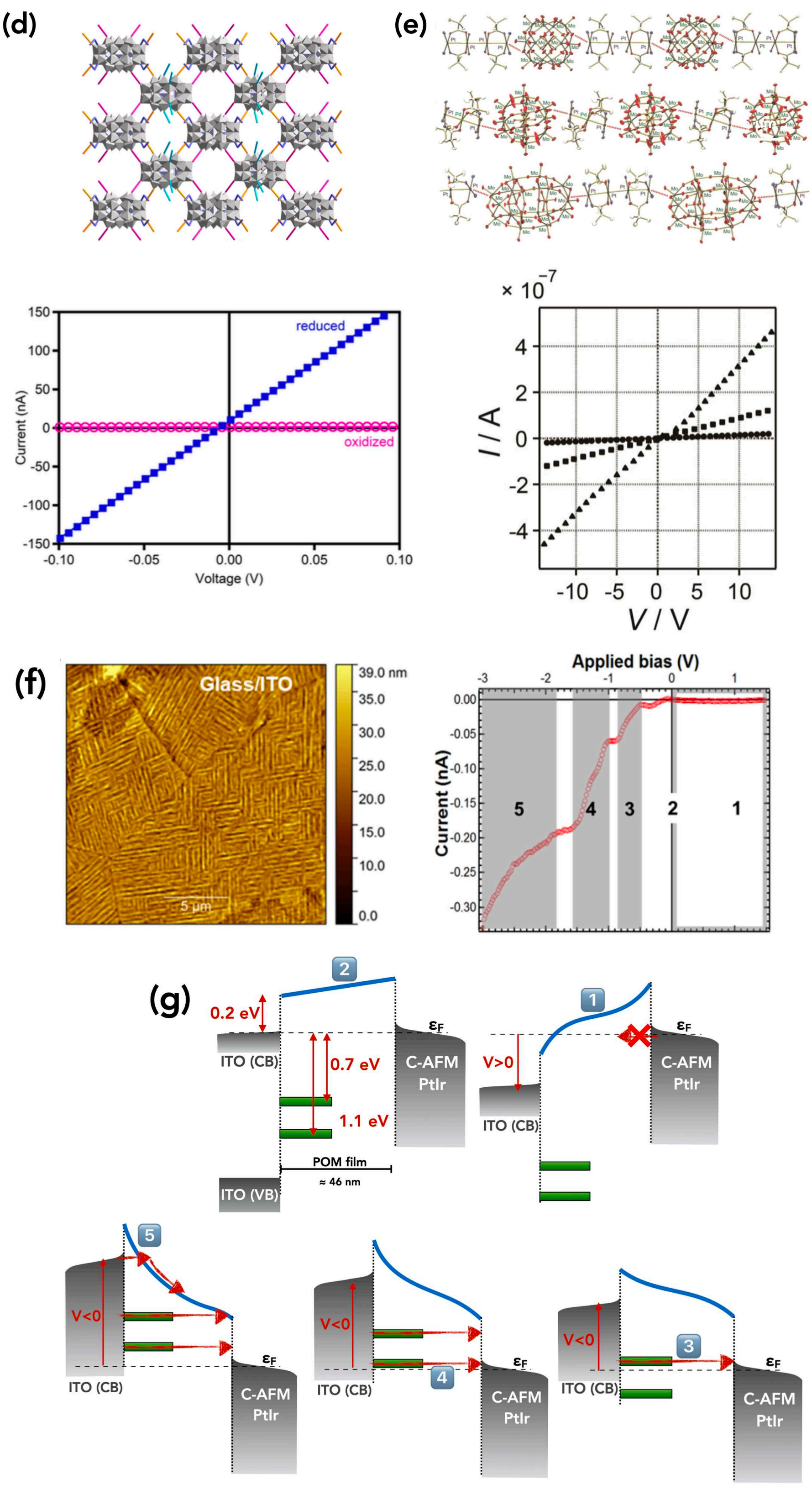

***Figure 2**. **(a)** Typical current-voltage (I-V, solid line) and conductance-voltage (dashed line) characteristics of Al-$H_3(PW_{12}O_{40})$/PMMA-Al planar device showing the three electron transport regimes (see text), with zoom on (i) I-V in the tunneling regime and (ii) the field-assisted Fowler-Nordheim tunneling regime, $ln(I/V^2)$ vs. 1/V plot. Reproduced with permission from ref. 139 Copyright (2003) American Institute of Physics. **(b-c)** Typical I-V characteristics for two n-type Si-$H_3(PW_{12}O_{40})$/PMMA-Al vertical devices with film thicknesses of 60 nm and 10 nm, respectively. Reproduced with permission from ref. 140. Copyright (2003) Elsevier. **(d)** Structural scheme of the $Co^{2+}$-linked $[NaP_5W_{30}O_{110}]^{14-}$ material and I-Vs of a single crystal in the oxidized and reduced states. Reproduced from ref. 151. Copyright (2019) American Chemical Society. **(e)** Crystal structures of $[\{PMo_{12}O_{40}\}\{Pt_4\}]_n$, $[\{PMo_{12}O_{40}\}\{Pt_2Pd\}]_n$ and $[\{P_2Mo_{18}O_{62}\}\{Pt_4\}]_n$ (top to down) and the corresponding I-Vs of Au/compressed pellet/Au of these materials: ● $[\{PMo_{12}O_{40}\}\{Pt_4\}]_n$, ■ $[\{PMo_{12}O_{40}\}\{Pt_2Pd\}]_n$, and ▲ $[\{P_2Mo_{18}O_{62}\}\{Pt_4\}]_n$. Reproduced with permission from ref. 152. Copyright (2024) John Wiley and Sons. **(f)** AFM topographic image and I-V curve (C-AFM) of a nanostructured (stripes) film of $K_6[P_2W_{18}O_{62}]$ deposited on indium tin oxide (ITO). Reproduced from ref. 150. Copyright (2019) American Chemical Society. **(g)** Suggested scheme of the electronic structure of the ITO/$K_6[P_2W_{18}O_{62}]$ stripe/C-AFM tip at various applied voltages shown in Fig. 2-f. A larger energy barrier is assumed at the POM/PtIr interface than at the POM/ITO interface, since the work function of PtIr (4.8-5.2 eV) is larger than ITO (4.2-4.8 eV). The red arrows symbolize the electron transport channel at negative voltages, redrawn from data in ref. 150.*

POMs in solid-state devices are intrinsically poor electron conductors because the delocalized d metal electrons are isolated within a single cluster with limited electronic interactions between neighboring anions in a solid-state film or crystal. To increase the intrinsic low conductance in POM-based devices, a strategy consists of connecting the POMs by judiciously selected moieties. Various chemical approaches are available using cationic metal ions, metal complexes coordinated with the O atoms at the periphery of the POMs,[87, 153] allowing the injected (*e.g.*, from the electrodes, from optical excitation) delocalized electrons to travel more easily from POM to POM, increasing the conductivity of the POM-based materials. Metal oxide framework single crystals made of Preyssler anions $[NaP_5W_{30}O_{110}]^{14-}$ linked by $Co^{2+}$ cations showed a remarkable large increase in conductivity (by ca. 500 - $10^3$) upon UV-light photo-reduction in the presence of a sacrificial reductant (here methanol).[151] The increase in conductivity was characterized by I-V measurements (Fig. 2-d) and impedance spectroscopy. This increase is due to delocalized electrons injected into the material during the photo-reduction. Up to 10 electrons per $[NaP_5W_{30}O_{110}]^{14-}$ anion (titration measurements) were added with no change in the crystal structure. Albeit this large increase, which is the highest reported value for POM-based films, the conductivity remains low at ≈ $10^{-4}$ S/cm for the reduced POMs. For comparison, the conductivity of a silicon crystal varies from $10^{-3}$ S/cm for a low-doped Si (dopant density ≈ $10^{13}$ $cm^{-3}$) to $10^3$ S/cm for a highly doped Si (≈ $10^{20}$ $cm^{-3}$).[154] Since the I-V curves were only recorded at very low voltages (-0.1 to 0.1 V) in the ohmic regime, no electronic structure of the device was proposed and it was not determined how the energy levels of the POMs molecular orbitals are modified upon reduction. In another approach, the POMs were bridged by multinuclear mixed-valence platinum complexes, leading to a $10^5$ times increase in material conductivity.[152] In this work, single crystals of Keggin-type $[PMo_{12}O_{40}]^{3-}$ or Dawson-type $[P_2Mo_{18}O_{62}]^{6-}$ bridged by tetranuclear $[Pt_4]^{5+}$ (for $[Pt_2(piam)_2(NH_3)_4]_2^{5+}$, piam = pivalamidate) or trinuclear $[Pt_2Pd]^{3+}$ (for $[Pt_2Pd(piam)_4(NH_3)_4]_2^{3+}$) were obtained and characterized in details (crystal structure, oxidation states, IR spectra, NMR, EPR, UV absorption, DFT calculations). Three single crystals were electrically measured (Fig. 2-e): $[\{PMo_{12}O_{40}\}\{Pt_4\}]_n$, $[\{PMo_{12}O_{40}\}\{Pt_2Pd\}]_n$ and $[\{P_2Mo_{18}O_{62}\}\{Pt_4\}]_n$ that have mixed-valence Pt(Pd) oxidation states and unpaired electrons on the metal $d_{z^2}$ orbitals, and in the case of $[\{PMo_{12}O_{40}\}\{Pt_4\}]_n$, two extra electrons delocalized on the $PMo_{12}$ framework. At room temperature, the conductivity ranges from $10^{-8}$ to $3x10^{-7}$ S/cm in the order $[\{PMo_{12}O_{40}\}\{Pt_4\}]_n$ < $[\{PMo_{12}O_{40}\}\{Pt_2Pd\}]_n$ < $[\{P_2Mo_{18}O_{62}\}\{Pt_4\}]_n$ (measured on pellets, the single crystals being too small for easy electrical connections) and $7x10^{-7}$ S/cm for a single crystal of $[\{P_2Mo_{18}O_{62}\}\{Pt_4\}]_n$. These values are ≈ $10^5$ times larger than those estimated for films made of the parent POMs without the metal complex bridges. These POM-platinum assemblies have a conductivity larger than other compounds (single crystals) of mixed-valence POMs ($[Mo^{V}_2Mo^{VI}_{16}O_{54}(SO_3)_2]^{6-}$) linked with tetra-alkylammonium cations (≈ $10^{-12}$ - $10^{-8}$ S/cm),[155] which is consistent with the electrical insulating behavior of saturated alkyl chains. A similar conductivity ($6.3x10^{-7}$ S/cm) was reported for single crystals of mixed-valence POMs ($[PMo^{V}Mo^{VI}_{11}O_{40}]^{4-}$ wired with cationic π-conjugated molecules (tetrathiafulvalene substituted by pyridinium)).[156] Higher conductivity ($10^{-4}$ S/cm up to ≈ 1 S/cm) was reported for a series of other crystalline TTF-type donor/POM hybrids,[157-159] in which the TTF-type donors and POMs form parallel columnar stacks and not a 3D framework of bridged POMs as in Refs. 152, 155, 156. In the former case, the high conductivity is ascribed to the preferential electron pathways through the π-conducting (TTF-type donor) columnar stacks. In the compounds reported in Refs. 152, 155, 156, temperature-dependent current measurements showed an Arrhenius behavior with activation energies in the window 0.3-0.6 eV. These energies were attributed to thermally activated electron hopping between adjacent POMs, the inter-POM distances being related to the chain length of the, insulating, tetra-alkylammonium cations in ref. 155. The improved conductivity reported in refs. 152 and 156 is to be ascribed to enhanced electronic interactions between the POMs and electro-active cations.

When a thin layer is deposited from solution on a solid substrate, the POMs are often not well organized in the film. Another important factor to optimize the conductivity is the structure and organization of the layer. For example, thick layers (46-150 nm thick) of $K_6[P_2W_{18}O_{62}]$ deposited on indium tin oxide (ITO) switch from insulating to conductive films upon structuration (induced by annealing) and the appearance of gap states below the POM LUMO.[150] The as-deposited films are insulating and the topographic AFM images show a rather disordered surface. After annealing (140 °C, 10 min in nitrogen atmosphere), the AFM images reveal the presence of a series of ordered stripes about µm long, with a height and width of ≈ 10 nm and ≈ 200 nm, respectively (Fig. 2-f). Local conductive-AFM (C-AFM) measurements on these stripes indicate a higher conductance. The UPS measurements have detected two gap states located at 0.7 and 1.1 eV below the Fermi energy. The C-AFM I-V curves are asymmetric (Fig. 2-f). At positive voltages (applied on the ITO, the PtIr tip being grounded), no current is detected due to the blocking behavior of the POM/PtIr tip interface and the POM film thickness (46 nm) that prevents electron tunneling injection from the tip to the ITO (Fig. 2-g). At negative voltages, the I-V shows 2 steps attributed to electron transport through the 2 gap states entering the energy window defined by the Fermi energy levels of the two electrodes, and then (at higher voltages) the current increases further due to the LUMO of the POMs that comes in resonance with the conduction band of the ITO electrode due to the energy band bending in the POM layer (Fig. 2-g). The physical origin of the gap states is not clear. The annealing-induced structuration ("stripes") of the $K_6[P_2W_{18}O_{62}]$ is a key factor (no gap state in the as-deposited film) but the nature of the counterions was also pointed out. Similar experiments with $Li_6[P_2W_{18}O_{62}]$ and $H_6[P_2W_{18}O_{62}]$ show that the density of the gap states is slightly reduced with $Li^+$ and even more with $H^+$. In this latter case, the POM layer is insulating. However the role and nature of the ITO substrate (presence of Sn and In) should have been examined by comparison with metal substrates (*e.g.*, Au, Pt). For instance, it was reported that POMs can be reduced in contact with Al and ITO electrodes.[160, 161] These features raise the question about the role of the substrate on the electron transport properties and device performances (*vide infra*, a brief discussion at the end of section 2.2.1).

**2.2. POM nanostructures, monolayers and single molecules.**

To better study the electronic properties of POMs, it is appropriate to work at the monolayer level and down to a single POM cluster. Monolayers (and few layers) of POMs have been deposited on various surfaces (Au, Si, $SiO_2$, ITO,...) by the Langmuir Blodgett (LB) technique,[88, 89] using a layer-by-layer (LbL) approach[56, 90-92] and by self-assembly (self-assembled monolayer, SAM).[94-96, 99, 101, 103, 104] The electronic properties of these devices have been characterized by gently contacting the monolayers with a Hg or eGaIn (eutectic GaIn) drop or by STM and conductive-AFM (C-AFM). In the former case, large-area electrode/POM/electrode are formed and thousands of POMs are measured in parallel. In the two latter cases, few molecules to few tens of molecules (depending on the size of the POM and of the STM or C-AFM tip) are electrically connected. To characterize a single molecule, there are also several approaches. The POM can be trapped in a nanometric gap between two electrodes made by e-beam lithography and electromigration or in a mechanically controlled break junction (MCBJ). In a sub-monolayer of molecules on metallic substrates, a single molecule can be contacted by STM (usually in an ultra-high vacuum or in a liquid environment) or using the STM-BJ variant (STM break junction). For details and the pros/cons of these various experimental techniques, see Refs. 4, 162-166. The metal/molecules/metal or semiconductor/ molecules/metal molecular junctions (MJs) are valuable devices to estimate the electrical conductance of the POMs and to determine the electronic structure of the MJs, i.e., to position the POM molecular orbitals with respect to the electrode Fermi energy.[167] This goal is achieved by analyzing the measured I-V curves with various charge transport mechanisms (coherent off-resonant tunneling, coherent resonant tunneling, and incoherent tunneling), see a review of these electron transport mechanisms in Refs. 38, 162, 163, 165, 168-171. In POM-based MJs, the electron transport is driven by the LUMO, i.e., by the non-bonding d orbitals of the metal atoms, while in MJs based on metal coordination complexes and organometallic molecules, it can be HOMO- or LUMO-driven, depending on the details of the dπ-pπ bonding interactions. In purely organic MJs, in most of the cases, HOMO-driven electronic transport dominates (π bonding orbitals) for donating molecules, and it is the LUMO (π* anti-bonding orbitals) for accepting molecules (albeit these general rules can be broken depending on the nature and strength of the metal/molecule interactions, *e.g.,* via the nature of the anchoring groups).

#### *2.2.1. Vertical devices.*

*Role of the transition metals and the central heteroatom.*

We first review several studies on the archetype Keggin POM $[PM_{12}O_{40}]^{3-}$ with M=W or Mo embedded in several MJs on Si substrates or Au electrodes with large-area top contacts (evaporated Al, Hg drop or eGaIn drop) or at the nanoscale with a C-AFM tip. The typical I-Vs are shown in Fig. 3 and the assessed electronic structures are summarized in Fig. 4. A monolayer of $H_3(PW_{12}O_{40})$ was immobilized by electrostatic interactions on a silicon/silicon oxide substrate functionalized by a SAM of APTES (3-aminopropyl triethoxysilane).[172] The Si substrate (n-type highly doped) is covered by a native ultra-thin layer of silicon oxide ($SiO_x$, usually non-stoichiometric, $x \leq 2$) and the top electrode was an Al pad evaporated through a stencil mask (large area MJs). The I-V (Fig. 3-a) was analyzed by the standard off resonant tunnel model (Simmons model[144, 145]) and an effective tunnel barrier $\Phi_T$ = 0.44 eV was deduced (Fig. 4-a). A value $\Phi_T$ = 0.26 eV (Fig. 4-b) was reported by the same authors for the same APTES/$PW_{12}$ MJs on a n-type Si substrate (medium doped) but with a thin layer of thermally grown silicon dioxide ($SiO_2$, 3.5 nm thick).[173] When the substrate is a p-type doped Si, a larger value was obtained, $\Phi_T$ = 0.93 eV (Fig. 4-c). Without the POMs, the tunnel barrier for the Si/$SiO_2$/APTES/Al MJ was found to be $\Phi_T \approx$ 1.6 eV.[173] Since the MJs is constituted by a stack of three layers, the measured effective tunnel barrier $\Phi_T$ is an average of the oxide barrier ($\Phi_{ox}$, not measured independently in these works), the APTES barrier ($\varepsilon_{C3}$) and the POM tunnel barrier assigned to the relative position of the LUMO ($\varepsilon_{POM}$) with respect to the Fermi energy of the electrodes (Figs. 3-a to 3-c). Thus, it is difficult to precisely determine the energy position of $\varepsilon_{POM}$, which is close to the Fermi level in these works. It is notable that the difference of $\Phi_T$ between the same MJs on p-type and n-type Si ($\approx$ 0.67 eV) can be explained by the shift of the Si Fermi energy (see Figs. 3-b and 3-c) that is 0.6-0.7 eV,[154] if we assume a standard doping level of $10^{15}$-$10^{17}$ $cm^{-3}$ for the two Si substrates (the exact value was not reported in refs. 172 and 173). Given these complex stacked device structures, it is difficult to distinguish the difference of the intrinsic electronic structure between a $[PW_{12}O_{40}]^{3-}$ and a $[PMo_{12}O_{40}]^{3-}$ POMs in these MJs.

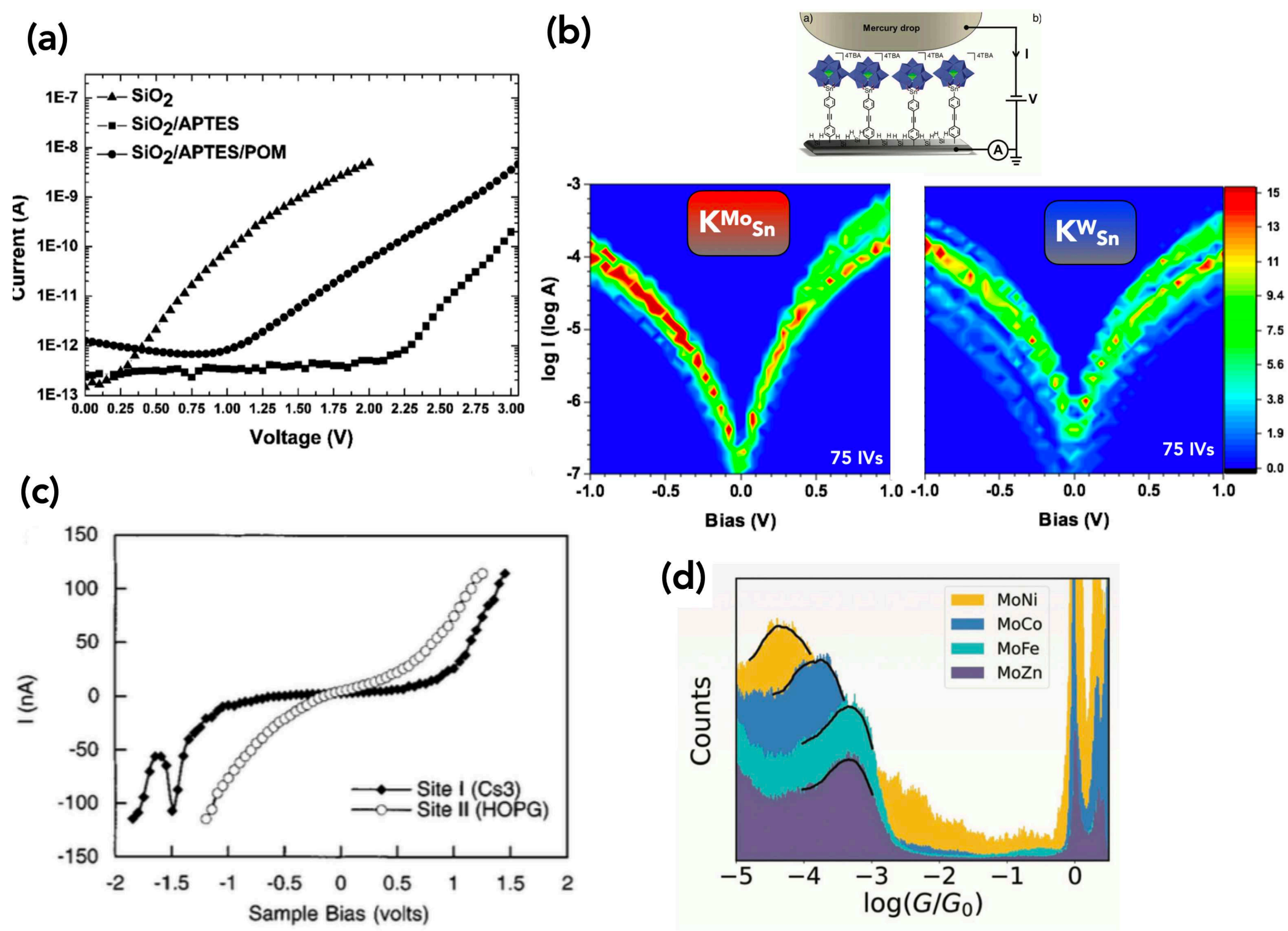

***Figure 3**. **(a)** Current-voltage (I-V) characteristics of the Si-$SiO_2$-$Si(CH_2)_3(NH_2/NH_3^+)$-$H_3(PW_{12}O_{40})$ MJs and comparison with the APTES functionalized and naked Si/$SiO_2$ substrate. Reproduced with permission from ref. 173. Copyright (2008) John Wiley and Sons. **(b)** 2D histograms (heat map) of the I-Vs measured on Si-$K^{W}_{Sn}$-Hg and Si-$K^{Mo}_{Sn}$-Hg MJs (where $K^{W}_{Sn}$ and $K^{Mo}_{Sn}$ stand for $TBA_4[PM_{11}O_{39}\{Sn(C_6H_4)C{\equiv}C(C_6H_4)\}]$ with M = W or Mo). Reproduced with permission from ref. 104. Copyright (2018) Royal Society of Chemistry. **(c)** Typical STM/STS I-Vs recorded on HOPG (open circles) and on HOPG/$Cs_3[PMo_{12}O_{40}]$ (dark diamonds). Reproduced from ref. 174. Copyright (1996) American Chemical Society. **(d)** Histograms of conductance (measured by STM-BJ) for $TBA_n[MMo_6O_{18}\{(CH_2O)_3CNH_2\}_2]$ (with a central metal atom M = $Fe^{III}$ or $Co^{III}$ with n=3; M = $Ni^{II}$ or $Zn^{II}$ with n=4), in units of the conductance quantum $G_0$=7.75x$10^{-5}$ S. Reproduced from ref. 175. Copyright (2021) Royal Society of Chemistry under Creative Commons CC BY-NC 3.0*

The same type of Keggin POMs was also directly chemisorbed on hydrogenated Si surfaces (Si-H surface, no oxide) by our groups.[104] Two POMs were synthesized with a terminal diazonium function: $TBA_3[PM_{11}O_{39}\{Sn(C_6H_4)C{\equiv}C(C_6H_4)N_2\}]$ with M = W or Mo, referred to as $K^{W}_{Sn}[N_2]$ and $K^{Mo}_{Sn}[N_2]$, respectively (K stands for Keggin), with TBA (tetrabutylammonium, $N(C_4H_9)_4^+$) as the counter-cation. A compact monolayer on highly doped n-type Si(100) substrate was obtained in the two cases (a density of *ca.* $6x10^{13}$ POM/$cm^2$, CV measurements)[101] free of gross defects (pinholes, aggregates), a surface rms (root-mean square) roughness of 0.15-0.19 nm (tapping AFM images) and a thickness of 2.7 and 3.2 nm for the $K^{W}_{Sn}$ and $K^{Mo}_{Sn}$ monolayers (ellipsometry). The MJs were contacted by a hanging Hg drop technique (large area MJs) to measure the I-V characteristics (Fig. 3-b) and analyzed by the off resonant tunneling model (Simmons model[144, 145]). The deduced effective energy barriers (Fig. 4-d) are $\Phi_T = 1.80 \pm 0.26$ eV and $1.60 \pm 0.35$ eV for the $K^{W}_{Sn}[N_2]$ and $K^{Mo}_{Sn}[N_2]$ MJs, respectively. This energy was attributed to the LUMO of the POM ($\varepsilon_{POM}=\Phi_T$) neglecting the weak tunnel barrier of the Hg oxide (experiments done in a glove box). This energy level shift is in agreement with the same trend observed for the electron affinity of the molecules in solution determined at -3.6 eV ($K^{W}_{Sn}[N_2]$) and -4.1 eV ($K^{Mo}_{Sn}[N_2]$) versus the vacuum level and with the known property that molybdates are easier to reduce than their analogous tungstates.[103, 176]

A similar trend was observed from STM/STS measurements (in air and at room temperature) on a series of POMs $H_n[XW_{12}O_{40}]$ and $H_n[XMo_{12}O_{40}]$ (with X = $P^{5+}$, $Si^{4+}$, $B^{3+}$, $Co^{2+}$ for the tungstate POM and X = $P^{5+}$, $Si^{4+}$ for the molybdate POM) deposited on highly oriented pyrolytic graphite (HOPG) and forming an ordered 2D array of POMs.[177] The STS measurements (Fig. 3-c) revealed the systematic presence of a negative differential resistance (NDR) peak in the I-V (only at negative voltages applied on the substrate), the voltage position $V_{NDR}$ of this peak increasing (less negative) with a less negative reduction potential of the POMs and the increase of the electronegativity of the heteroatoms.[177] The origin of the NDR peak was not clearly identified. If we assume that the NDR peak comes from a resonant electron transfer between an energy localized density of states of HOPG and the LUMO of the POMs, a more negative $V_{NDR}$ corresponds to a LUMO moving away from the Fermi energy of the electrodes (*i.e.*, lower electron affinity energy), i.e., a more negative reduction potential for the polyoxotungstate POMs compared to the polyoxomolybdates,[177, 178] as also reported for POMs chemisorbed on Si (*vide supra)*[104].

These two results demonstrate that the fingerprint of the chemical structure of the POM is clearly retained in the solid-state MJ and they unravel the role of the POMs in the electron transport properties.

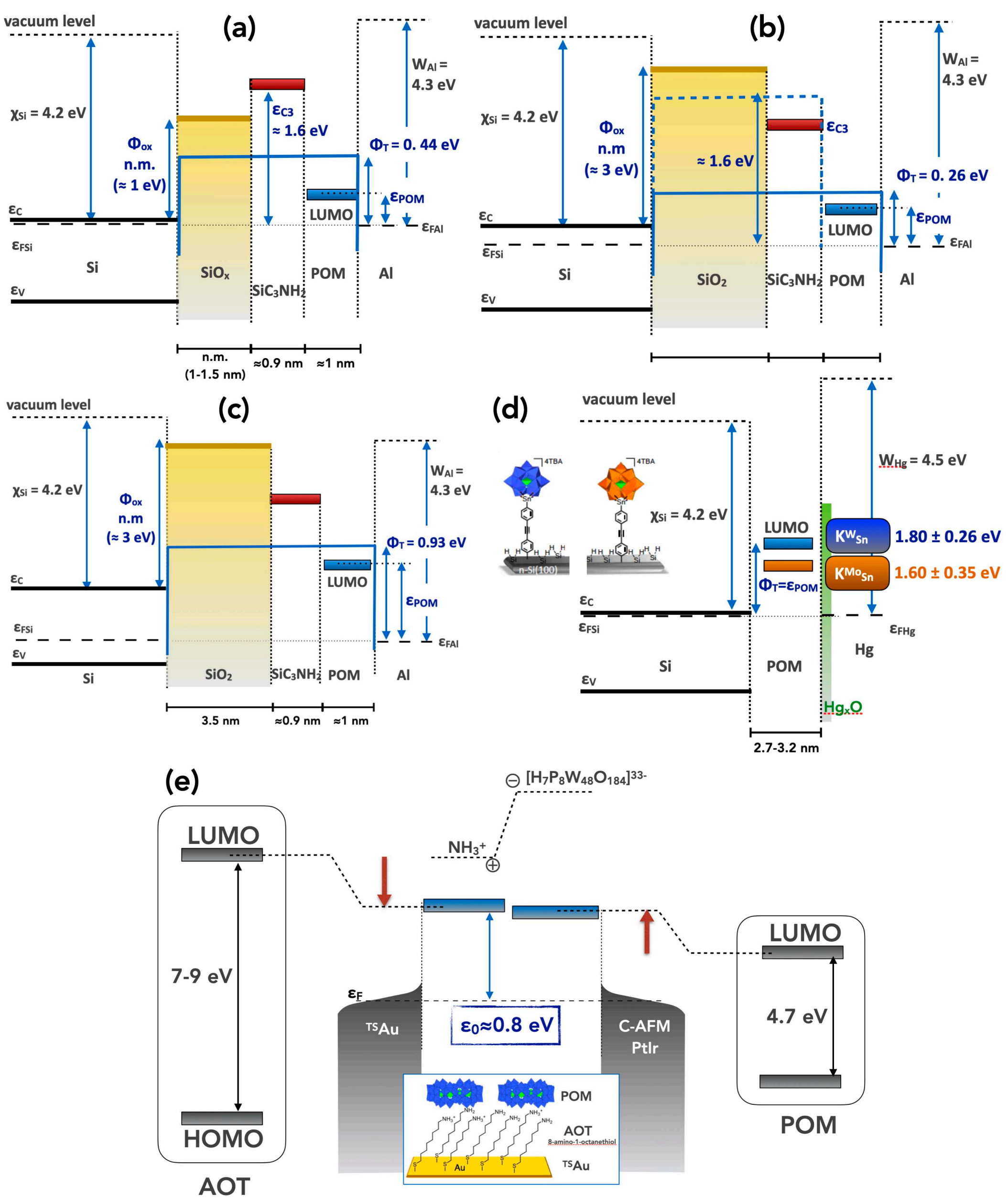

***Figure 4***. *Electronic schemes of silicon(or metal)-POM-metal MJs.* ***(a-c)*** *Si-SiO$_{x(x\leq 2)}$-Si(CH$_2$)$_3$(NH$_2$/ NH$_3^+$)-H$_3$(PW$_{12}$O$_{40}$) MJs. Data taken from Refs.172, 173. Several parameters (indicated as n.m.) were not measured nor given in these reports, the values in brackets are the usually admitted values. Color code: thick red line = LUMO of alkyl spacer, thick blue line = LUMO of POM, thick yellow line = conduction band of silicon oxide, solid and dashed blue lines represent the measured tunnel barriers.* ***(d)*** *Si-K$^{W}_{Sn}$-Hg and Si-K$^{Mo}_{Sn}$-Hg MJs (where K$^{W}_{Sn}$ and K$^{Mo}_{Sn}$ stand for*

*$TBA_4[PM_{11}O_{39}\{Sn(C_6H_4)C{\equiv}C(C_6H_4)\}]$ with M = W or Mo), data from ref. 104.* ***(e)*** *$^{TS}Au$-S-$(CH_2)_8$-($NH_2$/$NH_3^+$)-$K_{28}Li_5[H_7P_8W_{48}O_{184}]$-PtIr tip MJs, data from ref. 97.*

The role of the metal atoms in the electron transport has been observed for other POMs. The Anderson-type POMs $TBA_n[MMo_6O_{18}\{(CH_2O)_3CNH_2\}_2]$ (with a central metal atom M = $Fe^{III}$ or $Co^{III}$ with n=3; M = $Ni^{II}$ or $Zn^{II}$ with n=4) were measured at a single cluster level by STM-BJ.[175] The electrical conductance (Fig. 3-d) of the Au/POM/Au MJs increases with M (about a decade, from ≈ $6.1x10^{-5}G_0$ to $4.1x10^{-4}G_0$ ($G_0$ is the conductance quantum ≈ 77.5 µS) in the order G(Ni)<G(Co)<G(Zn)≈G(Fe).[175] The authors have also measured the Seebeck coefficient (or thermopower) of the MJs. They have found that the Seebeck coefficient also depends on M:[175] it is positive (≈ 15-17 µV/K) except for the Co-POM ≈ - 9 µV/K. A positive Seebeck coefficient is the fingerprint that the HOMO level is close to the Fermi energy (LUMO for a negative value of the Seebeck coefficient).[179, 180] This differs from the previous examples in which the electron transport is mediated by the POM LUMO, which is mainly developed on the metal addenda (Mo/W) $d_{xy}$ orbitals and which is the molecular orbital the closest to the electrode Fermi energy (*vide supra*). In Anderson-type POMs, the redox processes are first driven by the central transition metal cations, as confirmed by DFT calculations.[181] The HOMO mediation electron transport was ascribed to the amine anchoring groups as also observed for π-conjugated molecule MJs, which induce charge transfer at the molecule/electrode contact and shift of the molecular orbitals compared to the same molecular MJ with thiol anchoring groups.[182] The negative Seebeck coefficient observed for the Co-POM is explained by the presence of a dip in the energy-dependent electron transmission coefficient T(E) (destructive quantum interferences[165, 183]) near the HOMO that reverses the sign of the derivative of the logarithm of the transmission function near the Fermi energy (the Seebeck value is proportional to $\partial \ln T(E)/\partial E$).[184]

Finally, we note that the energy position of the LUMO, $\varepsilon_{POM}$, in our works (Figs. 2-d) is larger than in the earlier experiments (Figs. 2-a, 2-b and 2-c), albeit the polyoxotungstates belong to the same Keggin-type family. However, we note several differences. In our work, after the release of the diazonium function, the POMs bear a 4- total charge, instead of 3- in the earlier experiments. Another difference is that these earlier experiments have used evaporated Al top electrodes. It is known that Al can diffuse inside the monolayers of molecules,[185, 186] creating nanofilaments, which artificially increase the currents through the MJs and consequently induce lower tunnel barriers when analyzed by the tunneling models. One can also not exclude some reduction of the POMs by the aluminum top electrode, as it has been disclosed.[160, 161] The use of Hg drop or eGaIn drop electrodes suppresses this drawback.[163, 187-189] Another difference between these two sets of experiments is the presence, in the earlier experiments, of positively charged ammonium ions at the interface with the negatively charged POMs. Such interface dipoles can also modify the energy level alignment in the MJs.[190, 191] However, when $H_3[PW_{12}O_{40}]$ and $H_3[PMo_{12}O_{40}]$ POMs were deposited on positively charged APTES SAM on a highly doped Si substrate (covered with its native oxide) and the I-Vs were measured by Hg drop, LUMO energy levels at ≈ 1.7 eV and ≈ 1.4 eV were obtained, respectively.[192] Thus, the trends are consistent between the covalent and electrostatic methods to form the MJs. We have also used electrostatic interactions to deposit large ring-shaped $K_{28}Li_5[H_7P_8W_{48}O_{184}]$ POMs on an ultra-flat Au electrode (template stripped, $^{TS}$Au)[193-195] functionalized with a SAM of 8-amino-1-octanethiol (AOT) for which about 62% of the terminal groups are protonated ($-NH_3^+$) according to XPS measurements.[97] The choice of this POM was motivated by its high stability and high charge storage capability (up to 8 electrons), such characteristics being interesting for memory applications (*vide infra*). Analyzing the I-V curves (C-AFM measurements) with various models (Simmons tunneling model and an analytical molecular model derived from the Landauer formalism), we concluded that the LUMO of the AOT and the POM are almost aligned at ≈ 0.8 eV (Fig. 4-e)[97] above the Fermi energy despite their large difference of electron affinity (≈ -1 eV for alkylthiols[196, 197] and ≈ -4 eV for this POM[97]). The dipole at the interface AOT/POM (positive charges at the AOT amine groups and negative charges on the POMs) shifts the AOT LUMO downstairs and the POM LUMO upstairs, leading to this LUMO level alignment.

*Role of the linkers to the electrodes and the functional ligands.*

The nature of the linker between the POM and the electrode also plays a key role. To this end, we have compared the electron transport properties of MJs made with two POMs that differ by the chemical nature of the linker: $TBA_{4.4}[PW_{11}O_{39}\{Sn(C_6H_4)C\equiv C(C_6H_4)COOH_{0.6}\}]$ and $TBA_{3.4}[PW_{11}O_{39}\{O(SiC_2H_4COOH_{0.8})_2\}]$ (hereafter referred to as $K^W_{Sn}$[COOH] and $K^W_{Si}$[COOH]).[198] The COOH anchoring groups allow the formation of a monolayer on a highly doped n-type Si substrate covered by its native oxide (1.4 nm thick) and the top contact was a Hg drop. Following the same protocol as for $K^W_{Sn}$ and $K^{Mo}_{Sn}$ (*vide supra*) for the measurement and the analysis of the I-V curves (Fig. 5-a), we obtained $\Phi_T$ = 1.75 ± 0.12 eV and 1.39 ± 0.23 for the $K^W_{Sn}$[COOH] and $K^W_{Si}$[COOH] MJs, respectively (Fig. 5-c). We note a good agreement (1.80 eV and 1.75 eV) for the $K^W_{Sn}$ and $K^W_{Sn}$[COOH] MJs with the same linker but with a different nature of the anchoring group and the presence of the native oxide in the latter case. The effective tunnel barrier $\Phi_T$ is composed of the energy barrier of the oxide ($\Phi_{ox}$ ≈ 1.9 eV, measured for a reference sample without the POMs) and the energy of the POM LUMO (Fig. 5-c). Using a simple staircase energy barrier model,[97, 198] we ascribed the energy level of the POMs, $\varepsilon_{POM}$, at ca. 1.7 and 1.1 eV above the electrode Fermi energy for the $K^{W}_{Sn}$[COOH] and $K^{W}_{Si}$[COOH] MJs, respectively. The same LUMO shift of ca. 0.6 eV was also obtained from CV measurements and by DFT calculations and has been related to the difference in the POM total charge, -4 for $K^{W}_{Sn}$[COOH] and -3 for $K^{W}_{Si}$[COOH].[198]

When the POMs are functionalized by ligands on their periphery, the electronic properties of the POM-ligand complexes can be quite different, depending on the strength of the POM-ligand coupling. Monolayers of two dodecavanadate POMs (abbreviated as {$V_{12}$}) functionalized by phthalocyanine and lanthanide ($TBA_3[HV_{12}O_{32}Cl(LnPc)]$ or $TBA_2[HV_{12}O_{32}Cl(LnPc)_2]$, with Pc = phthalocyanine, Ln = lanthanide) were deposited on ultra-flat $^{TS}$Au substrates and electrically contacted with an eGaIn top contact (large area MJ).[199] Several MJs were compared: LnPc, {$V_{12}$}, Ln-{$V_{12}$}, PcLn-{$V_{12}$} and $(PcLn)_2$-{$V_{12}$} with a large variety of $Ln^{3+}$(Sm, Eu, Gd, Dy, Ho, Er, Yb, Y, Lu). The I-Vs (Fig. 5-b) were analyzed by the transition voltage spectroscopy (TVS) method to determine the energy position of the molecular orbital involved in the electron transport with respect to the electrode Fermi energy.[171, 200-202] The main conclusion was that the Ln has no significant effect on the electron transport in the presence of the Pc moiety and that the electron transport is HOMO-mediated, the HOMO being localized on the Pc moiety (DFT calculations in the gas phase) and the experimental values lie at around 0.17 eV below the Fermi level whatever the Ln (Fig. 5-d). This was rationalized because the Pc ligands are directly in contact with the electrodes (according to NEXAFS spectra) favoring electron transport through the π-conjugated pathway of Pc. On the contrary, without the Pc, the effect of Ln is no longer "shielded" and the electron transport is mediated via the LUMO of the Ln-{$V_{12}$} POM located at ca. 0.35 eV (for Ln = Gd, Ho or Yb) and slightly lower (≈0.26 eV) for Ln = Dy (Fig. 5-d). Albeit not well captured in the energetics of the MJs by the TVS method, the estimated LUMO being weakly dependent on Ln (Fig. 5-d), the shapes of the I-Vs for the Ln-{$V_{12}$} MJs significantly depend on the type of Ln with an asymmetric I-V (*i.e.*, more current at positive than negative voltages) more pronounced for the Gd-{$V_{12}$} MJ and even an inverted asymmetry ($I(V<0) > I(V>0)$) for the Dy-{$V_{12}$} MJ (Fig. 5-b). These differences in the I-V shapes are likely related to the details of the atomic configuration of the contact between the POM and the electrodes.[133] We also note that the energy positions (Fig. 5-d) are quite close to the Fermi level, but it is known that the use of an eGaIn electrode (compared to a metallic C-AFM tip for example) underestimates these values, likely due to hybridization of the molecular orbitals with oxide states in the unavoidable ultra-thin $Ga_2O_3$ oxide covering the eGaIn drop.[203]

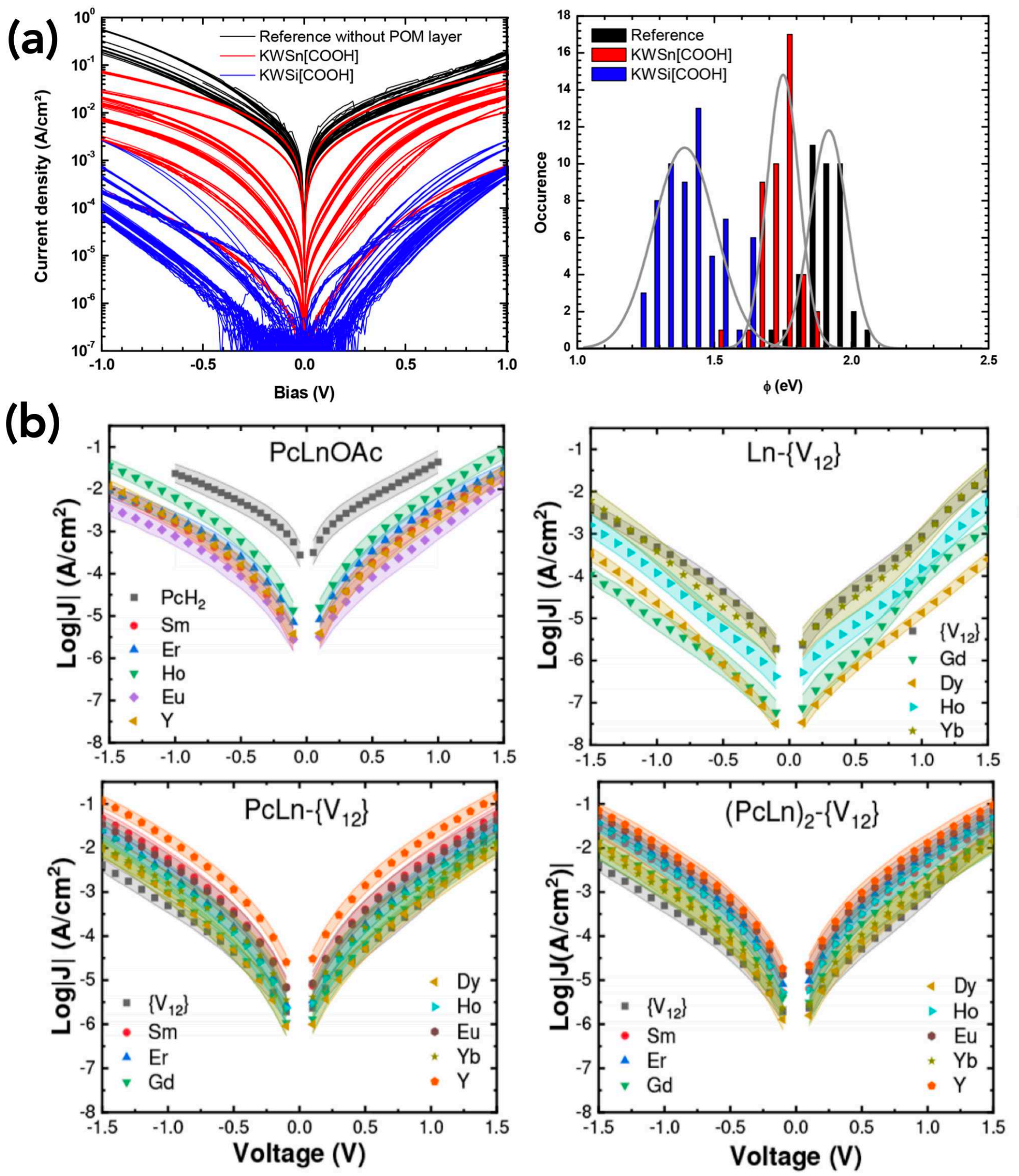

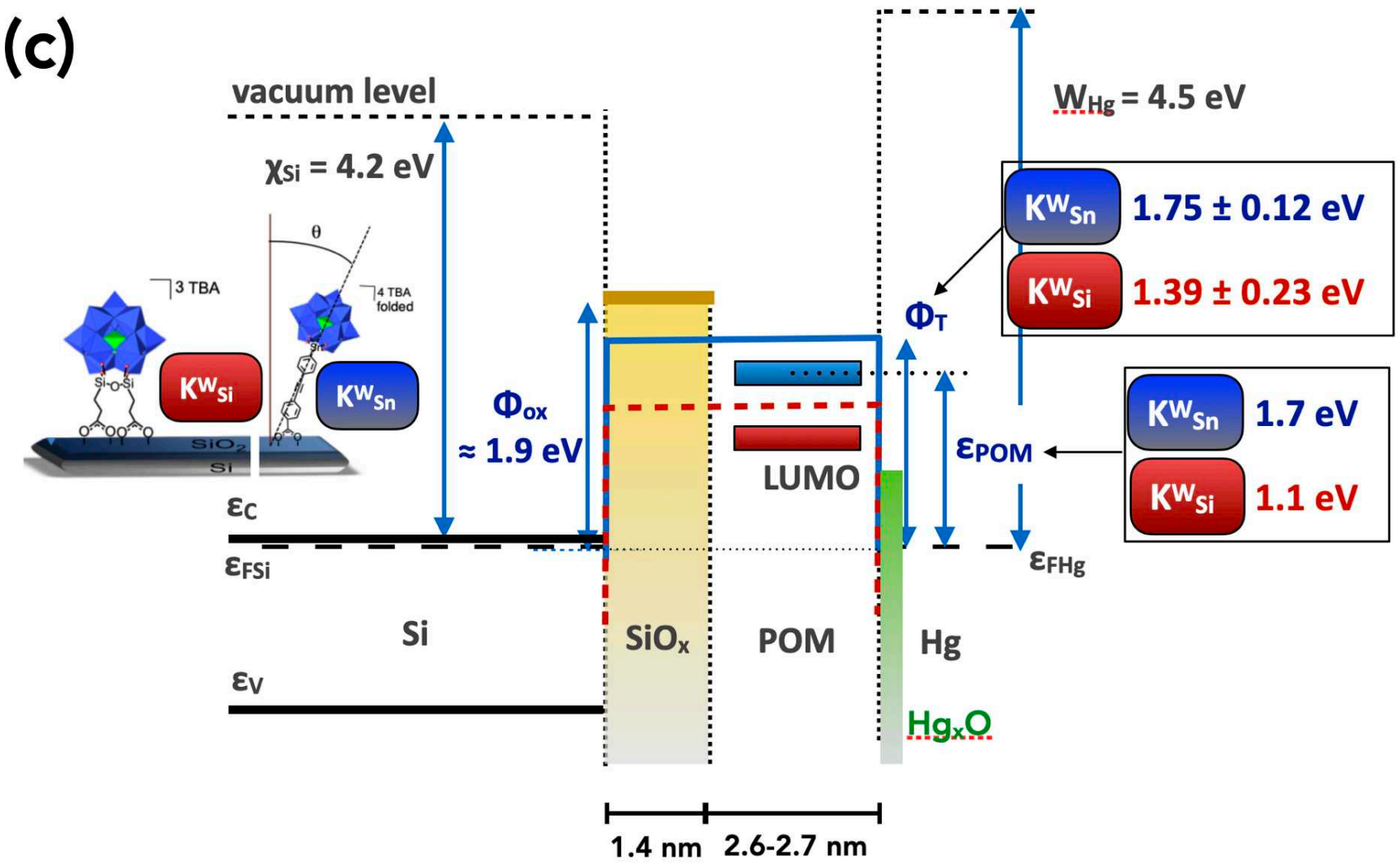

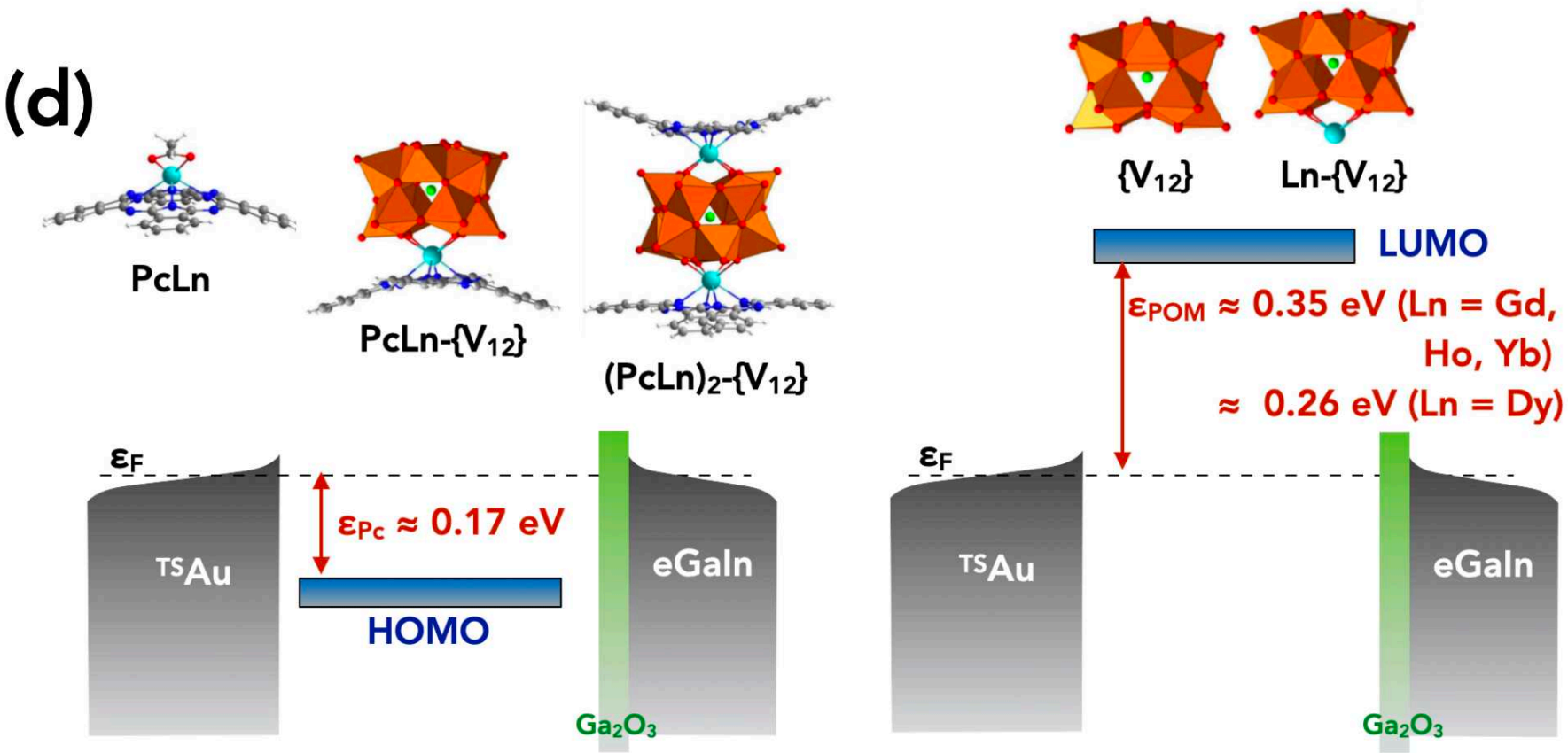

***Figure 5***. ***(a)*** *I-Vs of n-type Si/SiO$_x$-TBA$_{4.4}$[PW$_{11}$O$_{39}${Sn(C$_6$H$_4$)C≡C(C$_6$H$_4$)COOH$_{0.6}$}] and TBA$_{3.4}$[PW$_{11}$O$_{39}${O (SiC$_2$H$_4$COOH$_{0.8}$)$_2$}] (for short K$^W{}_{Sn}$[COOH] and K$^W{}_{Si}$[COOH])-Hg drop MJs and of the naked substrate, in red, blue and dark, respectively. Reproduced from ref. 198. Copyright (2020) American Chemical Society.* ***(b)*** *I-Vs of a series of dodecavanadate POMs (abbreviated as {V$_{12}$}) functionalized by phthalocyanine and lanthanide self-assembled on $^{TS}$Au and contacted by eGaIn. Reproduced from ref. 199. Copyright (2023) American Chemical Society.* ***(c)*** *Electronic structures of the Si/SiO$_x$-K$^W{}_{Sn}$[COOH]) and (K$^W{}_{Si}$[COOH]-Hg, data from 198.* ***(d)*** *Electronic structures of $^{TS}$Au-LnPc or {V$_{12}$} or Ln-{V$_{12}$} or PcLn-{V$_{12}$} or (PcLn)$_2$-{V$_{12}$}-Ga$_2$O$_3$-eGaIn MJs with {V$_{12}$} = [HV$_{12}$O$_{32}$Cl]$^{4-}$ and TBA counter-cations, Pc = phthalocyanine, Ln = lanthanide. The energy levels of molecular orbitals are calculated from TVS (transition voltage spectrocopy)*

*measurements with the simplified expression[202] $\varepsilon \approx 0.86\ V_{trans}$ using the average value of the ambipolar transition voltages $V_{trans}$ reported in 199.*

The ligands also have an important effect on the redox properties of the POMs; they are discussed in subsection "role of the redox states" (*vide infra*).

*Role of the counter-cations.*

Another important constituent of all the POM-based devices is the unavoidable counter-cation, since the POMs are anionic moieties. The POM-cation interactions play a key role in POM chemistry, assembly of POM-based materials and they can foster new properties and applications of POMs (see a review in ref. 87). These POM-cation interactions have recently attracted more attention in the electronic properties of POM-based devices.[134] The role of the cation is not just counterbalancing the charge of the anionic POM. The question is to clearly understand how the cations modify the electrostatic and energetics landscape in the MJs and whether or not they add an additional channel to transfer electrons between the electrodes. The I-V curves of a single $[W_{18}O_{54}(SO_3)_2]^{4-}$ POM connected by two Au electrodes were simulated by DFT combined with non-equilibrium Green's function method, for two counter-cations: $Cs^+$ and $TMA^+$ (tetramethyl amonium).[133] Compared to the I-V curve calculated without cations, the currents are increased with the cations, at positive bias the highest current is obtained with $TMA^+$, and at negative bias, $Cs^+$ cations induce the highest currents (Fig. 6-a). In other words, the I-V for the POM-TMA MJ is almost symmetric, while the one for the POM-Cs MJ shows a rectification behavior with more current at negative voltages. These calculations show that the LUMO levels of the POMs, with the counter-cations, are closer to the Fermi energy of the electrodes, increasing the electron transmission coefficient at the Fermi level and the current passing through the POM MJs (Fig. 7-a). However, the presence of the cations does not create new conduction channels in the MJs. The difference in I-V shape (symmetric with $TMA^+$ vs. asymmetric with $Cs^+$) is less clear. This may come from subtle differences in how the electric field landscape in the junctions is modified by the counter-cations (such types of effects have been observed in MJs with redox organic molecules[204-206]). On the experimental side, STM/STS measurements (in air and at room temperature) on a series of Keggin-type POM $(M^{n+})_{3/n}[PMo_{12}O_{40}]$ (with M = $H^+$, $K^+$, $Cs^+$, $Cu^{2+}$, $Co^{2+}$, $Zn^{2+}$, $Ba^{2+}$,$Mg^{2+}$, $Bi^{3+}$) monolayers deposited on highly oriented pyrolytic graphite (HOPG), revealed the systematic presence of a negative differential resistance (NDR) peak in the I-V (only at negative voltages applied on the substrate) - *vide supra* Fig. 3-c, the voltage position $V_{NDR}$ of this peak increasing (less negative) with a less negative reduction potential of the POMs and the increase of the metal counterion electronegativity.[174, 177] The same trend has been observed for Wells-Dawson POMs.[207] A more negative $V_{NDR}$ corresponds to a LUMO moving away from the Fermi energy of the electrodes, *i.e.*, a lower electron affinity energy (Fig. 7-b, note that the exact position of the LUMO cannot be precisely determined, since the exact potential profile in the MJ is not known and we assume that $\varepsilon_{LUMO} = \alpha|V_{NDR}|$ where α, the fraction of the potential seen by the molecule, is supposed independent of the POM). It was proposed that the more electronegative counterions facilitated the electron transfer between the POM anions via the counter-cations.[208]

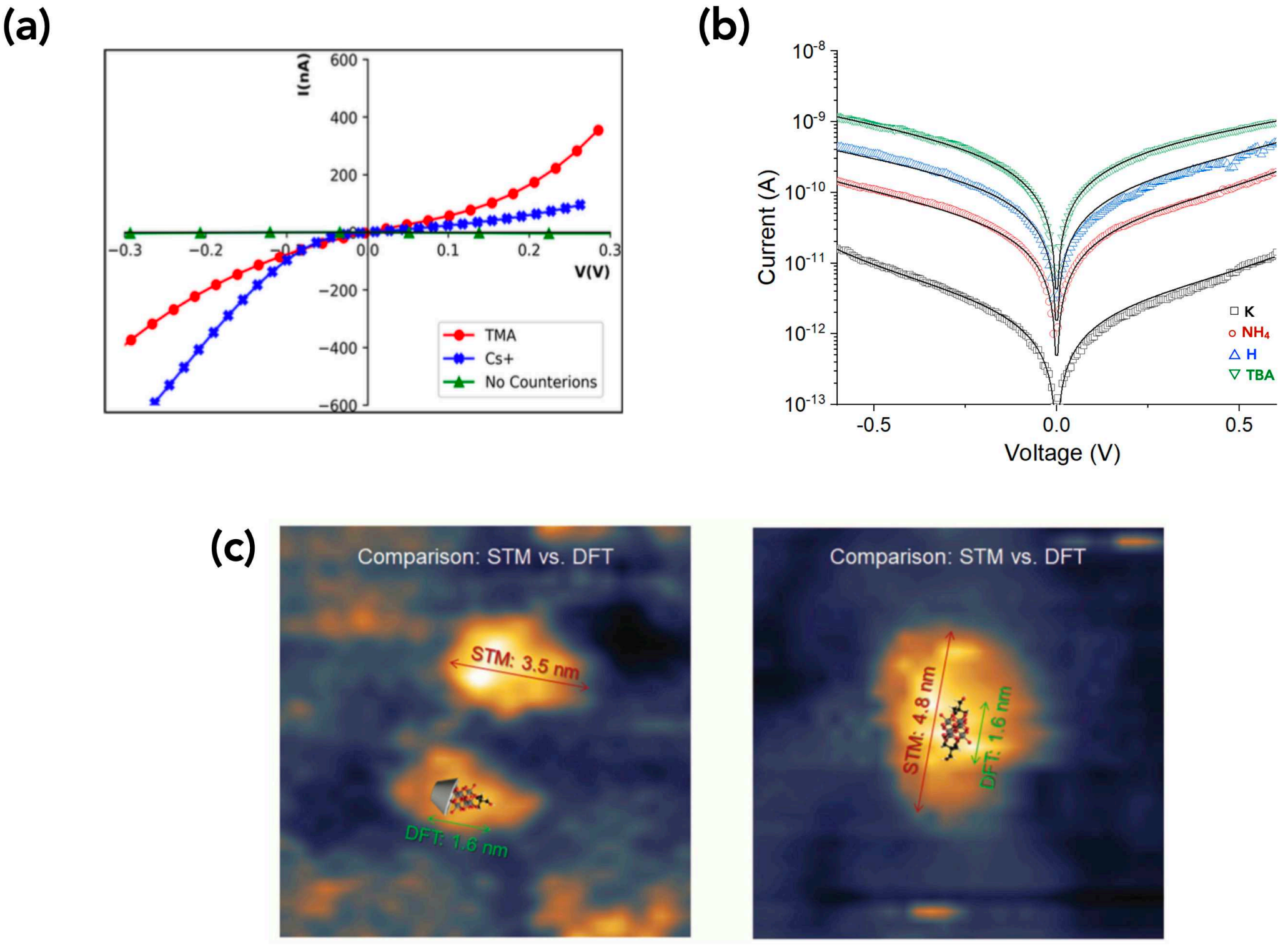

***Figure 6.*** ***(a)*** *Computed I-Vs of Au-[$W_{18}O_{54}(SO_3)_2$]$^{4-}$-Au MJs with $Cs^+$ and $TMA^+$ counter-cations. Reproduced from ref. 133. Copyright (2021) American Chemical Society.* ***(b)*** *I-Vs (mean current of a statistical distribution) of $^{TS}$Au-S-($CH_2$)$_6$-($NH_2$/$NH^+_3$)-($M^+$)$_{14-x}$[$Na\subset P_5W_{30}O_{110}$]-PtIr C-AFM tip MJs (with $M^+$=$H^+$, $K^+$, $NH_4^+$ or $TBA^+$). The symbols are the data, the solid lines are fits by the SEL (single energy level) model. Reproduced with permission from ref. 135. Copyright (2023) Royal Society of Chemistry.* ***(c)*** *STM images of the electron cloud density around the single Lindqvist-type hexavanadate structure without (left image, cyclodextrin capped) and with (right image) two*

*charge-balancing cations on gold and comparison with DFT calculated structures. Reproduced with permission from ref. 134. Copyright (2022) Taylor & Francis.*

Several studies have been reported more recently.[118, 135, 209-211] Preyssler POMs, $[Na\subset P_5W_{30}O_{110}]^{14-}$, were synthesized with four different counterions, $H^+$, $K^+$, $NH_4^+$ and tetrabutylammonium ($TBA^+$). Monolayers of these POMs were formed by electrostatic interactions on ultra-flat template stripped Au ($^{TS}Au$) functionalized with a positively charged SAM of 6-aminohexane-1-thiol hydrochloride ($HS-(CH_2)_6-NH_3^+/Cl^-$).[135] A significant increase in the currents (ca. a factor 100) in the following order $I(POM-K^+) < I(POM-NH_4^+) < I(POM-H^+) < I(POM-TBA^+)$ has been measured by C-AFM (Fig. 6-b). The analysis of the I-V curves with a single energy level (SEL) model (an analytical model derived from the Landauer formalism)[162, 212] shows that the energy position of LUMO of the POM $\varepsilon_{POM}$ (with respect to the Fermi energy) increases in the same order as the current, *i.e.*, $\varepsilon_{POM}(POM-K^+) < \varepsilon_{POM}(POM-NH_4^+) < \varepsilon_{POM}(POM-H^+) < \varepsilon_{POM}(POM-TBA^+)$, Fig. 7-c, from 0.42 to 0.66 eV. This feature is counterintuitive: a higher energy mismatch at the molecule/electrode interface, *i.e.*, a higher value of $\varepsilon_{POM}$, would have induced a lower electron transfer through the POM (*e.g., vide supra*, the DFT results[133]), and, consequently, a lower current. However, the same SEL model analysis reveals that another important parameter, the electronic coupling energy, $\Gamma$, also increases from ≈ 0.06 meV to ≈ 1 meV in the same counterion order (Fig. 7-c).[135] In this model, $\Gamma$ describes the strength of the hybridization between the molecule orbitals and the electron density-of-state in the metal electrode.[162, 212] A stronger molecule/electrode coupling induces a broadening of the molecular orbitals (FWHM = $2\Gamma$) and tends to increase the current through the MJs. In the present case, the evolution of the two parameters induces opposite trends of the current, the latter one (electrode coupling energy) being dominant. The suggested mechanism is related to POM/electrode interface dipoles. Due to steric hindrance, it is likely that the bulky $TBA^+$ cations are mostly located above the POM layers at the interface with the electrode, creating a large global dipole that shifts the POM LUMO upward. The smaller cations are more likely randomly distributed inside the MJs, around the POMs, resulting in a weaker average dipole and a weaker shift in the POM LUMO (Fig. 7-d). The cation-dependent evolution of the POM/electrode coupling energy $\Gamma$ remains to be understood. However, a recent study has revealed that the counter-cations can enhance the electronic coupling between two adjacent POMs.[211] This work showed that intermolecular electron transfers between two POMs (hybrid phthalocyanine (Pc)-lanthanide (Ln)-polyoxovanadate $[HV_{12}O_{32}Cl(LnPc)]^{3-}$ or $[HV_{12}O_{32}Cl(LnPc)_2]^{2-}$, with Ln = $Lu^{III}$ or $Dy^{III}$ for example) are mediated by the counter-cation (TBA) from the Pc moiety of one hybrid POM to the polyoxovanadate moiety of a neighboring hybrid POM when they are densely packed (in solution at high concentrations or in solid state).[211] This result suggests an enhanced POM-cation-POM electron conduction channel that remains to be clearly understood. Thus, we can suggest that a dense layer of TBAs between POMs and the metal electrode (Fig. 7-d) would enhance the POM-cation-electrode coupling and the overall current in the MJs. The POM/electrode hybridization mediated by the counterion is also substantiated by an STM study of a single cluster POM (Lindqvist-type hexavanadate $[V_6\text{-}(OH)_2]^{2-}$) deposited on an Au surface.[118, 119] The STM images showed that the electron cloud density at the POM/metal interface, with TBAs in the vicinity of a single POM, largely extends (several nanometers) the geometrical size of the POM (Fig. 6-c). This is the result of a counter-cation enhanced POM/metal hybridization. In the absence of TBA, the size of the electron cloud is reduced.[118, 119] Clearly, these results call for more detailed calculations and experiments to elucidate the role of counter-cations in the electron transport properties of POM-based MJs.

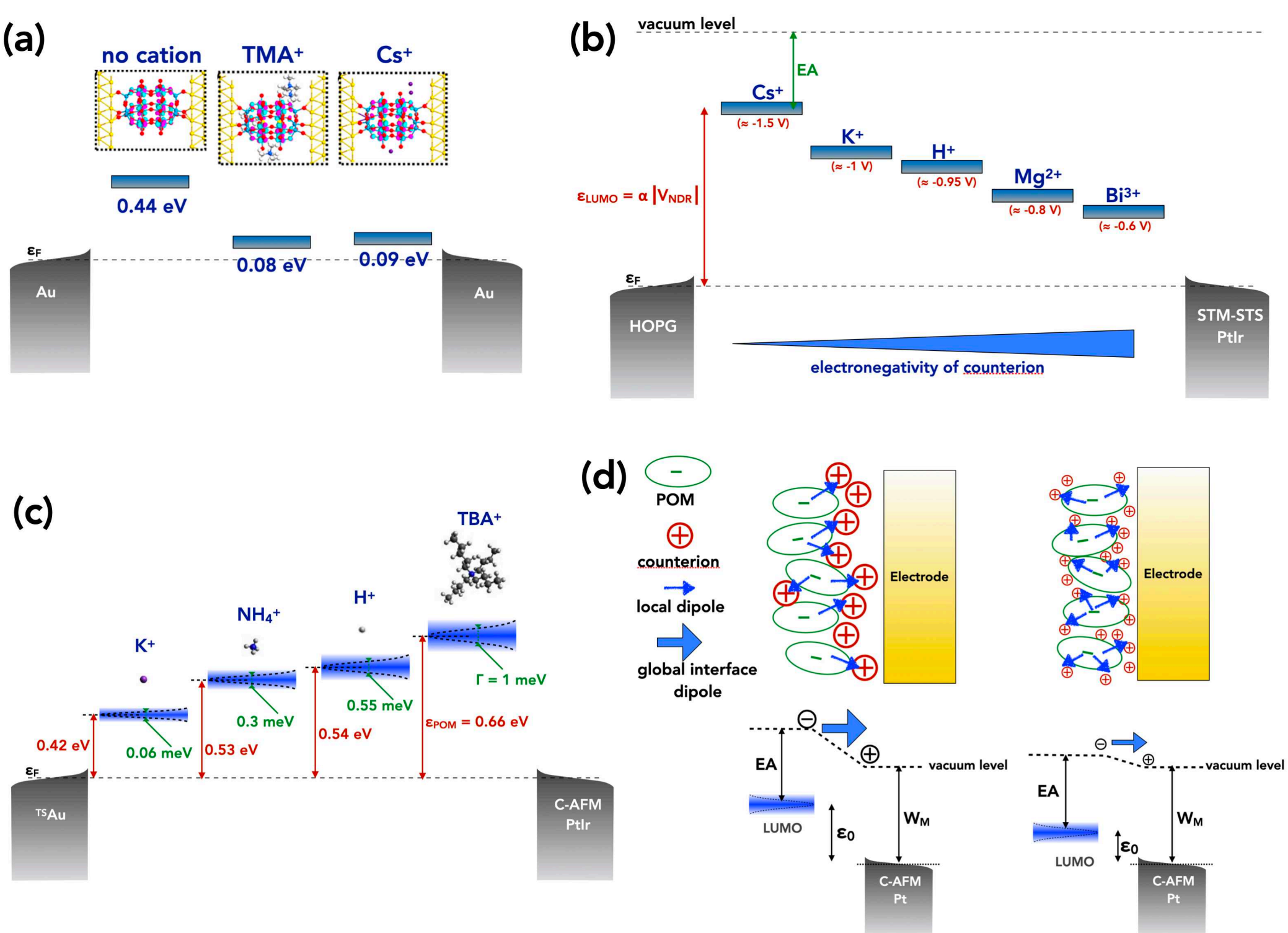

***Figure 7. (a)*** *DFT calculations of Au-$[W_{18}O_{54}(SO_3)_2]^{4-}$-Au MJs with $Cs^+$ and $TMA^+$ counter-cations, data from 133.* ***(b)*** *Electronic structure of HOPG-$(M^{n+})_{3/n}[PMo_{12}O_{40}]$-PtIr tip MJs (with M = $H^+$, $K^+$, $Cs^+$, $Cu^{2+}$, $Co^{2+}$, $Zn^{2+}$, $Ba^{2+}$,$Mg^{2+}$, $Bi^{3+}$). The LUMO levels are relatively scaled from the measured*

*$V_{NDR}$, assuming a constant α value (see text), $V_{NDR}$ is the voltage of the negative differential resistance peak (values in red and brackets), data from 174, 177.* ***(c)*** *Electronic structure of $^{TS}$Au-S-$(CH_2)_6$-$(NH_2/NH_3^+)$-$(M^+)_{14-x}$[Na⊂$P_5W_{30}O_{110}$]-PtIr tip MJs with several counter-cations: $H^+$, $K^+$, $NH_4^+$ and $TBA^+$, data from 135. Energy level $\varepsilon_{POM}$ (in red) and POM/electrode coupling energy Γ (in green).* ***(d)*** *Proposed model to explain the counter-cation dependence of the conductance of the POM-based MJs. Reproduced with permission from ref. 135. Copyright (2023) Royal Society of Chemistry.*

Other electronic properties of the POM MJs can also be influenced by the nature of the counterions. The number of redox states, which can be electrically switched in a single POM MJs (adding one electron at the time on the POM when injected from an STM tip, *vide infra* for more details), can be tuned by the nature of counter-cations balancing the negative trapped charges in the POM.[209]

*Role of the redox states.*

Last (but not least), the keystone property of POMs, namely their easy reduction and great electron storage capability (POMs are frequently referred to "electron sponges"), also plays a fundamental role in controlling the electrical properties of the POM-based MJs. Single hexavanadate clusters were successfully physisorbed on an Au(111) surface, preserving their structural and physico-chemical characteristics (as assessed by IR spectroscopy, XPS, high-resolution STM images) and the I-Vs were measured by STS in UHV at room temperature.[71] Two POMs were investigated : $TBA_2[V_6O_{13}\{(OCH_2)_3CCH_2SR\}_2]$, with R = $CH_3$ or $C_6H_5$. In the two cases, the I-V curves are non-linear (Fig. 8-a) and by using an adapted fitting procedure to cope with the large current noise, the authors showed (at V > 0 applied on the Au substrate) a staircase behavior with 3 steps in the voltage range 0 - 2 V (and the beginning of a 4th one near the upper voltage limit). The I-V behaviors are similar for the two POMs. The first current step arises at ≈ 0.6-0.7 V and the next ones are almost regularly spaced by ≈ 0.4 V (Fig. 8-a). These features were attributed to the successive one-electron reduction of the hexavanadate by the step-by-step filling of the vanadium d-band (Fig. 8-b), with up to 4 electrons per single POM injected from the STM tip when increasing the positive voltage at the bottom Au electrode. These current steps can be observed because the single POM acts as a quantum dot with a double-tunnel barrier structure (the vacuum gap between the STM tip and the POM and the weak physisorbed Au/POM bottom contact). In this case, the lifetime of the reduced states is commensurate with the time scale of the STS measurements (here a voltage scan rate of 1 V/s). The experimental analysis is supported by DFT calculations that show the presence of four peaks in the PDOS (projected density of states) of the vanadium d-band below 2 eV above the Fermi energy (Fig. 8-b)[71] and by further computational simulations, which reproduce the same trend of staircase I-V characteristics.[213] Furthermore, DFT calculations (gas phase) for almost the same POM ($[V_6O_{13}\{(OCH_2)_3CR\}_2]^{2-}$, with R = $OC_2H_4N_3$, $CH_2N_3$ or $O_3C_{29}H_{36}N_5$) but with different counter-cations ($H^+$, $Li^+$, $NH_4^+$ or $Mg^{2+}$) showed that the hexavanadate core can accept in the vacant vanadium d-orbitals at least 4 electrons and up to 9 (without the loss of its structural integrity and no charge leakage towards the lateral R groups induced by Coulomb repulsion) when counterbalanced by $H^+$.[209] This result is independent of the R functionality. A series of $TBA_2[V_6O_{13}\{(OCH_2)_3CR\}_2]$, with R = $CH_3$, $CH_2OH$, $NHCOCH_2Cl$ and $NHCOCH_2$-$OOCC_{10}H_{15}$, were deposited on Au (sub-monolayer coverage by drop casting with isolated single POM clusters), as well as the corresponding anions by ion soft-landing techniques (i.e., without the counter-cations).[119] The STS I-V measurements (in UHV) also showed 3 conductance steps as for the two $\{V_6\}$ POMs previously reported (*vide supra*[71]), without significant influence of the R substitution, nor the presence or absence of the counter-cations. All the MJs showed 3 conductance peaks between ≈ 1.2 and 2 V without clear and significant variations of the voltage peak positions, contrary to the redox properties of these POMs in solution (CV measurements) that slightly depend on the R functionality. Thus the role of the POM/Au interface seems to pin the electronic structure of the POM in the MJs, a mechanism also observed in organic MJs.[214-216] However, when the $\{V_6\}$ POMs are laterally equipped with organogold groups (phosphine-derivatized Au(I) moieties), the number of conductance steps of the STS I-V curves increased up to 5 (still between 0 and 2 V).[217] At the highest voltage (2 V), the size of the electron cloud (STM images) is larger than at lower voltages (1 V), which was attributed, with the help of DFT calculations,[218] to delocalization over the organogold moieties of the accepted electrons for the highest reduced states (+3 and +4 electrons). The multiple state resistive switching behavior was also observed for $TBA_3[HV_{12}O_{32}Cl(DyPc)]$ and $TBA_2[HV_{12}O_{32}Cl(DyPc)_2]$, with Pc = phthalocyanine) deposited on HOPG (sub-monolayer), the Pc moiety being used to immobilize the POM on HOPG via π-π interactions[219] (the same POMs were also studied on Au electrodes, with an eGaIn top electrode, *vide supra*[199]). Three conduction steps were also observed on the I-V traces for the PcDy-$\{V_{12}\}$, but at higher voltages (≈ 2.1, 3.2 and 4.0 V, and ≈2.1, 2.9 and 3.4 V for the $(PcDy)_2$-$\{V_{12}\}$ POM) than for the $\{V_6\}$ POM discussed above (Fig. 8-a). This difference has been related to the larger HOMO-LUMO gap for the PcDy-$\{V_{12}\}$ (2.4-2.6 eV) than that of the $\{V_6\}$ POM (1.0-1.2 eV). Another approach to observe the effect of the redox states on the molecule conductance is to add a third electrode to the substrate/molecule/tip MJ through an electrically gated ionic liquid (or electrolyte) environment (molecular transistor like configuration). In this configuration, the tip-substrate voltage is kept low (e.g., 200 mV) and the gated ionic liquid voltage is ranged as in classical CV experiments. When a pyridyl-capped Anderson-Evans POM, $TBA_3[MnMo_6O_{18}\{(OCH_2)CC_5H_4N\}_2]$, was anchored between an Au substrate and Au tip (via the pyridyl groups), 3 redox states were clearly observed between a bias range of 1.2 to -1.5V (vs. Pt electrode used as a reference electrode).[108] Both a reduced state (-4) and an oxidized state (-2) were observed with conductance ratios of about factor 10. These conductance modifications were well described using a Nernstian model[220, 221] assuming electron tunneling transport through the different redox states of the POM controlled by the electrochemical potential. However, contrary to other studies reporting a higher conductance for the reduced POM (*vide supra* refs. 71, 119, 210, 217, 219 and *vide infra* ref. 222), in this experiment the conductance of the reduced POM is lower than in the neutral state, which remains to be understood (probably because the redox state of the Mn and not that of the POM core is involved). Nevertheless, in all these studies, the clear attribution of each conductance step with successive redox states, as well as the role of ligands and counterions on the multiple resistance switching properties, require to be clarified/supported by further experimental studies and theoretical calculations on the complete metal-POM-metal MJs (and not only POM alone in the gas phase).

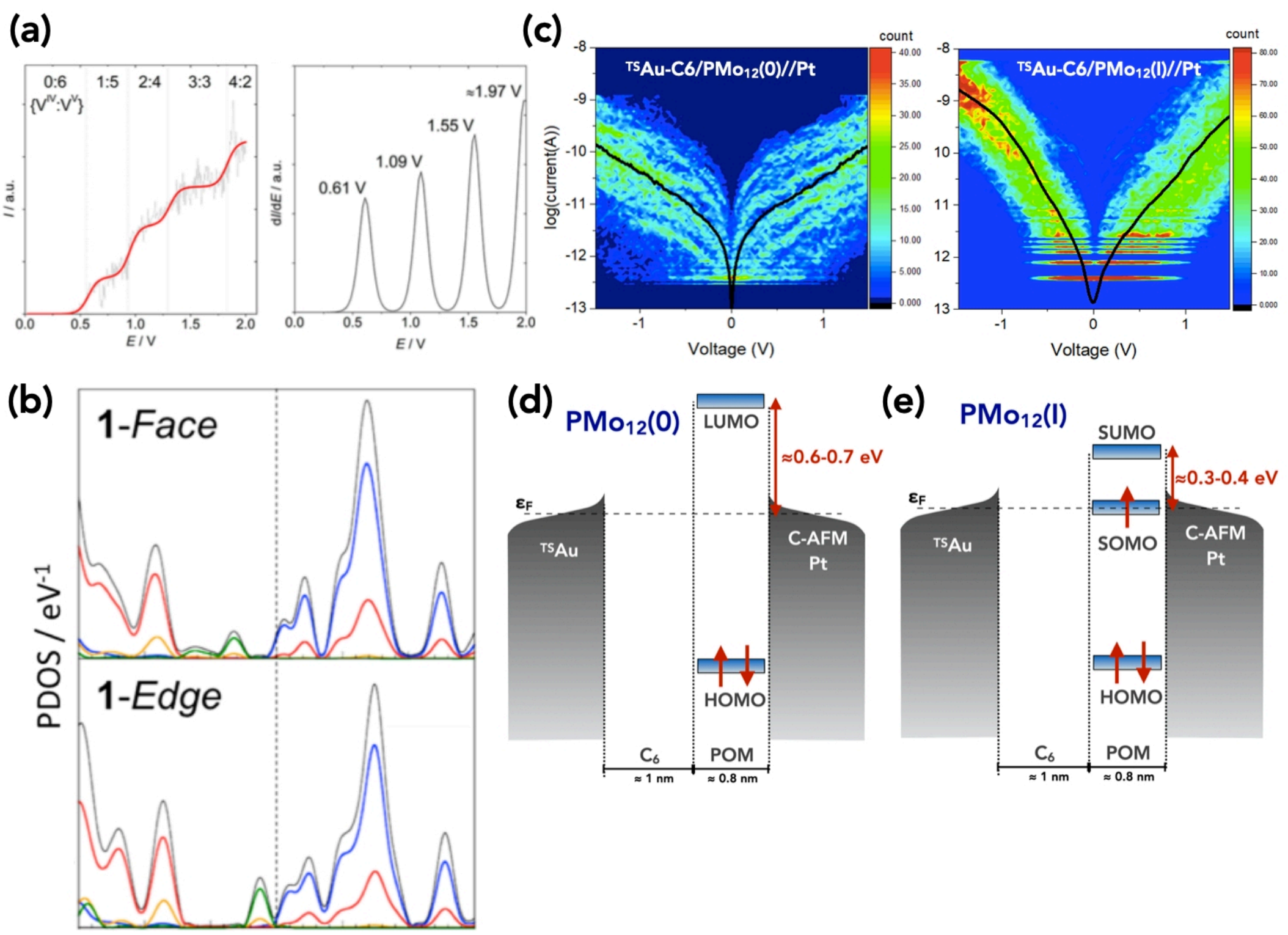

***Figure 8.*** ***(a)*** *Typical I-V of Au-TBA$_2$ [$V_6O_{13}${$(OCH_2)_3CCH_2SR$}$_2$]-Pt tip MJs and first derivative $\partial I/\partial V$. Reproduced from ref. 71. Copyright (2018) American Chemical Society.* ***(b)*** *Projected density of states (PDOS) of Au-TBA$_2$ [$V_6O_{13}${$(OCH_2)_3CCH_2SR$}$_2$]-Pt tip MJs (R = $CH_3$) with the POM connected to the electrodes by its face or edge: black = total PDOS, blue = d-V bands, red = O, yellow = C, green = S. Reproduced from ref. 71. Copyright (2018) American Chemical Society.* ***(c)*** *2D histograms (heat maps) of the I-V curves measured by C-AFM on the neutral $^{TS}$Au-S-$(CH_2)_6$-($NH_2$/$NH_3^+$)-TBA$_3$[$PMo_{12}O_{40}$]-PtIr tip and the one electron reduced (by UV light) MJs. Reproduced with permission from ref. 222. Copyright (2022) Royal Society of Chemistry.* ***(d)*** *Electronic structure of $^{TS}$Au-S-$(CH_2)_6$-($NH_2$/$NH_3^+$)-TBA$_3$[$PMo_{12}O_{40}$]-PtIr tip MJs for the neutral POM and* ***(e)*** *the one electron photoreduced (by UV light) POM. Reproduced with permission from ref. 222. Copyright (2022) Royal Society of Chemistry. The LUMO and HOMO of the alkyl ($C_6$) chain are omitted for clarity since the HOMO-LUMO gap is large (7-9 eV) and the MOs are far away from those of the POM.*

It is known that POMs can be photo-reduced, a mechanism used for the development of photochromic materials.[55] However, studies reporting the effect of photo-reduction on the electronic transport properties of POM-based devices are scarce. It was demonstrated a large increase in conductivity (ca. 500 - $10^3$) of single crystals made of Preyssler POMs $[NaP_5W_{30}O_{110}]^{14-}$ upon UV-light photo-reduction (*vide supra*, Fig. 2-d).[151] At the nanoscale, monolayers of $TBA_3[PMo^{VI}_{12}O_{40}]$ or $TBA_4[PMo^{VI}_{11}Mo^{V}O_{40}]$ (referred to as $PMo_{12}(0)$ and $PMo_{12}(I)$, for the non-reduced and monoreduced POM, respectively) were immobilized on ultra-flat $^{TS}Au$ electrodes prefunctionalized with 6-aminohexane-1-thiol hydrochloride ($HS-(CH_2)_6-NH_3^+/Cl^-$) and their I-V curves measured by C-AFM.[222] An increase in the conductance of these MJs by factor ≈ 10 was clearly evidenced with a modification of the I-V shape (Fig. 8-c): from symmetric I-V for the $PMo_{12}(0)$ MJs to an asymmetric one (more current at negative voltages, applied on the $^{TS}Au$ substrate) for the one-electron reduced $PMo_{12}(I)$ MJs. From the analysis of the I-V data set with the SEL model (*vide supra*), these features were ascribed to off resonant tunneling electron transport mediated by the POM LUMO located at ≈ 0.6-0.7 eV above the Fermi energy for the $PMo_{12}(0)$ MJs (Fig. 8-d) and to a resonant tunneling electron transport through the SOMO (almost aligned with the Fermi energy level) and the SUMO (at ≈ 0.3-0.4 eV from the Fermi energy) for the reduced $PMo_{12}(I)$ MJs (Fig. 8-e). The open-shell radical is stable in these MJs at room temperature likely because the SOMO orbitals are localized on the core Mo atoms of the POM and therefore "topologically" protected from a too strong interaction with the electrodes, which would have drastically diminished the lifetime of the one-electron reduced state. This result constitutes one of the rare examples of MJ based on open-shell molecules and stable at room temperature[223, 224] (while other examples required low-temperature and/or UHV conditions [225-227]). A careful analysis of large I-V data sets (500-600 I-V traces) with machine learning and clustering algorithms unveiled more details: for the pristine MJs (the one referred to as $PMo_{12}(0)$), a small fraction (≈ 18%) of the I-V traces corresponds to already reduced POM. Finally, the POMs can be photo-reduced in situ by exposure of the pristine $PMo_{12}(0)$ MJs to UV light without the need for a sacrificial reductant (we hypothesized that the electrons are supplied by the amine groups). The switch is reversible. The $PMo_{12}(I)$ MJs returned to the neutral state by resting in the dark at room temperature or upon moderate heating to accelerate the reoxidation kinetics.[222]

*Role of the substrate.*

The experiments to determine the ET properties of POM thin films, SAMs and single POM clusters have been done on various substrates, mainly metals (Au or Al), semiconducting substrates (Si covered by $SiO_2$, H-passivated Si, ITO), carbonaceous materials (HOPG, glassy carbon) to name the main. For the same POM (*e.g.*, Keggin type polyoxotungstate) it is clear from several studies (see Figs. 4-a to 4-d) that the energy position of the LUMO is not the same for MJs on an oxidized Si substrate and oxide-free Si (hydrogen-passivated Si) substrate. Among several possible reasons (as discussed above, see in section 2.2.1 at the end of subsection "Role of the transition metals and central heteroatom"), the presence of the oxide layer with oxide trap charges and Si/$SiO_2$ interface traps can influence the potential landscape in the MJs. This is a general concern in MJs as also clearly evidenced for simple archetype alkyl-chain MJs embedded in oxide-free (Au, Pt, Si-H) or slightly oxidized (Si/$SiO_2$, Al/$Al_2O_3$, eGaIn/$Ga_2O_3$, Hg/HgO) electrodes.[203] Moreover, it has been inferred, from electrochemical measurements in a wet environment, that the interfacial electron transfer depends on the nature of the electrode (glassy carbon or Si) for a given type of covalently assembled POM monolayers.[99, 101] Spontaneous reduction of the POMs has also been observed on Al and ITO substrates[161] which impact the ET properties, as well as gap states at the POM/ITO interface (see Fig. 2-g).[150]

### *2.2.2. Planar devices.*

The studies of the electronic properties of a few layers of POMs (nL, n=1 to fews) connected by two planar electrodes are scarce, mainly because the electron transfer from POM to POM is weak, resulting in low (even unmeasurable) currents. However, by optimizing the width and length of the gap between two planar electrodes, it was possible to measure the in-plane I-V characteristics of a monolayer of $H_3(PW_{12}O_{40})$ POMs immobilized on a thick insulating $SiO_2$ pre-functionalized with APTES (3-aminopropyltriethoxysilane).[92, 228] With these device geometries, measurable currents (> 0.1 pA) were recorded for applied voltages larger than a few volts, up to 0.1-1 nA at ≈ 20-30 V (Fig. 9-a). Adding several POM layers (by the LbL method) increases the current by offering alternative conduction paths to overcome the structural defects of a single layer. The study of the electron transport mechanism in these devices remained tedious and strongly dependent on the degree of experimental control of the precise structural organization of the POMs in the layers.[92] Basically, it was observed that POM-to-POM tunneling (temperature independent) dominates below 150 K and for the shorter device (L < 100 nm, Fig. 9-a), while temperature-activated hopping between adjacent POMs was the main mechanism above 200 K (with activation energy ca. 30 - 80 meV).[228] From a tunnel model, a tunnel barrier of 0.2 - 0.4 eV was estimated (Fig. 9-a), a value also consistent with the ones determined from the vertical MJs incorporating the same APTES/$[PW_{12}O_{40}]^{3-}$ system (*vide supra*, Figs. 3-a and 3-b).[172, 173]

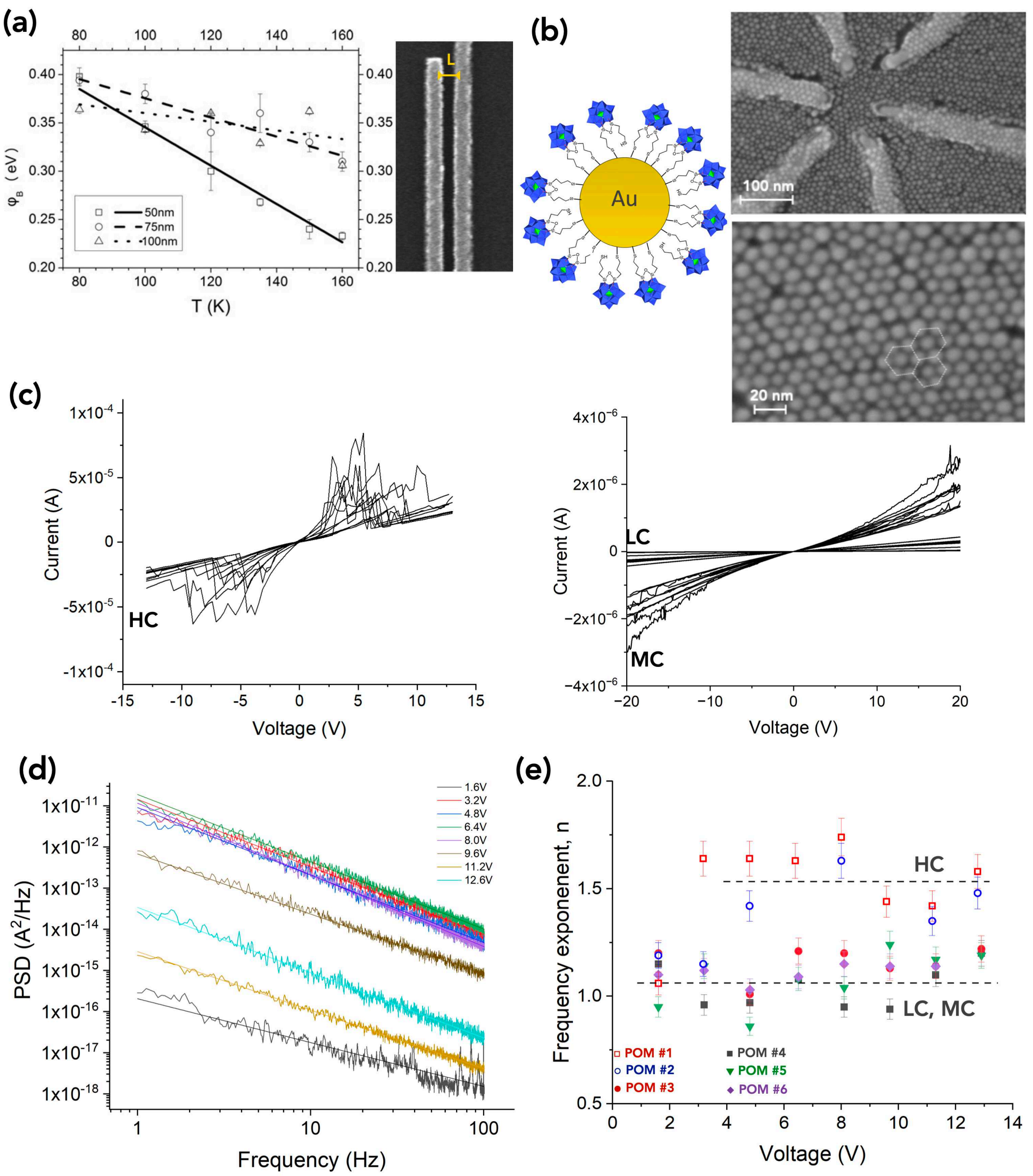

***Figure 9. (a)*** *Tunnel barrier energy for a monolayer of APTES/$H_3[PW_{12}O_{40}]$ between two planar electrodes (inset: SEM image) with a gap length L = 25 to 200 nm. Reproduced with permission from ref. 228. Copyright (2012) Elseviere.* ***(b)*** *SEM images at two magnifications of a monolayer of Au NPs (7-8 nm in diameter) chemically functionalized with $TBA_3[PW_{11}O_{40}(SiC_3H_6SH)_2]$ ($PW_{11}SH$ for short) connected by 6 Au electrodes. Scheme of the NP-POM (right).* ***(c)*** *I-Vs recorded between several pairs of electrodes (2 electrodes out of the 6, randomly selected) for three samples with three different (decreasing) densities of nanoparticles and several levels of current: high-, medium- and low-current (HC, MC and LC, respectively).* ***(d)*** *Low frequency noise measurement of*

*HC samples. Power spectrum densities versus frequency (PSD ∝ $1/f^n$, log-log scale) measured at several voltages. The slope is the frequency exponent n.* ***(e)*** *Summary of the frequency exponent, n, for all the samples. (b-e) Reproduced with permission from ref. 229. Copyright (2024) Royal Society of Chemistry.*

An efficient solution to increase the in-plane conductance of the POM-based device (as well as any other molecules) is to use small Au nanoparticles (NPs) as "relay stations" to bridge molecules together in a relatively large zone connected in the periphery by planar electrodes. These "nanoparticle-molecule-network" (NMN) form relatively well hexagonally organized 2D arrays of Au NPs linked by few molecules and they were demonstrated being versatile platforms to study the electron transport mechanism in molecular-based devices (*e.g.*, alkyl chains, π-conjugated small oligomers, molecular switches, spin crossover molecules).[230-233] Such NMNs were fabricated using Au NPs (7-8 nm in diameter) chemically functionalized with $TBA_3[PW_{11}O_{40}(SiC_3H_6SH)_2]$ ($PW_{11}SH$ for short). Well organized hexagonal arrays were obtained and they were connected by finger-shaped electrodes separated by ≈ 20 - 100 nm (Fig. 9-b).[229] Three families of POM-NMNs were fabricated with measured currents increasing with the density of NPs. Large currents (up to 10-100 µA at around 10 V) were recorded for the highly conductive POM-NMNs devices (Fig. 9-c). In that case, the I-Vs displayed stochastic fluctuations (abrupt increase/decrease of the current) and the low-frequency noise ($1/f^n$ noise, or flicker noise, Fig. 9-d) deviated from the usual behavior observed for the less conductive POM-NMNs (*i.e.*, a higher frequency exponent n ≈ 1.5 - 1.6, compared to n ≈ 1) - Fig. 9-e. This behavior was explained by a dynamic switching between the neutral and redox states (with a higher conductance) of the $PW_{11}SH$ POMs when a sufficiently high electron flux is passing through the POM-NMNs.[229] This dynamic behavior makes the POM-NMNs prone to be tested for the implementation of neuro-inspired computing devices such as reservoir computing systems (*vide infra,* section 5). Another approach to enhance the electron conductivity in a plane of POMs is to link the POMs together via electron transfer facilitators. One approach is to laterally functionalize the POMs with organogold moieties that extend the delocalization of the electrons (*vide supra*, ref. 217). Another possibility is to bridge the adjacent POMs by multinuclear metallic complexes. This was demonstrated to increase the conductivity of POM-based bulk materials (*vide supra*, ref. 152). The same approach, or using other types of electron transport facilitators, remains a challenge to be explored at the monolayer level in planar MJs.

### 2.3. Other electrical properties.

The POMs usually do not present a strong asymmetry of their chemical structures (unless functionalized with specifically designed ligands) and the POM-based MJs usually displayed almost symmetric I-V (or weak asymmetric behavior, typically a ratio $R \lesssim 10$ between the currents at voltages of opposite polarities).[97, 104, 113, 133, 135, 172, 174, 177, 198, 199, 222, 229, 234-236] In some cases, noticeable asymmetric I-Vs were observed due to specific device configurations or induced by associating the POMs with other molecular moieties, leading to molecular rectifying diode devices. In a single molecule STM experiment (at room temperature in ambient air), a rectification ratio $R \approx 250$ was observed for a $[DyP_5W_{30}O_{110}]^{12-}$ adsorbed on Au substrate when the STM tip is held at ≈ 1 nm above the POM.[237] The rectification behavior vanished ($R \approx 1$) when the tip was in contact with the POM (Fig. 10-a). The rectification effect is not intrinsic to the POM and it is due to the asymmetric electronic coupling of the POM with the two electrodes, as also demonstrated for organic MJs.[238-241] The electronic coupling energies differ by a factor $10^4$ between the weak coupling at the tip side and the strong coupling at the Au substrate side (Fig. 11-a). The same coupling energies are measured when the tip is in contact with the POM (Fig. 11-b). The energy position of the POM LUMO is also shifted closer to the Fermi energy of the electrodes due to an increase in the hybridization between the POM frontier orbitals and electronic density-of-states of the metal electrodes (Fig. 11-b).

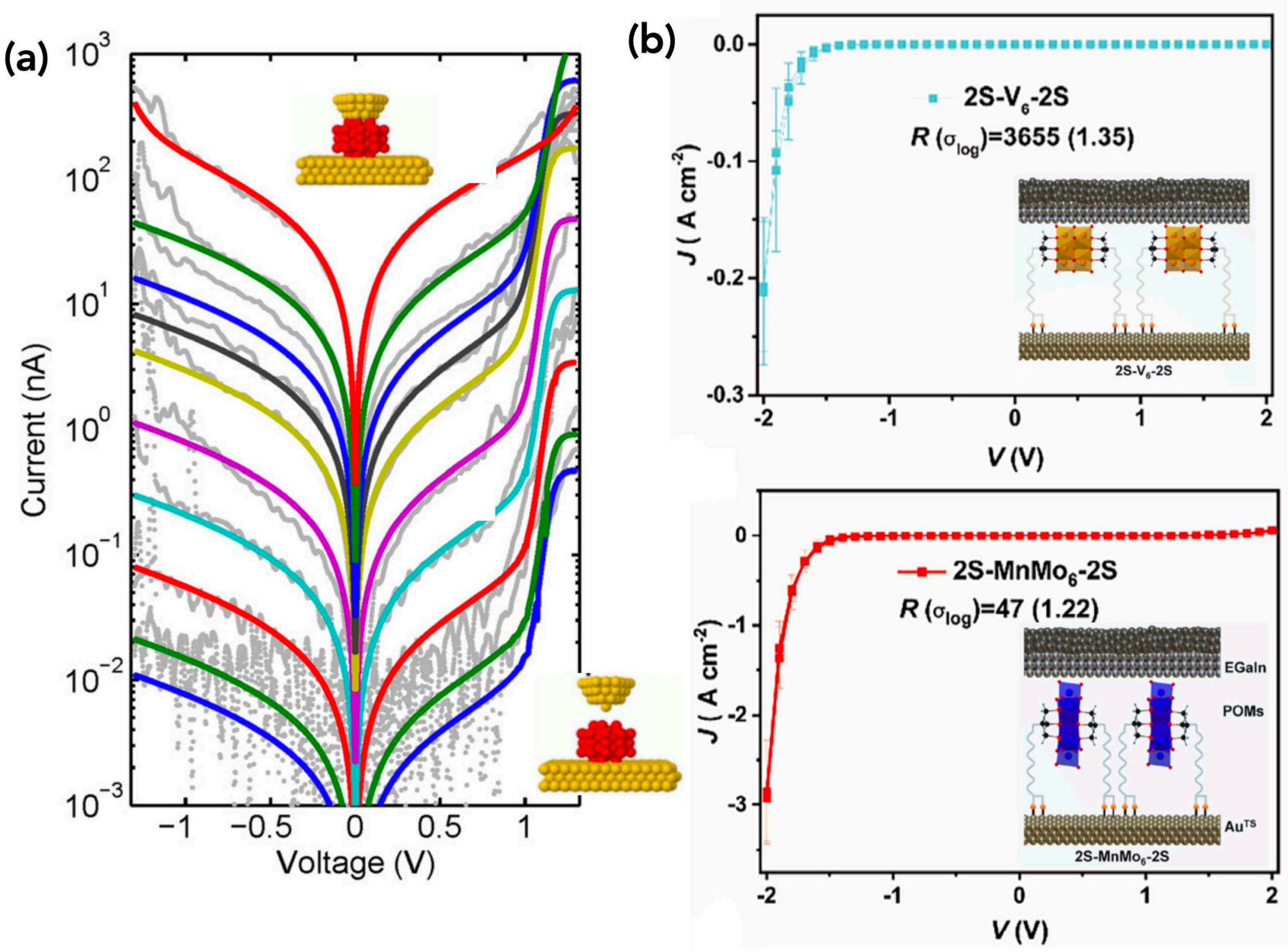

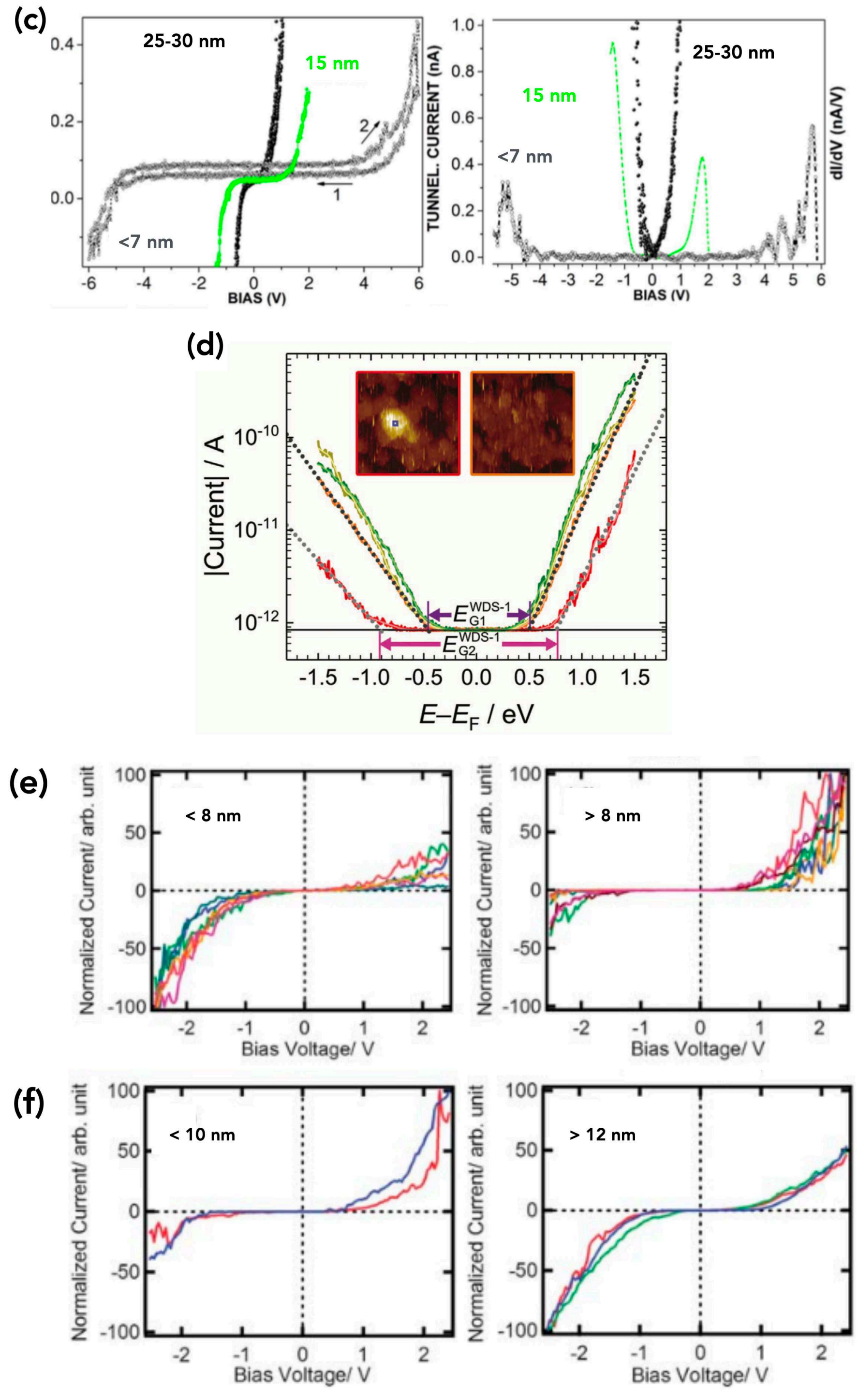

***Figure 10. (a)*** *I-Vs measured by STM/STS of a single $[DyP_5W_{30}O_{110}]^{12-}$ POM adsorbed on Au when the tip is moved from ca. 1 nm above the POM (lower current) to the contact (higher current). The gray lines are the data and the colored lines the fits by the single energy level (SEL) model.*

*Reproduced with permission from ref. 237. Copyright (2015) Institute of Physics.* ***(b)*** *I-V plots (average current density) of the Au-(2S-$V_6$-2S)-$Ga_2O_3$/eGaIn and Au-(2S-$MnMo_6$-2S)-$Ga_2O_3$/eGaIn MJs (scheme of the MJs above the plots). Reproduced from ref. 242. Copyright (2005) John Wiley and Sons.* ***(c)*** *I-Vs and ∂I/∂V of the nanocrystals (NC) of $H_3[PW_{12}O_{40}]$ deposited on ATPES functionalized Si/$SiO_x$ substrate measured by STM/STS for 3 sizes (diameters) of the NCs. Reproduced with permission from ref. 236. Copyright (2018) Institute of Physics.* ***(d)*** *I-Vs (absolute value) of Au-$TBA_5[HP_2V_3W_{15}O_{59}((OCH_2)_3C\text{-}CH_2SCH_3)]$-W tip MJs measured by STM/STS on one and two layers of POMs. Reproduced from ref. 113. Copyright (2022) John Wiley and Sons. Creative Commons Attribution-NonCommercial-NoDerivs License.* ***(e-f)*** *I-Vs measured by C-AFM on SWCNTs decorated with nanoparticles of $H_3[PMo_{12}O_{40}]$ (different sizes). Panel (e) is for semiconducting SWCNTs and panel (f) for metallic SWCNTs. Reproduced with permission from ref. 243. Copyright (2013) Royal Society of Chemistry.*

Very recently, a high rectification effect (R ≈ 3600) was obtained for a Au-POM-$Ga_2O_3$/eGaIn MJ based on a Lindqvist-type POM $TBA_2[V_6O_{13}\{(OCH_2)_3CCH_2OCO(CH_2)_4C_3H_5S_2\}_2]$ (2S-$V_6$-2S for short)[242] - Fig. 10-b. The $V_6$ core is equipped with two disulfide-terminated legs for anchoring to the Au surface. A smaller value, R ≈ 47 was measured for a MJ in the same configuration with an Anderson-type POM: $TBA_3[MnMo_6O_{18}\{(OCH_2)_3CNHCO(CH_2)_4C_3H_5S_2\}_2]$ (2S-$MnMo_6$-2S for short). Figure 11-c shows the electronic structures deduced from UPS and UV-vis spectroscopy. At low bias (|V| < 1 V), the electron transport is temperature-independent off resonant tunneling (mediated by the LUMO-Fermi energy barrier height). The high rectification is clearly observed at higher voltages (1.5 V < |V| < 2 V) with a temperature activated behavior of the current in this voltage window. A speculative voltage-driven formation of a Schottky-like barrier at higher bias is proposed,[242] but a true Schottky diode with the POM acting as an n-type semiconductor would have resulted in a forward current at positive voltages contrary to the reported measurements. Other mechanisms were not firmly discarded. When a redox switching at large bias is involved, the electron transport through the MJ can be described by a thermally activated combined Marcus-Landauer theory at large bias and essentially by an activationless coherent off resonant described by the Landauer model at low bias.[244-246] We also note that similar behaviors were reported for redox benzotetrathiafulvalene based MJs and explained by a transition between a Marcus (thermally activated) to inverted Marcus (activationless) charge transport behaviors depending on the voltage polarity.[247]

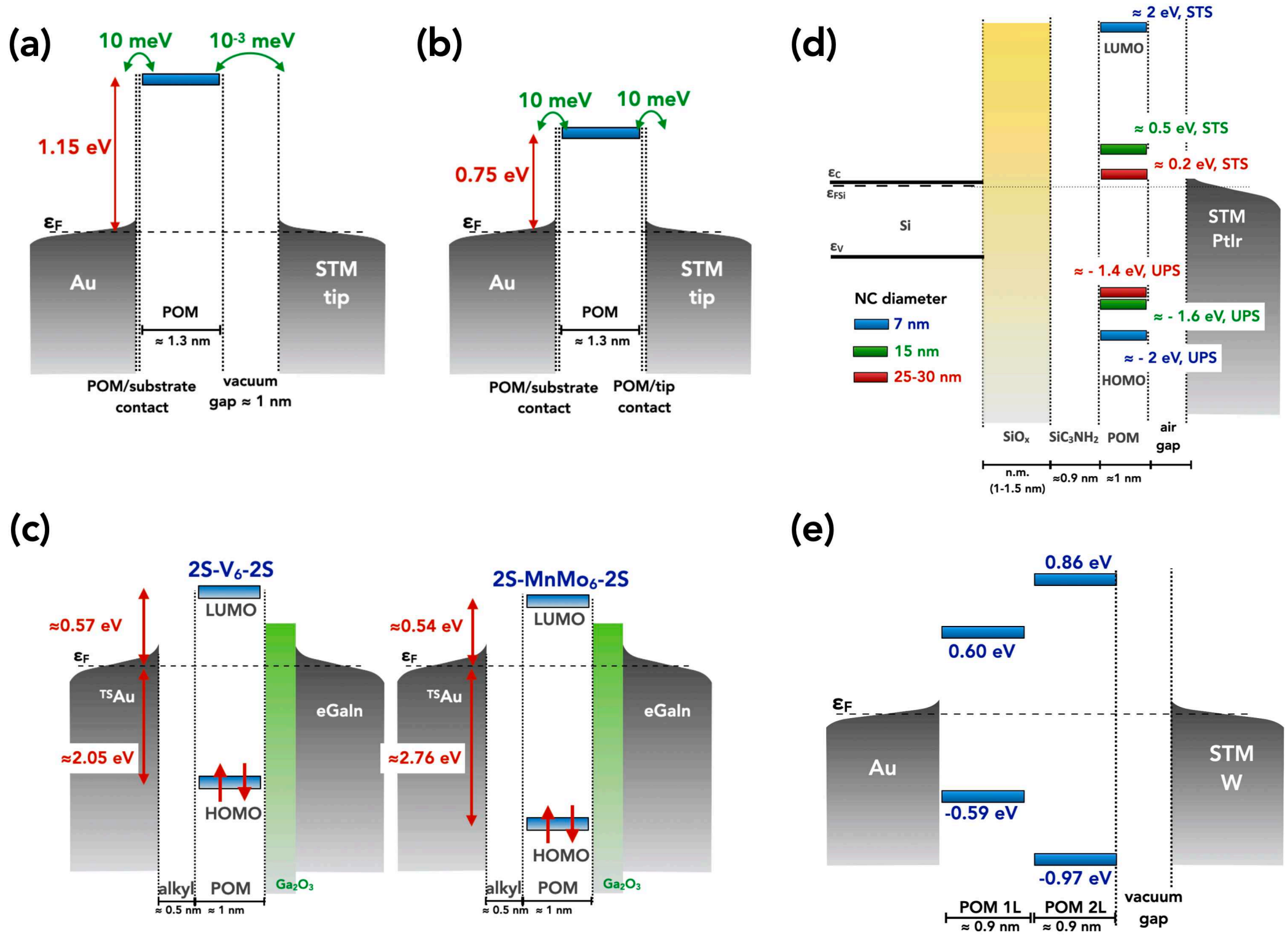

***Figure 11***. ***(a)*** *Electronic structure of Au-$[DyP_5W_{30}O_{110}]^{12-}$-STM tip when the tip is ≈ 1 nm away above the POM (molecular rectifying diodes) and* ***(b)*** *when the tip is contacting the POM (no current rectification) (data from 237).* ***(c)*** *Electronic structure of $^{TS}$Au-(2S-$V_6$-2S)-$Ga_2O_3$/eGaIn and $^{TS}$Au-(2S-$MnMo_6$-2S)-$Ga_2O_3$/eGaIn MJs showing a current rectification ratio of 3600 and 47, respectively (data from 242). The LUMO and HOMO of the alkyl chain are omitted for clarity, since the HOMO-LUMO gap is large (7-9 eV) and the MOs are far away from those of the POM.* ***(d)*** *Electronic structure of $n^{++}$Si-$SiO_x$-Si$(CH_2)_3$-($NH_2$/$NH_3^+$)-nanocrystal of ($H_3[PW_{12}O_{40}]$)-PtIr tip as a function of the POM nanoparticle diameter (blue 7 nm, green 12 nm and red 25-30 nm). The LUMO energy is calculated from the current threshold of the $\partial I/\partial V$ versus V curves, assuming that about half of the applied voltage V is lost in the tunnel barrier on each side of the POM nanoparticle (the height of the air gap and the thickness of the native oxide are not given; the value in brackets is the usually accepted thickness). The HOMO energy is measured by UPS (data from 236). The LUMO and HOMO of the alkyl ($C_3$) chain are omitted for clarity, since the HOMO-LUMO gap is large (7-9 eV) and the MOs are far away from those of the POM.* ***(e)*** *Electronic structure of Au-$TBA_5[HP_2V_3W_{15}O_{59}((OCH_2)_3C-CH_2SCH_3)]$-W tip MJs with one and two layers of POMs, data from 113.*

Several types of nanostructures (nanoparticles, nanorods...) have been synthesized based on POMs and/or mixed with other compounds. Nanocrystals (NC) of POMs ($H_3[PW_{12}O_{40}]$) have been formed in solution (diameters ≈ 7-25 nm depending on POM molarity and pH, the NC diameter increasing with the POM molarity in the precursor solution), and thin films of these NCs formed on highly doped n-type Si functionalized by the SAM of APTES.[236] The I-V of these films of POM NCs were measured by STM/STS (in air at room temperature) - Fig. 10-c. Combined with UPS measurements, it was shown that the HOMO-LUMO gap decreases when increasing the POM NC diameter (Fig. 11-d),[236] a trend confirmed by UV-vis absorbance. This behavior has been accounted for by the combination of Coulomb repulsion (increasing the HOMO-LUMO gap) in the smallest NCs and a larger extension of the electronic states (increase of the electronic coupling between NCs in the film) that decrease the HOMO-LUMO gap of the films made with the largest NCs. Another dimensional effect was obtained in a multilayer of POMs. It was observed (STS in UHV) that the HOMO-LUMO gap of the 1st layer of POM (here $TBA_5[HP_2V_3W_{15}O_{59}((OCH_2)_3C-CH_2SCH_3)]$) deposited on Au is reduced (1.19 eV) compared to the one of a second layer (1.83 eV) starting to grow on top (Fig. 10-d).[113] The HOMO-LUMO gap (1.83 eV), Fig. 11-e, of the second layer is close to the calculated ones for analog POMs tris(alkoxo)-ligated fully oxidized Wells-Dawson POMs,[248, 249] while XPS measurements on the POMs in the 1st layer showed that the vanadium atoms in the $V_3$ cap are reduced. Thus, the smaller HOMO-LUMO gap was associated with POMs spontaneously reduced by electron transfer from the Au electrode.

When POM-based nanoparticles are associated with single-walled carbon nanotubes (SWCNT), the current rectification direction (*e.g.*, more current at positive voltages than negative ones or vice versa) depends on the POM-NP height (2-16 nm) and the electronic nature (metallic *vs.* semiconducting) of the SWCNT.[243] SWCNTs and $H_3[PMo_{12}O_{40}]$ were mixed in solution and drop cast onto a $SiO_2$/Si substrate forming a network of SWCNTs, each SWCNT being randomly decorated by $PMo_{12}$ NPs with naked SWCNT in between. A gold electrode was evaporated on one side of the SWCNT/$PMo_{12}$ network and the I-Vs were recorded at different locations of a hybrid SWCNT/$PMo_{12}$ by connecting this wire with a C-AFM tip. On semiconducting SWCNTs, the current is more important at negative voltages applied on the Au electrode (referred to as a negative rectification with a ratio $R^- = I(V<0)/I(V>0)$ up to ≈ 100) for the smallest $PMo_{12}$ NPs (< 8 nm) and inverted (positive rectification with $R^+ = I(V>0)/I(V<0)$ up to ≈ $10^3$) for the largest ones (heights 8 - 12 nm), Fig. 10-e. This behavior is completely reversed for a metallic SWCNT, *i.e.*, a moderate positive rectification ($R^+$ ≈ 10) for the small $PMo_{12}$ NPs (6-10 nm) and a moderate negative rectification ($R^-$ ≈ 10) for the biggest ones (12-16 nm), Fig. 10-f. This behavior reversal between semiconducting and metallic SWCNT is attributed to the inversion of charge distribution in the $PMo_{12}$ NP/SWCNT structure: $PMo_{12}$ NPs positively (negatively) charged relative to semiconducting (metallic) SWCNT, respectively, as revealed by surface potential measurements using Kelvin probe force microscopy (KPFM). This is like inverting the n-doped and p-doped sides in a p-n semiconductor junction. In addition, the created dipole at the POM/SWCNT interface also induces a drift of the I-V curves, which also contributes to the rectification behavior. When the size of the $PMo_{12}$ NPs increases, the number of charges stored in the NPs and the charge transfer at the POM/SWCNT are modified, thus modifying the interface dipole and the direction of the rectification behavior. Metallic NPs (Pt) with a small diameter (average values: 1.6 and 2 nm) decorated with Keggin-type POMs $TBA_n[XW_{11}O_{39}\{O(SiC_3H_6SH)_2\}]$ where the central heteroatom is X=P, Si or Al, were self-assembled to form nanorods (≈ 100 nm in diameter, 0.5 - 1 µm long).[250] Increasing the POM charge n=3, 4 or 5 (for X=P, Si and Al, respectively) decreases the polarizability (i.e., the propensity to modify the electron cloud distribution upon the application of an electric filed) as observed by the decreasing measured relative dielectric constant ($\varepsilon_r$ ≈ 4, 3.4 and 3.1 for X=P, Si and Al, respectively).[251] These nanorods were deposited on an Au substrate and their I-Vs measured (at several locations) by C-AFM (Pt tip). First, the level of current increases from Al to Si to P, indicating a decrease in the energy position of the LUMO with respect to Fermi energy of the electrodes (*vide supra* for the tuning of the electronic structure of POM MJs with the nature of the central heteroatom and DFT calculations[252]). The precise experimental determination of $\varepsilon_{LUMO}$ is difficult from the experimental I-Vs of NP-molecule networks, since the electron transport depends on the combination of the NP and molecule electronic properties. Beyond this increase in the current, the shapes of the I-Vs were also modified. The I-Vs were found to follow a power law $I \propto V^{\alpha}$ usually observed in metallic NP networks,[253-255] with α depending on the topology of the current pathways in the network and the electron transport mechanisms between neighboring NPs. In this case, α increases (from ≈ 2 to 3) with the charging (Coulomb) energy of the POM-NP hybrids, which globally increases when decreasing $\varepsilon_r$ (since the NP size and the inter-nanoparticle distance are almost identical in the three systems).[250] These results illustrate the role of the molecule polarizability in the electronic properties of MJs in general.[256]

## 3. POM-BASED MEMORY DEVICES.

The ability of POMs to store several charges (electrons) within a reasonable voltage window (see section 2.2.1, "Role of the redox states") has prompted studies to evaluate their incorporation in non-volatile capacitive memories, either in two terminals devices (hybrid POM/semiconductor or insulating films) or in three terminal transistor-like devices where the POMs are used as floating charge nodes in place of conventional polysilicon floating gates. The advantages of POMs are: (i) better thermal stability than other redox active organic molecules that make them compliant with silicon CMOS technology (at least at the back-end-of-line (BEOL) level); (ii) multiple and easily accessible redox states, which allow to target multi-bit storage memories. Alternatively, the redox-dependent electrical conductance of POMs has also stimulated many works to develop resistive switching (RS) memories. Only a few mini-reviews have been published[80-82] on POM-based memories. Here, we provide an extensive review both of capacitive (based on charge trapping in a floating gate) and resistive switching (based on change in the conductance/ resistance of the active layer) memories.

### 3.1. Charge trapping (capacitive) memories.

In capacitance memory, charges are trapped/detrapped in an insulating film, modifying the capacitance-voltage characteristics (voltage shift) when the film is sandwiched between two electrodes (two terminals (2T) memory cells) or the current circulating in an underneath semiconductor channel in a transistor configuration (three terminal (3T) memory cells). Archetype Keggin POMs $H_3(PW_{12}O_{40})$ have been embedded in the insulating stack of metal-insulator-semiconductor (MIS) capacitors, with increasing complexity of the stack structure, and electrical behavior and memory cell performances were partly evaluated. The memory performances of the devices discussed in this section are summarized in Table 1. In the simplest case, the MIS capacitor is n-Si/$SiO_2$(6.5 nm)/APTES(≈ 1 nm)/POMs(aggregates)/IPA(n.g.)/Al.[74] The values in brackets are the measured thickness of the layer (n.g. stands for not given in the publication). The silicon substrate is n-type moderately doped (1-2 Ω.cm, *i.e.* 2-5x$10^{15}$ $cm^{-3}$), APTES is aminopropyl triethoxysilane used to provide positive charges at the $SiO_2$ surface and to facilitate the electrostatic deposition of the POM anions, IPA stands for an oligolayer of isopentylamine. The role of this IPA layer is to passivate the stacks and to insulate the POM (charge nodes) from the metal gate (Fig. 12-a). Scanning electron microscope images revealed that the POMs do not form a homogeneous layer, but uniformly distributed aggregates (mean diameter ≈ 17 nm, height 5-14 nm). Since the LUMO of the POM is near the Fermi energy of the electrodes (see Figs. 3-a to 3-c, for the electronic structures of the same POMs embedded in similar molecular junctions and discussion in Section 2.2.1), the POMs can be easily charged/ discharged by the electrons injected from the electrodes, which give rise to large hysteresis in the capacitance-voltage (C-V) and the current-voltage (I-V) curves, as well as a negative differential resistance when resonant electron transport occurs via the POM LUMO, within a weak voltage window (-2 to 0 V) - Fig. 12-a. However, only the maximum charge density of the memory cell was reported (≈ 0.14 - 0.26 µC/cm$^2$) whatever the IPA insulating layer is present or not. Pulsed transient capacitance measurements showed that the IPA layer reduced the charging/discharging rate of the POMs and likely increased the data retention time (which was not measured).[74] The same POMs were embedded in a more complex stack based on high-k dielectric materials, these high-k dielectrics being widely used in nanoelectronics to reduce leakage current (thicker dielectric layer) at equivalent capacitance of the device. Several cell structures were fabricated and characterized: (i) n-Si/$SiO_2$(3.1 nm)/APTES(≈ 1 nm)/POMs/IPA(n.g.)/$Ta_2O_5$ (25, 80 or 145 nm)/ Al (P-Ox for short), (ii) n-Si/$SiO_2$(3.1 nm)/$Ta_2O_5$ (25, 80 or 145 nm)/APTES(≈ 1 nm)/POMs/IPA(n.g.)/ Al (Ox-P for short), (iii) n-Si/$SiO_2$(3.1 nm)/$Ta_2O_5$ (25, 80 or 145 nm) APTES(≈ 1 nm)/POMs/ IPA(n.g.)/$Ta_2O_5$ (25, 80 or 145 nm)/Al (Ox-P-Ox for short), see Fig. 12-b.[257] In these devices, the $Ta_2O_5$ layer is a porous nanocrystalline structure (ca. 7 8% of porosity) and the POMs diffuse inside the layer forming a hybrid $Ta_2O_5$/POM charge storing layer (note that oxygen vacancies in the $Ta_2O_5$ may also contribute to the trapping of electrons). The best memory cell performances were obtained for the Ox-P-Ox device with a 25 nm thick $Ta_2O_5$ layer. A programming window of 4 V is obtained with a write pulse voltage of 20 V and a write pulse time of 100 ms (Fig. 12-b). The programming window is the shift of the flat-band voltage $V_{FB}$ measured by high frequency (1 MHz) capacitance-voltage curves after each successive applied write pulse. The data retention time (here in this review, defined as the time at which the programming window has decreased by 50% after the application of a single write pulse) is ca. 300 s (Fig. 12-b). The charge density is estimated at ca. 30-50 µC/cm$^2$.[257] Increasing the $Ta_2O_5$ layer thickness induces a large charge trapping (a larger hysteresis of the capacitance-voltage curves), likely due to an increase in the negatively charged oxygen vacancies, but at the price of the need for higher write pulse voltages and times. The memory cells with the P-Ox and Ox-P structures can be written faster (100 µs) but with a weaker programming window (ca. 1 V), Fig. 12-b. Adding an insulating and blocking layer of $Al_2O_3$ (20 nm thick) on top of the $Ta_2O_5$/POM charging layer increases the programming window up to 7 V (Fig. 12-c) but the write pulse voltage threshold increases from 6 V to 10 V (Figs. 11-b and 11-c).[258] $Al_2O_3$ has a larger band gap (6-8 eV) than $Ta_2O_5$ (4.9 eV) and it is a more efficient blocking layer to reduce the relaxation of the trapped electrons. Remarkably, the same programming window (at a write pulse of 20 V) is independent of the pulse write time from 10 ns to 100 ms. Admittance spectroscopy measurements confirmed the electron trapping/detrapping by the POMs with time constants in the 10-100 ns range. The data retention time is about $10^3$ seconds and the maximum charge density is estimated between ca. 250 and 750 μC/cm$^2$ (depending on how many electrons are trapped in individual POM, assumed to be 1 to 3 electrons per reduced POM). Without the $Ta_2O_5$ layer, i.e. with the $Al_2O_3$ layer on top of n-Si/$SiO_2$/APTES/POM structure, the programming window is reduced to ≈ 4 V (Fig. 12-c), the maximum charge density is weaker (≈ 100 - 300 μC/cm$^2$) but, surprisingly, the data retention data is extrapolated to ≈ 10 years (Fig. 12-c). The reason is not clear: the weaker performance of the $Ta_2O_5$/POM system is attributed to a fast decrease in the first 10 s, tentatively explained by internal charge mitigation in the hybrid $Ta_2O_5$/POM material.[258] The best memory cell performance data are summarized in Table 1. The erase functionality cannot be tested with these n-Si capacitor structures, since the injection of minority carriers (i.e., holes) is not available, contrary to a transistor cell for which both carriers are available from the transistor channel and the source-drain regions of opposite doping polarities. We note that all these devices were fabricated with technology processes that are compatible to co-integration with Si CMOS circuits (at least at BEOL level).

POMs have also been embedded in three terminals (3T), transistor memory devices. A solution of the Dawson-like $[W_{18}O_{54}(SeO_3)_2]^{4-}$ POMs (Fig. 13-a) was drop-cast on a 3T transistor device displaying a lateral geometry and consisting of two source and drain electrodes, a Si nanowire channel (≈ 4-5 nm width, ≈ 60 nm height and ≈ 200 nm long) covered with 4 nm thick $SiO_2$ and a side-control gate at a distance ≈ 60 nm from the channel (Fig. 13-b).[73] The POMs were distributed over the entire device, with a large density of $2x10^{15}$ cm$^{-2}$. They were charged and discharged by applying a ± 20V voltage at the gate. The programming window is measured by the threshold voltage shift $\Delta V_T$ on the transfer characteristics (drain current $I_D$ versus gate voltage $V_G$ curves) between charged and uncharged POMs (Fig. 13-c). After a write pulse at -20 V/1 s, which negatively charges the POMs, a positive shift $\Delta V_T \approx 1.2$ V (at low $I_D \approx 10^{-10}$ A) to ≈ 2.5 V (at $I_D \approx 10^{-7}$ A) was measured. The memory is erased by applying a symmetric +20 V pulse, which discharges the POMs, and causes the $I_D$-$V_G$ almost returned to its initial behavior (Fig. 13-c). A minimum pulse time of 0.1 seconds is required to induce a measurable $\Delta V_T$. In terms of performance, the write/erase times of 0.1 - 1 s are long, partly due to the large POM density and the non-optimized lateral gate geometry. The information was retained for at least 336 hours and it is expected a longer data retention time, since no significant loss of charge has been observed over this time period.[73] Simulations combining DFT and mesoscopic device modeling were carried out to evaluate the potentiality of POMs in realistic flash memories. The simulated devices used a POM layer as a charge floating gate. Similar results were obtained for both $[W_{18}O_{54}(SeO_3)_2]^{4-}$ and $[W_{18}O_{56}(WO_6)]^{10-}$ POMs in their one- or two-electron reduced forms, the two POMs giving identical $I_D$-$V_G$ characteristics, the more reduced the POM the higher the threshold voltage shift $\Delta V_T$. The differences in the local charge distribution in the POMs are not sensed by the larger channel area.[73] However, the variability of the position and number of POMs in the gate dielectric plays an important role in the variability of the programming window of the memory cell, exceeding the effect of the random dopant fluctuations usually encountered in nanoscale MOS devices.[259, 260] The simulations also showed that a FDSOI (fully depleted silicon-on-insulator) flash memory cell is less affected by these POM fluctuations than a bulk silicon memory.[261] The $[W_{18}O_{54}(SeO_3)_2]^{4-}$ POM is templated with two selenite $[Se(IV)O_3]^{2-}$ moieties that are prone to be oxidized to form a $[Se(V)_2O_6]^{2-}$ dimer. However, these authors showed that once the Se are oxidized and formed a Se-Se bond, the memory can no longer be switched between the two states, remaining in the erase state.[73]

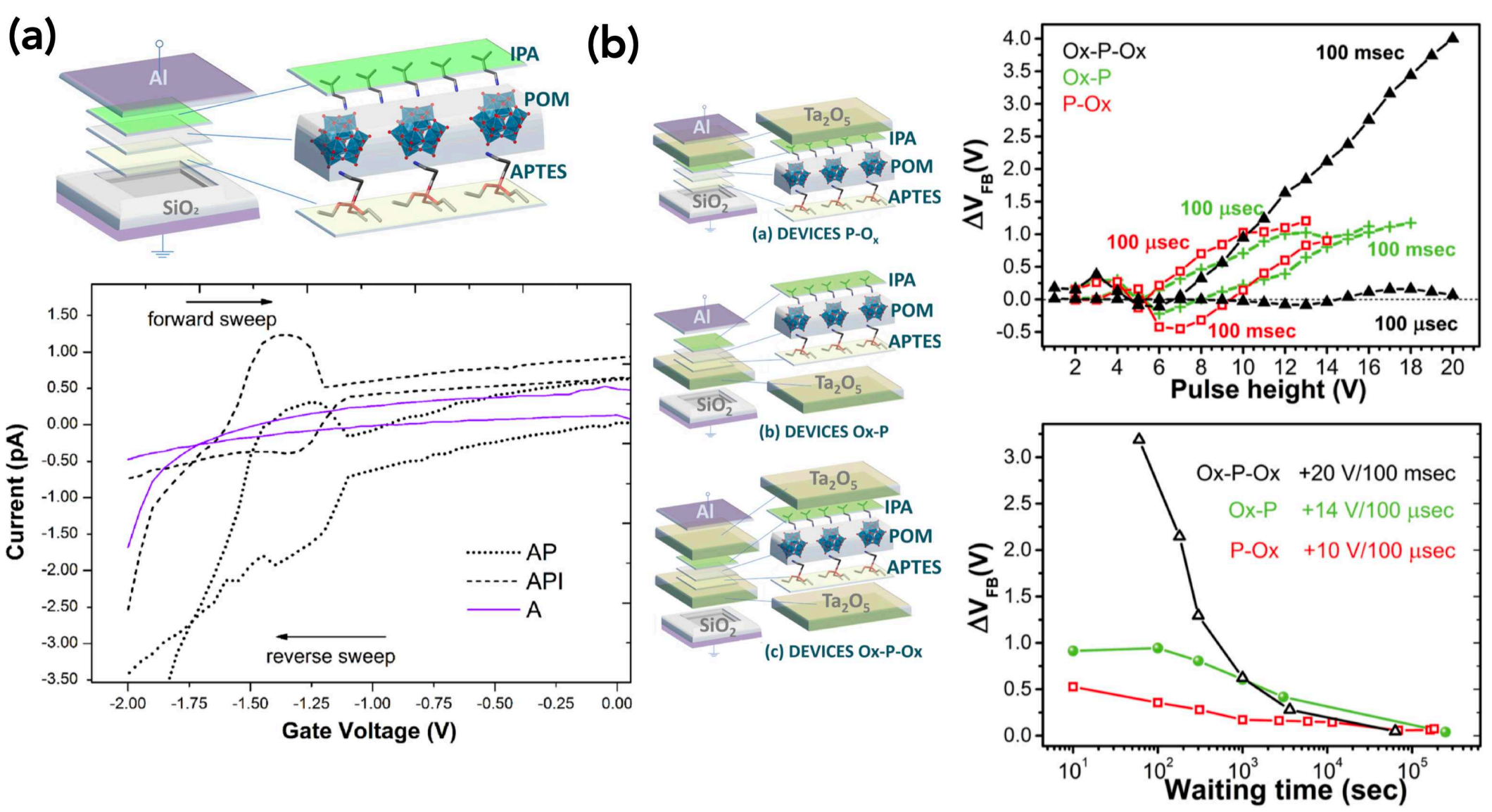

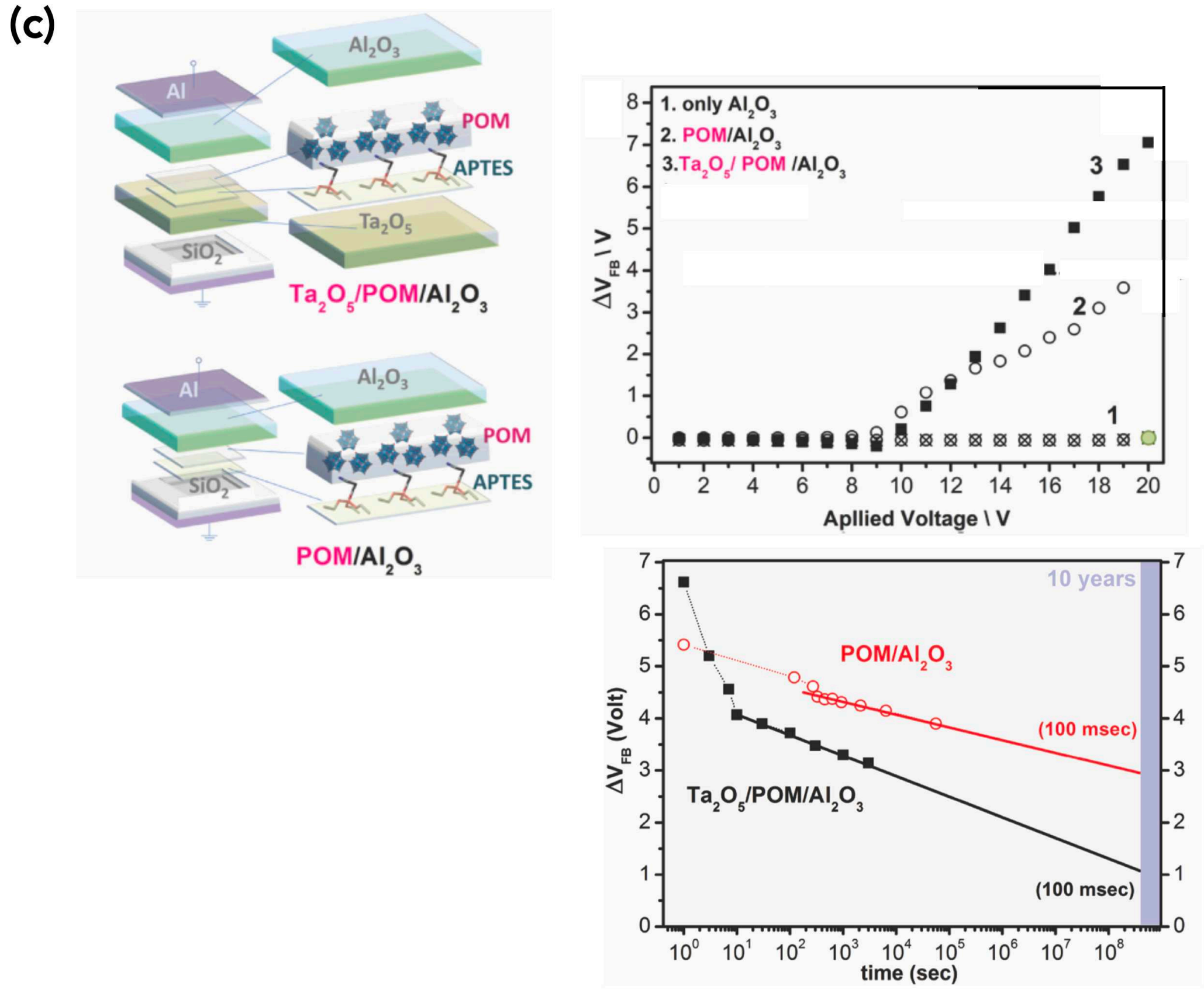

***Figure 12.*** *Schemes of the structures of the 2T MIS capacitor memory cells incorporating $H_3[PW_{12}O_{40}]$ POMs and typical electrical memory behaviors:* ***(a)*** *Si/$SiO_2$/APTES/POMs/IPA/Al (see text for details), current-voltage (I-V) curves for the devices with (labeled API) and without (AP) the IPA top layer. The purple curve (A) is for the Si/$SiO_2$/APTES/Al structure (no POM). Reproduced with permission from ref. 74. Copyright (2014) American Institute of Physics.* ***(b)*** *Cells with a hybrid $Ta_2O_5$/POMs core in three different structures (see text for details). Evolution of the programming window $\Delta V_{FB}$ versus the amplitude of the write pulse at a pulse time of 100 µs and 100 ms. Data retention behavior: the decrease in the programming window is measured versus time at rest after a single write pulse (as indicated in the panel). Reproduced from ref. 257. Copyright (2016) American Chemical Society. Data retention time is defined in this review as the time at which $\Delta V_{FB}$ has lost 50% of its initial value.* ***(c)*** *Cells with an insulating $Al_2O_3$ layer on top of the stack. The write and data retention behaviors are presented as in panels (b). Reproduced with permission from ref. 258. Copyright (2019) John Wiley and Sons.*

| POM | Cell architecture | Write | | Erase | | Programming windows | Retention time (s) | Endurance (W/E cycles) | Charge density (μC/cm²) | Ref. |
|---|---|---|---|---|---|---|---|---|---|---|
| | | voltage | time | voltage | time | | | | | |
| $[PW_{12}O_{40}]^{3-}$ | 2T, MIS capacitor Si/$SiO_2$/APTES/POMs/IPA/Al | | | | | | | | ≈ 0.14-0.26 *(c)* | 74 |
| | 2T, MIS capacitor Si/$SiO_2$/$Ta_2O_5$/POMs/$Ta_2O_5$/Al | > 6 V | 100 µs 100 ms | | | 4 V *(a)* | ≈ 300 | | ≈ 30-50 *(c)* | 257 |
| | 2T, MIS capacitor Si/$SiO_2$/$Ta_2O_5$/POMs/$Al_2O_3$/Al | > 10 V | 10 ns | | | 7 V *(b)* | ≈ $10^3$ | | ≈ 250-750 *(d)* | 258 |
| | 2T, MIS capacitor Si/$SiO_2$/POMs/$Al_2O_3$/Al | | | | | 4 V *(b)* | ≈ $3.2x10^8$ (10 yrs) | | ≈ 100-300 *(d)* | |
| $[W_{18}O_{54}(SeO_3)_2]^{4-}$ | 3T, Si nanowire/$SiO_2$/POMs transistor, lateral gate | - 20 V | 0.1 - 1 s | + 20 V | 0.1 - 1 s | 1.2 - 2.5 V | ≈ $10^6$ | | ≈ 320 *(e)* | 73 |
| $[PW_{12}O_{40}]^{3-}$ | 3T, pentacene/PMMA/POMs/ rGO/$SiO_2$/Si back gate | ($e^-$) + 50 V ($h^+$) - 50 V | 1 s | ($e^-$) - 50 V ($h^+$) + 50 V | 1 s | ($e^-$) 22 V ($h^+$) 18 V | ≈ $10^3$ | | | 262 |
| Benchmarking | Flash memory (floating-gate and charge-trap) | ≈ 5-10 V | ≈ 1 - 50 µs | ≈ 10-20 V | ≈ 1 ms | ≈ 3-8 V | ≈ $3.2x10^8$ (10yrs) | ≈ $10^5$-$10^6$ | | see text |

*(a) at 20V/100 ms write pulse. (b) at 20V/100 ns to 100 ms write pulse. (c) calculated from the measured voltage shift of the electrical characteristics ($Q=C\delta V$, C the device capacitance). (d) calculated from the measured density of "electron traps" (i.e. POMs) $D_T$ assuming n = 1 to 3 electrons per reduced POMs, $Q=nqD_T$, q the electron charge). (e) from an estimated density of POMs and assuming one electron reduced POM.*

***Table 1.*** *This table summarizes the best memory cell performances for various POMs embedded in several two terminal (2T) and three terminal (3T) charge (capacitance) memories (the structures given in the table are simplified, see text for details). When data were not available, the table is light gray filled. None of these studies examined the endurance of the memory cells. In the benchmarking line, the voltages are in absolute values, usually the erase voltages use reverse polarities than the write voltages. The values give typical ranges; the exact performances depend on the type of flash memory (i.e., floating gate or charge trap) and on the architecture (NAND or NOR flash).*

The properties of a field-effect transistor (FET) incorporating a POM-rGO (reduced graphene oxide) charge trapping floating gate were also investigated.[262] The floating gate was built on a stack of alternate layers of GO and $H_3[PW_{12}O_{40}]$ POMs, respectively, deposited on 3-aminopropyltriethoxysilane (APTES) and poly(allylamine)hydrochloride (PAH) and assembled by the electrostatic layer-by-layer technique on a Si-back gate covered with 100 nm thick $SiO_2$ (Fig. 13-d). After POM-mediated UV-photoreduction of GO to rGO, a polymethyl methacrylate (PMMA) film was spin-coated on the top of the POM-rGO assembly as the tunneling dielectric layer. Finally, p-type pentacene (30 nm) was used as the transistor channel material between Au source and drain electrodes, deposited by thermal evaporation (Fig. 13-d). Contrary to most flash memories, which exhibit unipolar charge trapping behavior, the POM-rGO based transistor showed an ambipolar behavior, trapping both electrons and holes, depending on the polarity of the programming operation (programming/erasing bias pulses of +50/-50V for 1s, a high operation voltage due to the 100 nm thick $SiO_2$ gate dielectric), Fig. 13-e. The electron/hole trapping was confirmed by measuring the surface potential of POM/rGO stacks with KPFM (Kelvin probe force microscopy) upon injection of charges from the probe tip. The use of a double floating gate (rGO and POM) enlarged the programming windows (≈ 18 - 22 V) compared to POM or rGO alone or POM/GO hybrid (≈ 8 - 15 V). Data storage (less than 50 % of loss) was maintained after $10^3$ s.[262] This device was used as an artificial synapse (see Section 5). Finally, we note that none of these studies examined the endurance (number of write/erase cycles without a significant loss of performance) of these memories (Table 1).

Compared with today's performance of FLASH memories (both floating gate and charge trap),[263-265] we conclude that several POM-based memories display write/erase voltages and programming windows on a par with the existing flash memories. However, the write/erase times generally remain below the state-of the art. Moreover, the endurance of these POM-based memory cells remains to be evaluated.

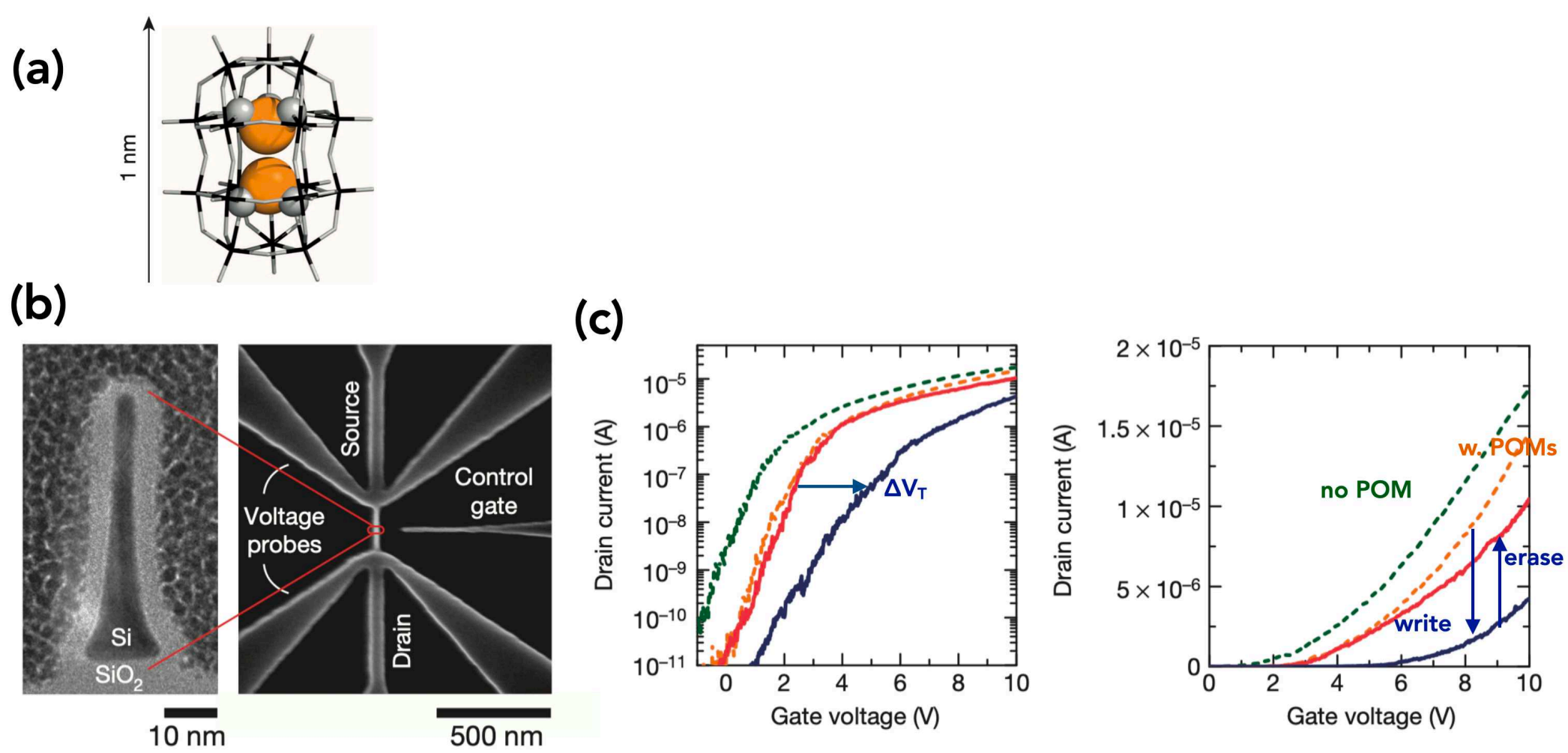

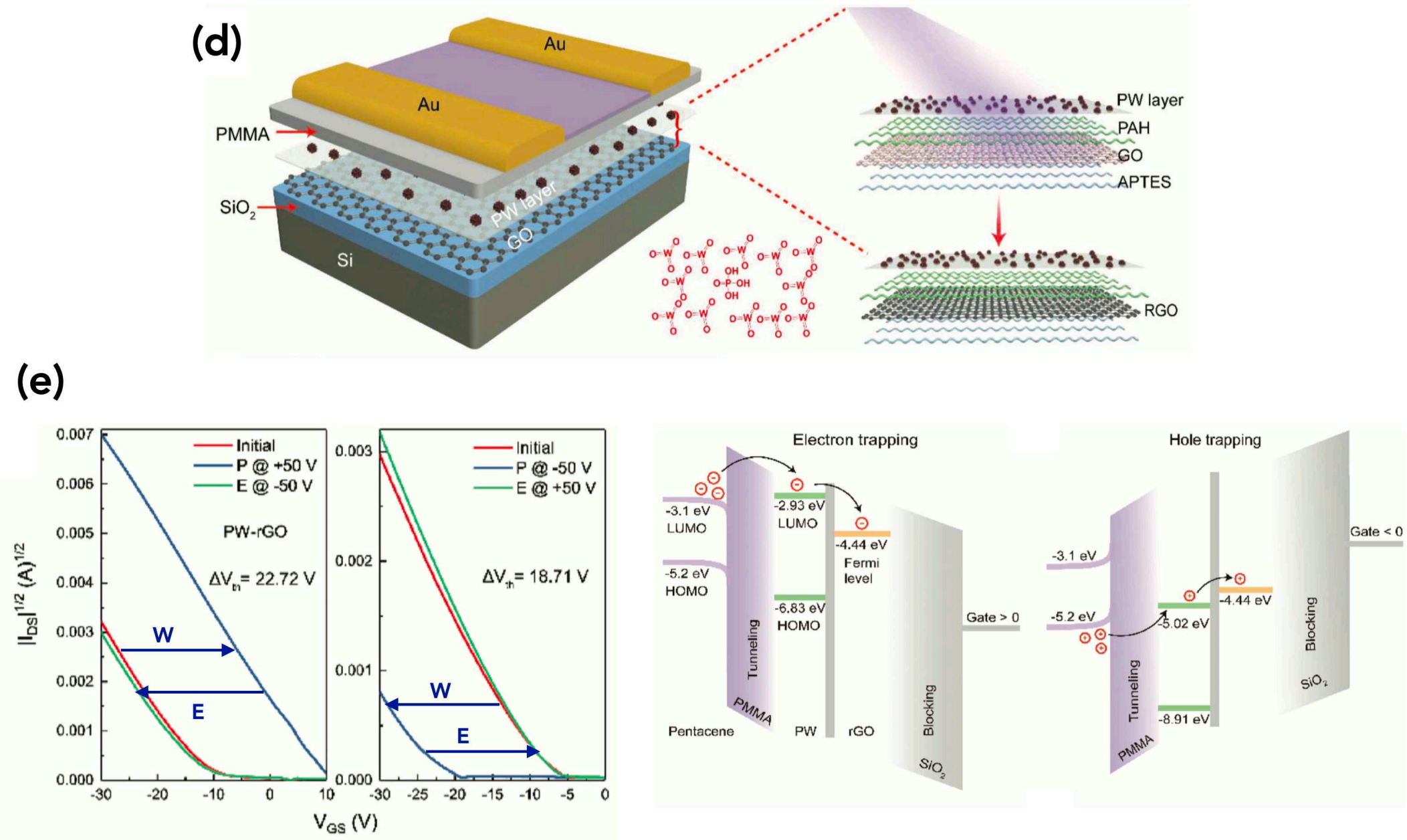

***Figure 13.*** ***(a)*** *Structure of the Dawson type* $[W_{18}O_{54}(SeO_3)_2]^{4-}$ *POM.* ***(b)*** *Cross-section TEM (transmission electron microscope) image of the Si/*$SiO_2$ *nanowire and top-view SEM (scanning electron microscope) image of the device.* ***(c)*** *Typical drain current - gate voltage (*$I_D$*-*$V_G$*) of the memory cell in log-scale (left) and lin-scale (right): dashed green (no POM), dashed orange (after deposition of the POMs), solid blue (after a write pulse - 20 V/1 s), solid red (after an erase pulse + 20 V/ 1 s). (a-c) Reproduced with permission from ref. 73. Copyright (2014) Springer Nature.* ***(d)*** *Structure of the pentacene transistor with the rGO/POM* $[PW_{12}O_{40}]^{3-}$ *in the gate dielectric stack.* ***(e)*** *Drain current - gate voltage characteristics (plotted as* $|I_D|^{1/2}$*) for the initial state (red line), after a write pulse (blue line) and an erase pulse (green line) - voltage pulse indicated in the panel (all for 1 s) in the case of electron trapping (left panel) and hole trapping (right panel). Schemes of the electronic structure of the memory cell. (d-e) Reproduced with permission from ref. 262. Copyright (2018) John Wiley and Sons.*

### 3.2. Resistive switching (RS) memories.

Since the redox state of the POMs significantly modifies its electrical conductance (see Section 2.2.1), several works have explored the possibility of using them as the building blocks for resistive switching (RS) memories and to evaluate their performances. The general memory cell structure consists of a film of POMs included in a matrix material and sandwiched between two electrodes (mainly vertical structure). The current-voltage (I-V) shows a hysteric behavior with a high-resistance state (HRS) and a low-resistance state (LRS), with a switching at a set voltage ($V_{SET}$) and a back switching at a reset voltage ($V_{RESET}$). The memory is ambipolar when $V_{SET}$ and $V_{RESET}$ have opposite polarities (otherwise they are unipolar). The HRS/LRS current ratio is measured at a lower voltage than the set and reset voltages. If the current value of the HRS/LRS states evolves upon successive applied voltage sweeps, or depends on the pattern and sequence of applied write pulses, the device can be classified as a memristor[266, 267] (memory resistor, i.e., the resistance at a time t depends on the device past history). The device is an NVM (non-volatile memory) if the ON state is retained for a given time (data retention time) after the voltage is turned off to 0 V. In certain cases, the memory can be switched from the HRS to LRS only once during the first voltage sweep and then it remains in this state (WORM memory, write-once read many times). The best memory cell performances are given in Table 2.

Anderson-type POMs $TBA_3[Mn^{III}Mo_6O_{18}\{(OCH_2)_3CNH_2\}_2]$ were covalently inserted into PMMA, forming a hybrid copolymer and a film (70 nm thick), which was spin-coated onto an ITO electrode and contacted by a top Pt electrode.[75] This example departs from those that will be published in the following years by several points: (i) the shaping process of the POM leading to a covalent hybrid polymer, thanks to the tris ligand, which was further extended with a methacrylate group and copolymerized with methyl methacrylate; (ii) the localization of the redox processes on the central Mn cation and not on the polyoxomolybdate core; (iii) the attribution of the resistive switching behavior to a change in the redox state of the Mn, with three oxidation states Mn(IV), Mn(III) and Mn(II) involved, giving rise to a ternary memory. Initially in the HRS, the device switches to the LRS at $V_{SET}$ = -1.35 V (applied to the gate to inject electrons from Pt electrode), Fig. 14-a. This ON state is retained after the voltage is turned off (0 V). The device is switched back to the initial state at $V_{RESET}$ = +1.5 V. However, if the positive voltage sweep is stopped at 1 V, an intermediate state is observed (medium resistance state, MRS). The resistance ratios (measured at 0.3V) are MRS/LRS ≈ 4 and HRS/LRS ≈ 20. This behavior is reproducible for 50 cycles without noticeable fatigue. The data retention of the 3 states was measured up to $10^4$ without any loss of data (these data are summarized in Table 2). This multi-level resistive switching was ascribed to the multi-redox states of Mn in the POM. In the HRS, the Mn atoms are in their highest valence state (Mn(IV)) and electrons injected from the top gate at $V_{SET}$ induced a reduction to Mn(II). In the RESET process, Mn(II) oxidized to Mn(III) if the applied voltage is +1V or Mn(IV) if the applied voltage is +1.5 V.[75] The description of this RS system thus follows those also reported for devices based on metal coordination complexes, organometallics and purely organic molecules.[38, 39, 122, 124] The charge transport mechanism was not addressed. The density of POMs in the insulating PMMA is not accurately known, but the high current density (up to ≈ 10 $A/cm^2$) for a 70 nm-thick film suggests that POMs are close enough to allow electron tunneling and hoping between POMs.

Crossbar memory cells were measured for a blend of $H_3[PW_{12}O_{40}]$ with PMMA. A 35 nm thick film was spin-coated between ITO and Au electrodes.[268] A bipolar NVM memory effect was observed with $V_{SET}$ = 1.2 V (applied to ITO, electrons injected from Au to reduce the POMs) and $V_{RESET}$ = -1.7 V (Fig. 14-b). The HRS/LRS ratio is ca. 600 (at a reading voltage of 0.5 V). These performances are maintained up to 100 cycles and the data retention time was tested up to $10^4$ s.[268] The ET mechanism is likely electron tunnel hopping between adjacent POMs, the average neighboring distance being ≈ 2-3 nm (from TEM image), controlled by the charge states of the POMs. A stack of the same $H_3[PW_{12}O_{40}]$ POMs and flakes of GO (graphene oxide) sandwiched between ITO bottom electrode and Al gate initially showed a WORM (write-once-read-many-times), and then a bistable RS behavior when the GO was turned to reduced graphene oxide (rGO) by POM-assisted UV photocatalytic reduction,[269] Fig. 14-c. The WORM memory switches from HRS to LRS during the first voltage sweep at $V_{SET}$ ≈ 4 V (bias applied to the top electrode) and it remains in the LRS states for all the subsequent voltage sweeps (-5 to 5 V). The HRS/LRS ratio is ≈ $10^4$. With the rGO, a bistable RS is observed with $V_{SET/RESET}$ ≈ 3.2/-3.4V, a HRS/LRS ratio of ≈ $10^3$ and endurance tested up to 200 cycles and a retention time estimated at $4 \times 10^4$ s. In both cases, the HRS/LRS ratios are greatly enhanced compared to devices with POM or GO alone (HRS/LRS ≈ 10). The transition from WORM to bistable NVM is tentatively explained by a larger energy barrier at the GO/POM than at the rGO/POM interface, which blocks the detrapping of electrons in the GO to go back to the gate electrode, or would have required $V_{RESET}$ < -5 V (not tested) to observe a bipolar RS switching in the ITO/GO/POM/Al devices.

Another approach used hydrothermal techniques to combine POMs with metal-viologen complexes or simple viologen cations. The POMs have either been found in the cavities of metal-organic-frameworks (MOF) such as $[GeW_{12}O_{40}]^{4-}$,[270] $[SiW_{12}O_{40}]^{4-}$,[271] or part of oligomeric, 1D or 2D assemblies ($[Co_4(H_2O)_2(B\text{-}\alpha\text{-}PW_9O_{34})_2]^{10-}$ in ref. 272, β- and γ-$[Mo_8O_{26}]^{4-}$ in ref. 273 or $[MnMo_6O_{18}L_2]^{3-}$ in ref. 274). In these approaches, mixing the POMs with host materials possessing thermochromic properties brings enhanced thermal stability to the hybrid materials and allows the memories to work at higher temperatures than the usual limit for many electronic devices (> 125 °C). Bipolar NVM behavior has been measured at RT and 150°C for Keggin-type $[GeW_{12}O_{40}]^{4-}$ POMs inserted in the cavities of a metaloviologen MOF, $[Co_2(bpdo)_4(H_2O)_6]_n^{4n+}$ with

bdpo = 4,4'-bipyridine N,N' dioxide. Memory cells were fabricated by spin-coating the POM@MOF film (2.5 µm thick) on ITO and contacted by a network of Ag nanowires.[270] The typical averaged parameters are $V_{SET}$ = 1.77 V, $V_{RESET}$ = -3.42 V (independent of the temperature), but a higher HRS/LRS ratio ≈ $3.5x10^3$ was measured at 150°C than at RT (≈ 100), at a reading voltage of -0.6 V. No performance deterioration was noticed after 100 set/reset cycles. The same group reported an unusual behavior for a NVM device based on another POM ($[Co_4(H_2O)_2(B\text{-}\alpha\text{-}PW_9O_{34})_2]$) forming 1D or 2D polymers mixed with metalloviologen complexes.[272] In the temperature range 30 - 60 °C, no RS has been observed. When heated at 150° C, a bipolar NVM behavior appears ($V_{SET}$ = 0.75 V, $V_{RESET}$ = -0.63 V, HRS/LRS ratio ≈ 27, stable up to 20 cycles), Fig. 14-d. These approaches, combining POMs and cations with metal-viologen complexes, increase the complexity, since both partners are redox active. Controlled experiments have shown some synergy, each partner taken individually having no or poor RS performance.[274] This amazing transition from a resistive switching silent system in the low temperature range of 30-150°C to a NVM behavior in the high temperature range of 150-270°C has been ascribed to thermal-induced reversible structural changes from 1D oligomers formed between $[Co_4(H_2O)_2(B\text{-}\alpha\text{-}PW_9O_{34})_2]^{10-}$ POMs and cobalt-(4'4'-bipyridine-N,N'dioxide) complexes to 2D materials (Fig. 14-d).[272] At 270°C, an RS switching behavior is still observed, but with a degraded behavior (switching less "abrupt", weak hysteresis (i.e. RS) at V < 0 and the hysteresis is anti-clockwise (while clockwise at 150 °C).[272] This feature was not explained and the charge transport mechanism remains to be further investigated, the suggested SCLC mechanism being not convincing demonstrated. Furthermore such hybrid hetero-systems are light responsive, opening perspectives for multi-addressable photonic memory devices, optically programmable and electrically erasable.[271]

The role of the host matrix can also be optimized to enhance the RS performance. The Anderson-type POM, $[MnMo_6O_{18}L_2]^{3-}$ incorporating 2 ligands L = 2-(hydroxymethyl)-2-(pyridine-4-yl)-1,3-propanediol), has been combined with methylviologen ($MV^{2+}$), ($MV^{2+}$)/$Cu^{2+}$ and ($MV^{2+}$)/($Cu_2I_3$)$^-$ and films (thickness ≈ 600 -960 nm) of these materials were spin-coated on ITO and top contacted with Ag paste.[274] The best results were obtained with the iodocuprate (see Table 2), in particular with HRS/LRS ratio of ≈ 230 (vs. 65 and 13 with ($MV^{2+}$)/$Cu^{2+}$ and $MV^{2+}$, respectively). The 3 compounds clearly displayed different structures (determined by XRD): discrete POM-MV association, 2D layer of POM-Cu and 1D chain of POM-$Cu_2I_3$. The precise reasons for this performance enhancement remain to be clearly understood. The proposed mechanism[274] involving the formation and rupture of conducting filaments of oxygen vacancies can be discarded (see discussion below). The 1D chain organization might favor electron transfer from POM to POM and facilitate their reduction.

| POM | Cell type | Cell architecture | set voltage [d] | reset voltage [d] | HRS/LRS ratio | Retention time (s) | Endurance (set/reset cycles) | Ref. |
|---|---|---|---|---|---|---|---|---|
| $[Mn^{III}Mo_6O_{18}\{(OCH_2)_3CNH_2\}_2]^{3-}$ | NVM, bipolar 3 states | ITO/PMMA-POM hybrid/Pt | -1.35 V | 1 V [a]<br>1.5 V | ≈ 5 [b]<br>≈ 20 [c] | $10^4$ | 50 | 75 |
| $[PW_{12}O_{40}]^{3-}$ | NVM bipolar | ITO/PMMA-POM blend/Pt | 1.2 V | - 1.7 V | ≈ 600 | $10^4$ | 100 | 268 |
| $[PW_{12}O_{40}]^{3-}$ | WORM | ITO/POM/GO/Al | 4 V | | $10^4$ | | | 269 |
| | NVM bipolar | ITO/POM/rGO/Al | 3.2 V | -3.4 V | $10^3$ | $4x10^4$ | 200 | |
| $[GeW_{12}O_{40}]^{4-}$ | NVM bipolar | ITO/POM@MOF/ Ag(nanowires) | 1.77 V | -3.42 V | ≈100 (at RT)<br>≈ $3.5x10^3$ (at 150°C | | 100 | 270 |
| $[Co_4(H_2O)_2(B\text{-}\alpha\text{-}PW_9O_{34})_2]^{10-}$ | NVM bipolar if T>150°C | ITO/POM-MV/Ag | 0.75 V | -0.63 V | 27 | | 20 | 272 |
| $[MnMo_6O_{18}L_2]^{3-}$ | NVM bipolar | ITO/POM-MV/Ag | 1.4 V | -2.5 V | 13 | | 20 | 274 |
| | | ITO/POM-MV-Cu/Ag | 1.4 V | -3.6 V | 65 | | 84 | |
| | | ITO/POM-MV-$Cu_2I_3$/Ag | 1 V | -3.5 V | 230 | $10^4$ | 700 | |
| $[P_2W_{17}O_{62}(SiR)_2]^{6-}$ | WORM | ITO/cross-linked POMs/Au | -1.1/-1.6 V | | ≈ 100 | $10^4$ | | 275 |
| | NVM bipolar | ITO/cross-linked POMs/Au | -2 V | 4 V | ≈ 10 | | | |
| $[PW_{12}O_{40}]^{3-}$ | memristor unipolar | ITO/POM-P3HT/Ag | 0.54 V | 0.06 V | ≈ $10^6$ | 200 ms | 300 | 276 |
| | NVM bipolar | ITO/POM-P3HT/Au | ≈ 3-4 V | ≈ -2 V | ≈ $10^4$ | | | |
| Benchmarking | RRAM (oxide) | | ≈ 0.5 V | | $10^9$-$10^{10}$ | ≈ $3.2x10^8$ (10yrs) | ≈ $10^{12}$ | see text |
| | RRAM (PCM, 2D, perovskite,....) | | ≈ 1.5 V | | $10^7$-$10^8$ | ≈ $1.2x10^8$ (4yrs) | ≈ $10^9$ | |
| | RRAM (organic) | | ≈ 0.5 V | | ≈ $10^9$ | $10^5$-$10^6$ | ≈ $10^{12}$ | |

*(a) intermediate state medium resistance state (MRS). (b) resistance ratio HRS/MRS. (c) resistance ratio HRS/LRS. (d) the polarity of the set/reset voltages depends on the terminal connections (i.e., substrate or top gate grounded) and on which electrode is the easiest electron injector.*

***Table 2.*** *This table summarizes the main memory cell performances for various POMs embedded in RS memory cells. When data were not available, the table is light gray filled. The last 3 lines give the best performances for 3 classes of RRAM (resistive random access memory) studied in today's nanoelectronics and under development (see text for a short discussion).*

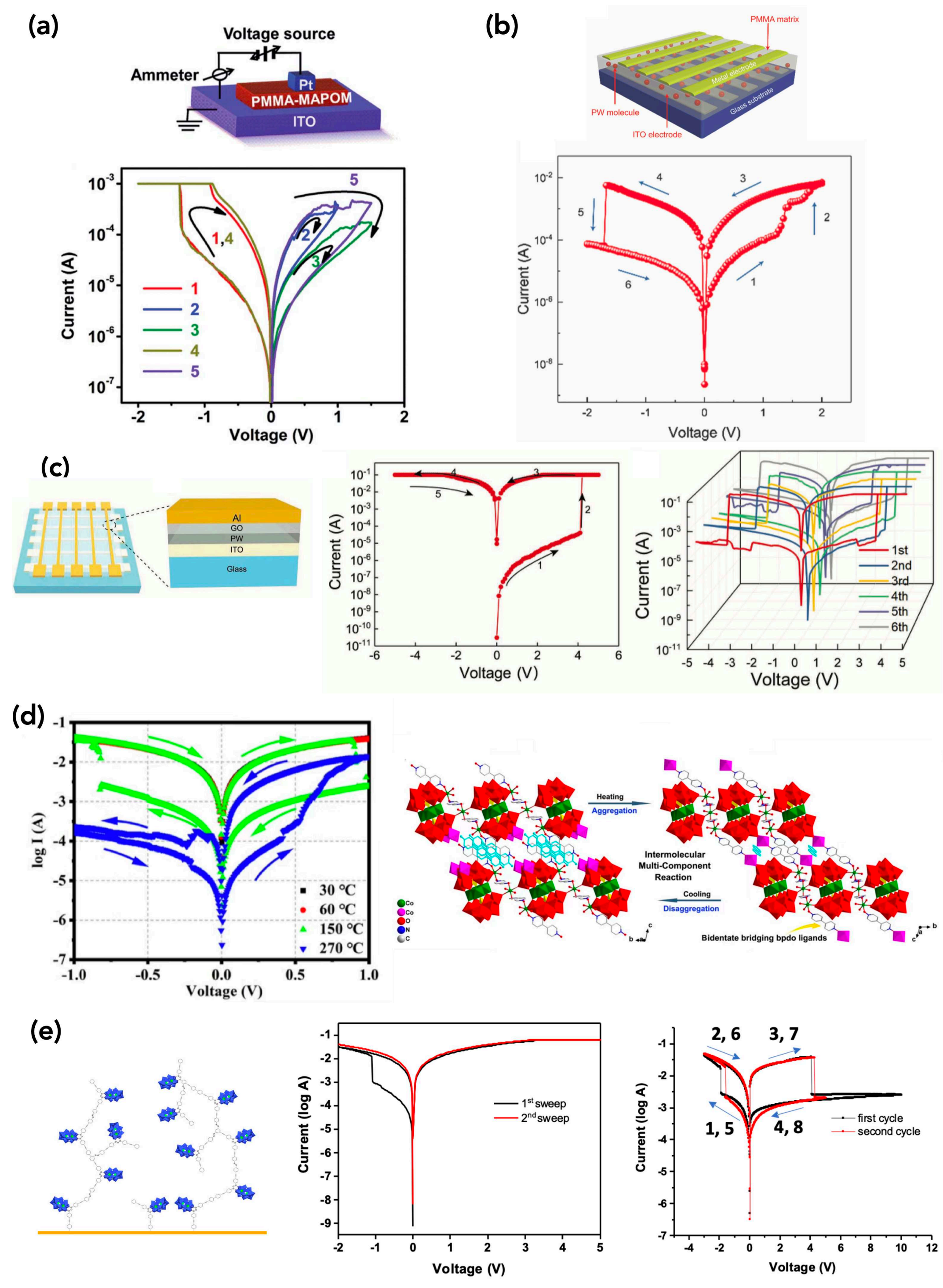

***Figure 14**. **(a)** Scheme of the ITO/PMMA-POM hybrid/Pt (POM is $[Mn^{III}Mo_6O_{18}\{(OCH_2)_3CNH_2\}_2]^{3-}$) and I-Vs showing the three states depending on the applied positive voltages (1 or 1.5 V). Reproduced with permission from ref. 75. Copyright (2014) Royal Society of Chemistry. **(b)***

*Structure of the ITO/PMMA-$[PW_{12}O_{40}]^{3-}$blend/Pt device and a typical RS switching I-V curve. Reproduced with permission from ref. 268. Copyright (2019) Royal Society of Chemistry.* ***(c)*** *Scheme of the ITO/$[PW_{12}O_{40}]^{3-}$/GO(or rGO)/Al memory cells and examples of the WORM (with GO), middle panel, or bipolar NVM (with rGO), right panel, behaviors. Reproduced with permission from ref. 269. Copyright (2019) John Wiley and Sons.* ***(d)*** *I-V curves of $[Co_4(H_2O)_2(B\text{-}\alpha\text{-}PW_9O_{34})_2]^{10-}$ mixed with a metalloviologen polymer (MV) showing no RS at 30 and 60°C and bipolar NVM after heating >150 °C. View of the structural and dimensional transition from 1D oligomers formed between $[Co_4(H_2O)_2(B\text{-}\alpha\text{-}PW_9O_{34})_2]^{10-}$ POMs and cobalt-(4'4'-bipyridine-N,N'dioxide) complexes to 2D materials after heating. Reproduced with permission from ref. 272. Copyright (2021) John Wiley and Sons.* ***(e)*** *Illustration of the POM-polymer cross-linked network grown on ITO. POM is $[P_2W_{17}O_{61}]^{6-}$. Typical I-V of the dominant WORM behavior (middle panel) and bipolar NVM (≈ 5 % of the devices, right panel). Reproduced with permission from ref. 275. Copyright (2024) Royal Society of Chemistry.*

More recently, another route was proposed where an extended and cross-linked covalent network of Dawson-type $[P_2W_{17}O_{62}(SiR)_2]^{6-}$ POMs functionalized with short oligo(phenyleneethynylene) tether is directly formed on ITO by electrochemical grafting of diazonium end-groups, resulting in films with a controlled thickness between ≈ 8 and ≈ 26 nm (depending on the number of electrografting cycles).[275] Completed with a top Au electrode, the devices showed a WORM behavior with a low $V_{SET}$ ≈ -1.1 to -1.6 V and a HRS/LRS ratio of ≈ 100 and a data retention time tested up to $10^4$ (Fig. 14-e). The HRS to LRS switch is ascribed to the reduction of POMs leading to more conducting films, as also observed in molecular-scale junctions (see section 2.2.1).[71, 222] Note that in some cases (about 5 % of the devices) a bistable RS was also observed. This batch-to-batch variability as well as the non-reversible switching (WORM memory) is not clear and is still under investigation. In this study as well as in all the reported results of POM-based NVM (see above), the role of the counterions (*e.g.*, low mobility of $TBA^+$) should be examined in more detail (see section 2.2.1 and Refs. 135, 206, 276 showing that counterions play a role in the conductance and hysteresis loop in molecular-scale devices).

While in these previous works, the host matrix is an insulating or low-conducting material, $H_3(PW_{12}O_{40})$ POMs have also been incorporated into an archetypal organic semiconductor P3HT (poly(3-hexylthiophene)).[277] The POMs were functionalized with cationic surfactant tetrakis(decyl)amonium bromide (TDABr) to make them soluble in organic solvents, allowing a uniform blend with P3HT. ITO/POM-P3HT (38 nm thick)/Ag crossbar memory cells were fabricated and tested. The I-V curves displayed a unipolar RS behavior with a low $V_{SET} \approx 0.54$ V and $V_{RESET} \approx 0.06$ V. The HRS/LRS is higher than $10^6$. Moreover, the memory effect is volatile (< 200 ms) and the device response depends on the pattern and time sequence of the applied write pulses. This memristive behavior is further used to develop neuromorphic applications (see Section 5). The switching mechanism is attributed to POM-accelerated $Ag^+$ diffusion and the subsequent formation/breaking of Ag metallic filaments. This behavior is absent for the reference ITO/P3HT/Ag device. When an Au electrode is used (ITO/POM-P3HT/Au), a bipolar NVM behavior is observed.[277]

These results call for remarks and concerns. Several papers make the hypothesis that two electrons are trapped in the POMs,[268, 270, 271] referring to the POM spontaneous reduction observed from an Al-metallic electrode.[160] Yet, this depends on the relative Fermi energy of the metallic electrode and the POM LUMO levels, varying with the POMs but lying in the range -4.5 to -5.5 eV. The Al work function is rather low ($W_F \approx -4.3$ eV) as that of Ag ($W_F \approx -4.3$ to -4.7 ev) but Au has a higher work function ($W_F \approx -5.3$ to -5.5 eV), making the spontaneous reduction process from Au top electrodes much less likely to occur. Also note that in the RS devices the POMs are not necessarily in direct contact with the top electrodes. Pieces of evidence of the POM reduction are provided by XPS and KPFM.[268] The reduction of the POM metal centers during the device operation is relevant given the electronic acceptor character of the POMs and will be consistent with the charge trapping mechanism involved in memory based on metal coordination complexes, organometallics or purely organic molecules.[38, 39, 122, 124] This would also be consistent with the observation of conductive pathways.[268] However, what is more questionable is the transposition of the resistive switching mechanism generally admitted for 3D bulk metal oxides based on the formation and disruption of conductive filaments due to the nanoscale oxygen vacancy migration.[278, 279] This conceals that the formation of oxygen vacancies is also associated with the reduction of the metal centers, which is also proposed in redox active molecule assemblies, yet without the release/incorporation of molecular oxygen. However, if the formation of an oxygen vacancy in POMs is possible locally under specific conditions (Mars-van Krevelen mechanism[280, 281]), the formation in a POM thin film of a percolating oxygen vacancy conducting pathway is highly unexpected. The measured conductance values would then correlate with the molecular redox states.[276, 282] This, in turn, raises the question of the charge compensation mechanism, which is still unclear. Formation of image/mirror charges on the electrodes has been proposed[276] but this lack of understanding of the mechanisms stabilizing the different molecular states involved in the switching hinders further development and large-scale applications. If the charges resulting from the redox events are not stabilized, low conductance ON/OFF ratio is expected (low efficiency), low switching speed and low endurance.[283] When Ag top electrodes are used, conduction through the formation/disruption of Ag filaments, like in the case of electrochemical memories, is another possibility.[284, 285] It has been ruled out by SEM and EDS experiments in ref. 274 but considered the main mechanism in refs. 273 and 277. Several works[268-270, 272, 274] discussed the I-V and RS behaviors of POM-based systems with models established for bulk oxides or materials with low charge mobility (i.e., space charge limited current, SCLC[286-288]). This bulk-limited electron transport mechanism is strictly valid if the energy barriers at the electrodes (*i.e.*, energy mismatch between the Fermi energy of the electrodes and the frontier orbitals of the POMs) are weak (or ideally in case of ohmic contact at the electrode interfaces). This is not the case in these works and the real situation is likely a mix of SCLC and interface-limited injection mechanism such as thermionic emission. As a consequence, we note that in many of these works, not all the fingerprint characteristics of the SCLC I-V behaviors are clearly observed (i.e., ohmic at low voltage, $I \propto V^n$ for the "trap-SCLC" regime with $n \geq 2$ and "trap-free" SCLC, $I \propto V^2$ above a given trap-free limit voltage[288]). Careful temperature-dependent and film thickness-dependent electron transport measurements in both the HRS and LRS are required to distinguish these different electron transport mechanisms. Moreover, the reported energy structures in these devices are based on calculations done for the POM alone, neglecting charge transfers and interface dipoles that inevitably arise when the POMs are embedded with other materials. Similarly, the key role of the counterions, which modifies the energy landscape[133] in the devices (see section 2.2.1) is not included in these works.

Finally, we briefly benchmark the POM RS-memory performance with other RRAM (resistance random access memory) technologies (the last 3 lines in Table 2). We reported the highest on/off ratio, the longest retention time, the larger endurance and the smallest set/reset voltages taken from recent reviews and reports.[263, 276, 289-295] The on/off conduction ratios of POM-based devices remain modest compared to oxide RRAM and non-oxide RRAM, but in some cases approach those reported for organic memories (i.e. < $10^{5-6}$), albeit some works, in these latter cases, also reported larger on/off ratios (up to $10^{7-9}$).[122, 124, 296] The data retention times and the endurance (number or set/reset cycles) are far from the state-of-the art of the other RRAM technologies. The set and reset voltages of the POM-memory are in the same range as the other RRAM technologies and, in several cases, are compatible with low-voltage electronics.

## 4. SPINTRONICS AND QUANTUM COMPUTING.

Many POMs possess interesting magnetic properties, which were intensively studied in solution and in the solid-state, down to a single molecule magnet (SMM), see relevant reviews in refs. 25, 69, 297-302. Compared to other metal coordination compounds, POMs in spintronics and quantum computing offer interesting perspectives, owing to their advantageous molecular oxide structures: (i) robustness to be integrated in solid-state devices; (ii) as inorganic ligands, POMs offer the possibility to accommodate magnetic ions at very precise specific sites of their structure, with a controlled environment that keep the magnetic centers well insulated, and with a composition that is almost free of nuclear spins, thus minimizing the main source of decoherence of the spin qubit; (iii) high-symmetry coordination environment to control the magnetic properties of specific ions, (iv) POMs can accept numerous electrons without altering their structure, possibly leading to mixed-valence systems with extended delocalized electronic spins.[69, 298, 299, 301] Numerous studies and results were reported for magnetic POMs in solution, powder or crystals, characterizing the spin-lattice relaxation time ($T_1$) and the spin coherence time ($T_2$).[299, 300, 302-309] In particular, POMs (and molecules in general) can be optimized by chemically controlling the sensitivity to decoherence sources (*e.g.*, dipolar and hyperfine interactions with electrons and nuclear spins, the effects of coupling with molecular vibrations). For example, coherence has been increased in magnetically diluted samples (e.g., in a diamagnetic host material) or using nuclear free ligands and environments. Another approach consists of playing with the energy-level structures of the molecule and of operating the qubit in an optimal condition (the so-called atomic clock transition) where the qubit is, at the first order, almost invisible to the magnetic field. Some examples for POMs are given in the following and more results on molecular qubits can be found in more specialized reviews).[18, 68, 69, 298, 301] Note that a compromise will be required between the use of magnetically diluted systems and the need for qubit coupling in scalable networks for efficient manipulation for processing quantum algorithms and quantum information. This raises the need to explore the processes and methodologies (e.g., combining POMs with MOFs or hybrid materials) to develop suitable platforms for scalable networks of POM qubits (see a brief discussion in section "conclusions and perspectives").

Reports on spin-polarized transport and spin-related devices (e.g., spin valves, magnetic tunnel junctions) remain scarce. In these devices, two ferromagnetic electrodes are separated by a thin non-magnetic film. Changes in the magnetoresistance are measured depending on whether or not the spin-polarized electrons can be injected and travel or tunnel (in case of very thin films, typically a few nanometers) through the non-magnetic barrier: a high(low) resistance is observed when the spins in the electrodes are oriented antiparallel(parallel), respectively.[310, 311] Spin-polarized electron injections in organic semiconductors were successfully achieved[312-314] (see a review in refs. 315-317) as well as through organic monolayers sandwiched between ferromagnetic electrodes.[318, 319] Similarly, POM-based spin valve devices were studied. Thin films of $[PMo_{12}O_{40}]^{3-}$ were deposited on $La_{0.7}Sr_{0.3}MnO_3$ (LSMO) substrate by spin coating (the POMs were mixed with long counterion alkyl chains, dimethyldioctadecylamonium (DODA), to avoid aggregation). Homogeneous films of $DODA_3[PMo_{12}O_{40}]$ with a thickness 80-100 nm were obtained. The molecular spin valve (MSV) was realized by the evaporation of Co or $MoO_x$/Co electrode as the top electrode (figures 14-a and -b).[320] A magnetoresistance (MR) of ca. 6-7 % (at 50 K and low voltages, ± 0.1V) was reported, which is maintained to ≈ 4% at ± 3.5 V in the case of the LSMO/$DODA_3[PMo_{12}O_{40}]$/$MoO_x$/Co MSVs. Albeit not completely understood, this feature is ascribed to the better energy level alignment between the POM HOMO and the Co Fermi energy (Fig. 15-b). The thin (3 nm) $MoO_x$ film, known as hole injection layer in organic electronics, increases the Co electrode work function and favors bipolar transport through the frontier orbitals, as also observed for organic MSVs.[321, 322] However, these MR performances remain modest compared to metal-containing organic counterparts (e.g., ≈ 40%,[313] ≈ 300 %,[314] and up to ≈ 440%,[323] all the experiments at T ≤ 10 K),[124] and these experiments were not pursued up to now. Magnetic POMs are also widely studied from the perspective of implementing qubits (quantum bits) for quantum computers. Only a few magnetic POMs have been embedded in metal/POM/metal MJs. A current-blockade effect, known as ground-state spin blockade (GSSB) was reported for a pyridine-functionalized $[Mn(III)Mo_6O_{18}\{(OCH_2)_3(CNHCOC_5H_4N\}_2]^{3-}$ Anderson-type POM in a single-molecule transistor device (Fig. 15-c).[324] This effect arises when the spin difference between the subsequent charge states is larger than 1/2 and the energy cost of sequential electron tunneling (SET) between the states is forbidden by the spin selection rules (spin state transition ΔS = 0 at constant charge state, thus 1/2 by adding or removing an electron). Here, the two charge states (marked as N-1 and N, Fig. 15-c), which correspond to $Mn^{4+}$ and $Mn^{3+}$, respectively, have a total spin S=1 and S=5/2. Since the SET is forbidden, the 2D conductance map (Fig. 15-c, bottom left) does not show Coulomb blockade diamonds, with slanted conductance SET lines crossing at V = 0 mV (the charge degeneracy point is missing). By applying a magnetic field of 8 T, the GSSB is canceled by a ground state transition, making the excited high-spin state of the N-1 state $Mn^{4+}$ (S=2) the ground state and thus reducing the spin difference to 1/2.[324] The conductance map (Fig. 15-c, bottom right) looks now like the usual

Coulomb blockade diamonds with slanted SET conductance lines crossing at the charge degeneracy point at V = 0 mV.

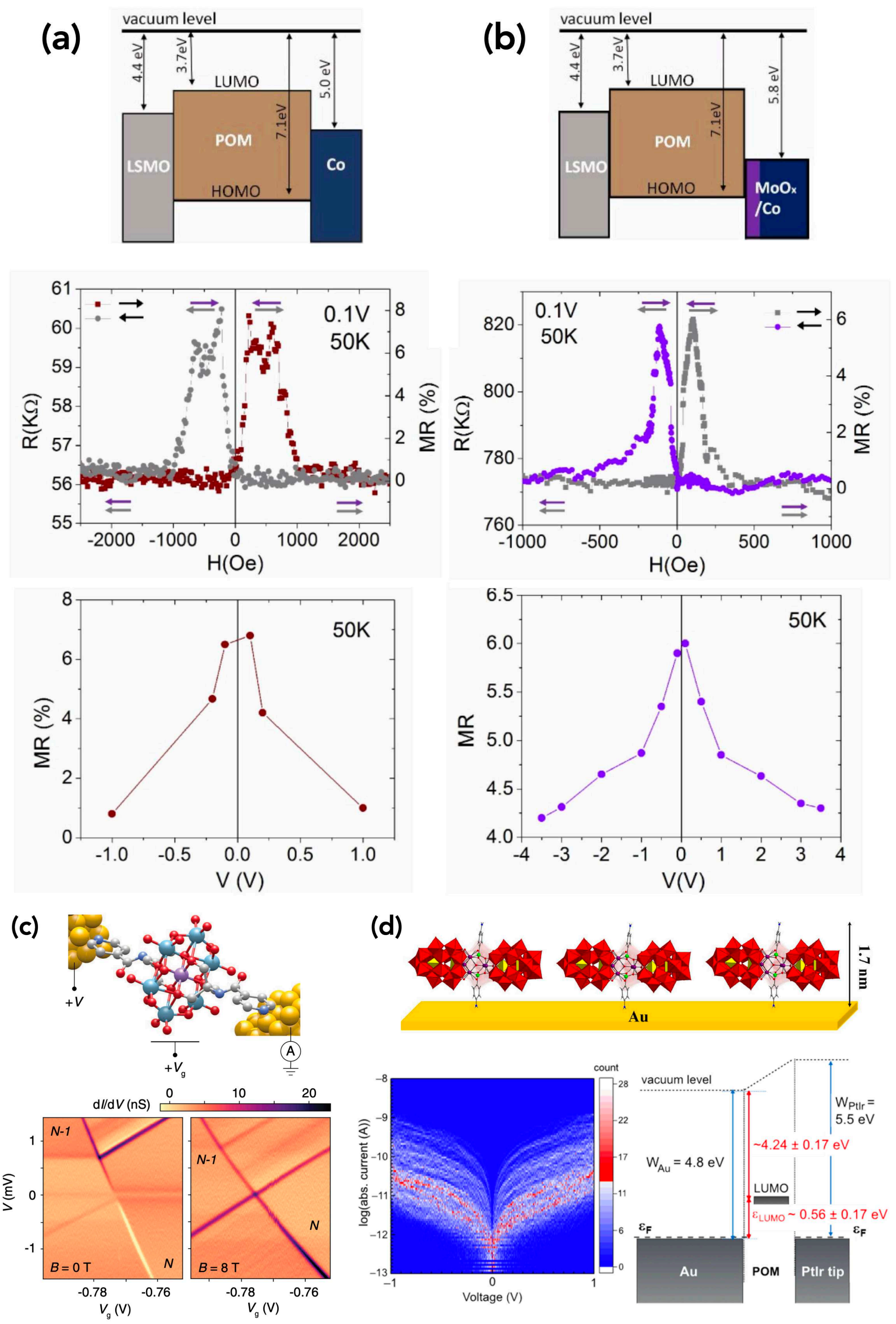

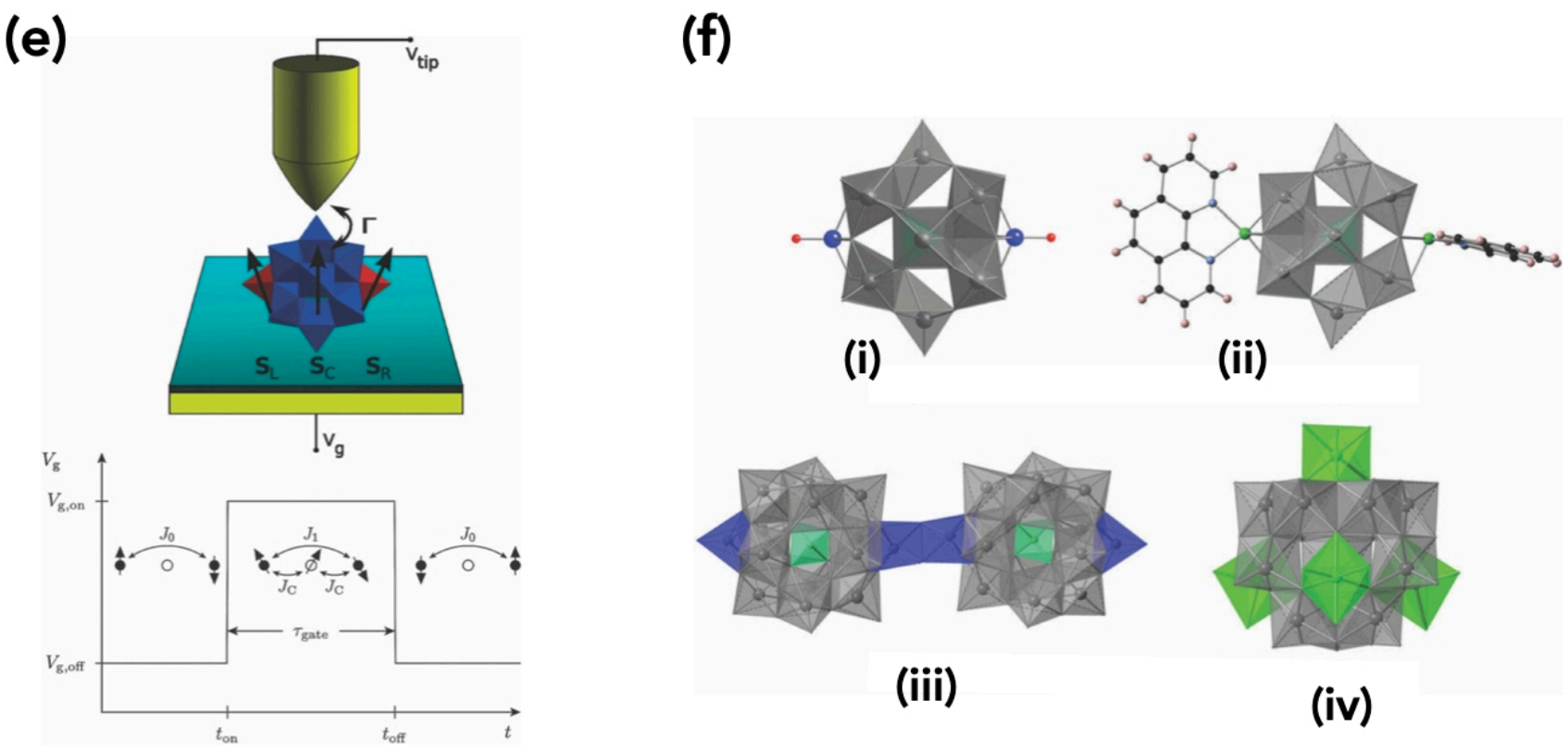

***Figure 15.*** *Energy scheme, resistance vs. magnetic field and amplitude of the magnetoresistance vs. the applied voltage at 50 K of* ***(a)*** *LSMO/DODA$_3$[PMo$_{12}$O$_{40}$]/Co and* ***(b)*** *LSMO/DODA$_3$[PMo$_{12}$O$_{40}$]/MoO$_x$/Co molecular spin valves. (a-b) from ref. 320.* ***(c)*** *Scheme of the [Mn(III)Mo$_6$O$_{18}${(OCH$_2$)$_3$(CNHCOC$_5$H$_4$N}$_2$]$^{3-}$ POM in a single-molecule transistor configuration and conductance (∂I/∂V measured at 40 mK) maps at 0 and 8 T. The slanted darkest lines correspond to the sequential electron tunneling conductance (scale above the plots). Reproduced with permission from ref. 324. Copyright (2019) American Physical Society.* ***(d)*** *Sketch of the {Co$_9$(P$_2$W$_{15}$)$_3$} self-assembled monolayer on Au, 2D map of the current-voltage curves of the Au/{Co$_9$(P$_2$W$_{15}$)$_3$}/PtIr molecular junction (left) and energy diagram of the device (right). Reproduced from ref. 325. Copyright (2017) American Chemical Society.* ***(e)*** *Scheme of the single molecule all-electric two-qubit gating and readout. The [PMo$_{12}$O$_{40}$(VO)$_2$]$^{q-}$ is connected between a bottom gate electrode (covered with a thin insulating tunnel barrier) and the tip of an STM and a typical gating sequence: a voltage pulse is used to inject electrons in the POM core that causes the swap of the vanadyl spins. Reproduced with permission from ref. 326. Copyright (2007) Springer Nature.* ***(f)*** *Other POM candidates for which the complexity of the qubit network is extended using Ni$^{II}$ (S = 1) or with four magnetic sites: (i) [PMo$_{12}$O$_{40}$(VO)$_2$]$^{q-}$, (ii) [PMo$_{12}$O$_{40}${Ni(phen)}$_2$]$^{2-}$, (iii) [Si$_2$Mo$_{24}$O$_{80}$(VO)$_4$]$^{8-}$, and (iv) [Mo$_{12}$O$_{30}$(μ$_2$-OH)$_{10}$H$_2$(Ni(H$_2$O)$_3$)$_4$]. Reproduced with permission from ref. 298. Copyright (2012) Royal Society of Chemistry.*

Beyond a single molecular qubit, efficient and practical quantum computing requires scaling up to systems with thousands of qubits. This objective remains an open challenge, which is still far from being attained, since it involves the individual detection and manipulation of each molecular spin. Anyway, form a chemical and technological point of view, a critical step to be achieved is that of controlling the position of the POMs in the device, since electronic and magnetic interactions between them can perturb the expected behavior (e.g., due to the quantum decoherence effect).[298] This requires the immobilization of a large number of POMs on the surface (electrode) without the loss of their magnetic properties. A possible approach is to embed the POM, $[Fe_4(H_2O)_2(FeW_9O_{34})_2]^{10-}$, into an amorphous gelatin biopolymer or a crystalline diamagnetic metal-organic framework (MOF), this latter being deposited on pyrolytic graphite electrodes. Through a series of characterizations, the magnetic properties of the POM@gelatin and POM@MOF compounds were found close to those of the pure POM, especially the presence of clear resonant quantum tunneling of magnetization.[327] The same POMs were also successfully grafted (non-covalently) on carbon nanotubes, while preserving their magnetic bistability,[328] which opens the opportunity to integrate magnetic POMs in carbon nanotube devices and circuits. Large magnetic POMs $Na_{25}[Co^{II}_9(H_2O)_6(OH)_3(p\text{-}RC_6H_4AsO_3)_2(\alpha\text{-}P_2W_{15}O_{56})_3]$ with R = H or $NH_2$, for short $\{Co_9(P_2W_{15})_3\}$, have been successfully assembled as a chemisorbed monolayer on an ultra-flat Au electrode via a novel organoarsonate ligand (Fig. 15-d).[325] The extensive physicochemical characterizations of this POM/Au system revealed a stable and homogeneous self-assembled monolayer. The electron transport properties of the Au/$\{Co_9(P_2W_{15})_3\}$/PtIr were measured by C-AFM (Fig. 15-d) to determine the energetics of the molecular orbital involved in the ET: the LUMO is measured at 0.56 ± 0.17 eV above the electrode Fermi level. These results render magnetically functionalized POMs accessible for further spintronic experiments and devices. For instance, spin manipulation/switch by the application of an electric field is a prerequisite for applications in quantum devices and computing.[326, 329] In a seminal proposal (concept and basic theory) of POMs as spin qubits,[326, 329] the $[PMo_{12}O_{40}(VO)_2]^{q-}$ has two $(VO)^{2+}$ groups with spin S = 1/2 sitting at diametrically opposite positions of the $PMo_{12}$ core. Depending on the number of electrons in the $PMo_{12}$ core, the localized spins on the two $(VO)^{2+}$ and the delocalized spin on the $PMo_{12}$ can be modified (strong coupling for an odd number of electrons, weak coupling for an even number). Thus, if this POM is connected between a bottom electrode and an STM tip (Fig. 15-e), a voltage pulse can inject electrons in the $PMo_{12}$ core, inducing a "swap" quantum gate, *i.e.*, it swaps the spin states between the two spin qubits (depending on the values of the exchange coupling strength $J_1$, $J_C$ and the POM-tip electronic coupling energy Γ, Fig. 15-e bottom panel).[326] For a half-time pulse duration, a square-root-of-swap √SWAP quantum gate (with entanglement of the swapped and non-swapped states) is obtained.[330] A √SWAP is one of the fundamental quantum gates. Several other POM candidates extending these design rules were also suggested (Fig. 15-f).[298] From an experimental point of view, the electrical control of spin qubits at the nanoscale has been little investigated.[331-334] In this context, POMs also provide ideal examples with the demonstration of an enhanced spin-electric coupling in $[Ho(W_5O_{18})_2]^{9-}$ crystal (that is a voltage control of the spin clock-transition frequency).[335]

## 5. NEUROMORPHIC DEVICES.

In neuromorphic computing systems, memory and logic are co-located in memory processing units, enabling reduced latency and power consumption.[61-63] Neuromorphic computing is thus emerging as a promising approach to support AI hardware, notably through the development of artificial neural networks for machine learning.[64] These systems require the emulation of biological synaptic functions, which can be achieved using arrays of memristive cells or memristors.[61, 266, 267, 336, 337] Neuromorphic devices based on unconventional technologies (instead of silicon CMOS) have attracted a growing interest since the year 2010 or so,[61, 67, 338] including molecular-based systems.[339-344] The dynamic properties of POMs (*e.g.*, redox switching, noise, memristive behavior, see Sections 2.2.1, 2.2.2 and 3) make them appealing for neuromorphic device applications.

The simple neuromorphic device is an artificial synapse that mimics the biological synapse plasticity. The signal transmitted through the synapse is modulated by the synaptic weight that depends on the combination and time sequences of a series of pre- and post-synapse pulsed signals (*e.g.*, number or/and frequency of pulses, time synchronization). Depending on these parameters, the synapse weight is increased (facilitation behavior) or decreased (depression behavior). In an artificial synapse, the synaptic weight is the electrical resistance or conductance of the device. Several examples of short-term plasticity (STP) have been obtained with POM-based memristors.[262, 271, 277] Figure 16-a shows pulsed-pair facilitation (PPF) and pulsed-pair depression (PPD) obtained with pentacene/PMMA/$[PW_{12}O_{40}]^{3-}$/rGO/$SiO_2$Si transistors.[262] PPF and PPD are typical examples of STP for which the response of the device at the second pulse is amplified or decreased depending on the time elapsed between the pulses. STP behaviors were also reported for $[SiW_{12}O_{40}]^{4-}$ embedded in metalloviologen MOF[271] and for ITO/$[PW_{12}O_{40}]^{3-}$-P3HT/Ag memristor (Fig. 16-b).[277] In this latest case, a crossbar array of 100 devices was trained to perform several simple pattern recognition tasks (recognize 3 different emojis, handwritten digit classification).[277] We note that another type of plasticity very useful in biological synapses, STDP (spike time-dependent plasticity, the synapse weight is amplified when pre- and post-synapse pulses are time coincident) was not reported, unlike in other nanotechnology-based artificial synapses, *e.g.* in refs. 340, 345-347.

A more complex neuromorphic engine is referred to as "reservoir computing" (RC). The concept of RC was proposed at the beginning of the 2000s.[348, 349] RC is a peculiar type of recurrent neural network (RNN) in which the training to compute a given task is simplified compared to other RNNs. The main part of RC is the reservoir, a network of nodes and links (weight $W_{res}$) randomly connected (Fig. 16-c). The building blocks of the network must be characterized by large variability, strong non-linear responses and complex dynamics.[348-352] The signals of some output nodes are read by an output layer (basically a simple perceptron),[353] which is trained (updating the weights $W_i$ with an appropriate learning algorithm) to perform a given information processing task. Contrary to multi-layer feed-forward neural networks and/or convolution neural networks, where all the hidden layer weights need to be trained and adjusted, RC is a simplified computation system at the hardware level because only the output layer weights $W_i$ must be trained, while the reservoir weights ($W_{res}$) remain fixed. The implementation of hardware RC was tested using a variety of physical devices and technologies (see a review in ref. 352) including nanoscale materials and devices (see a specific review in ref. 354). One of the mandatory conditions for an efficient RC is great variability of the $W_{res}$ values.[348-350, 352] Note that the output layer can be implemented physically or most of the time by a software algorithm.

Dense and random networks of carbon nanotubes (CNT) complexed with POMs $[PMo_{12}O_{40}]^{3-}$ or functionalized porphyrin-POMs $(H_4TPP)_2[SV_2W_{10}O_{40}]$ were used to implement RC (Fig. 16-d).[355-357] Several tasks were demonstrated. The RC was trained to pass the benchmark task of NARMA-10 (Nonlinear Auto-Regressive Moving Average) time series predictions, a standard and widely used univariate time series forecasting developed to test the performances of recurrent neural networks and RC.[351, 358] The system was also trained to generate Boolean functions (OR, AND, NOR, NAND, XOR, XNOR) with accuracy > 90%, or trained as a waveform generator (accuracy 99.4%). The success of classification tasks needs devices featuring random fluctuations (*i.e.*, random telegram noise) and low-frequency noise, $1/f^n$ noise (see Section 2.2.2, Fig. 9), with a frequency exponent $n > 1.2$, while the task failed with devices featuring only white noise, or with $n < 0.2$, Fig. 16-e.[357] These dynamic behaviors were associated with redox switching of the POMs in the RC network upon charge injection and accumulation,[355] as also observed for polyoxometalate/Au nanoparticle networks (Fig. 9);[229] thus, this later platform is likely also suitable for physical reservoir computing.

Since the redox-dependent conductance of POMs can be switched electrically (*vide supra*, sections 2 and 3 - *e.g.,* Figs. 8-a, 14) and optically (*vide supra*, section 2 - *e.g.*, Fig. 8-c), the two stimuli have been combined at different time scales to implement a dual behavior (volatile memory with optical pulses and non-volatile memory with quasi-static voltage sweep) in the same hybrid polymer/POM device (POM is $[Sm(PW_{11}O_{39})_2]^{11-}$). These devices were implemented in RC and recurrent neural network circuits for speech and motion recognition.[359] A similar dual opto-electrical memristive behavior of a $[PMo_{11}VO_{40}]^{4-}$ thick film (≈ 500 nm) was combined with the typical color change of POMs under redox process to develop a biomimetic device combining color tuning, memory and external-stimulus adaptative behaviors.[360]

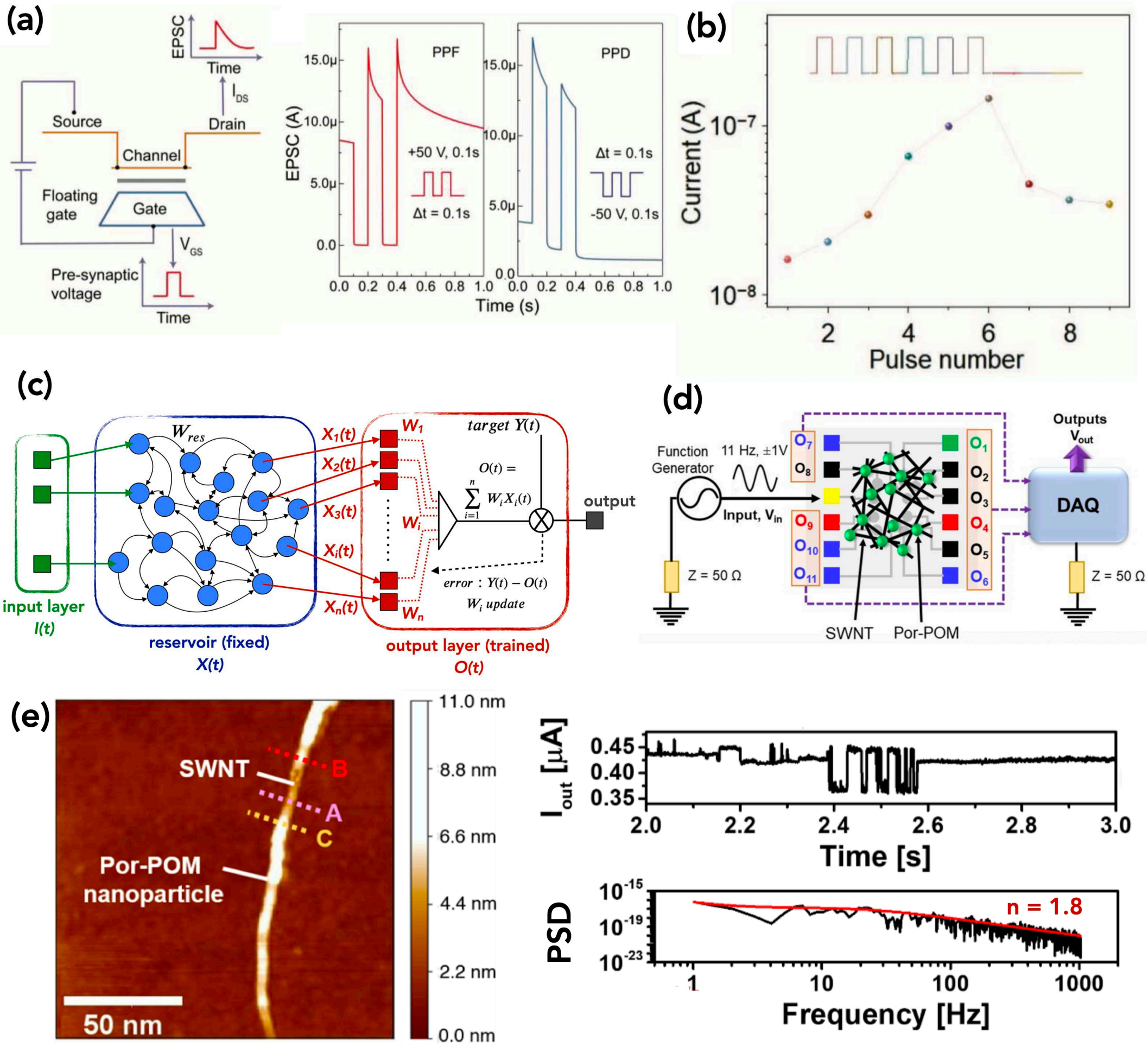

***Figure 16.*** ***(a)*** *Scheme of the electrical connections of the synaptic $[PW_{12}O_{40}]^{3-}$/rGO/pentacene transistor and typical examples of the PPF (paired-pulse facilitation) and PPD (paired-pulse depression) behavior. Reproduced with permission from ref. 262. Copyright (2018) John Wiley and Sons.* ***(b)*** *Shot-term plasticity of a ITO/ $[PW_{12}O_{40}]^{3-}$-P3HT/Ag two-terminal memristor: facilitation upon a sequence of pulses and depression when the artificial synapse device is at rest. Reproduced with permission from ref. 277. Copyright (2022) John Wiley and Sons.* ***(c)*** *The basic scheme of a reservoir computing (RC) system that is made of an input layer, a reservoir and a trained output*

*layer. The reservoir is a randomly interconnected (black lines) network of nodes (blue circles). The transfer functions of the links are characterized by weight $W_{res}$ that are kept fixed. Several $X_i(t)$ reservoir outputs are read and weighted ($W_i$) as a linear combination to generate and output O(t), which is compared to the target Y(t). The error Y(t)-O(t) is minimized by updating the weights $W_i$ using a learning algorithm. Reproduced with permission from ref. 229. Copyright (2024) Royal Society of Chemistry.* ***(d)*** *Schematic diagram of a RC engine based on a network of carbon nanotubes and porphyrin-POM, $(H_4TPP)_2[SV_2W_{10}O_{40}]$ (DAQ is the digital acquisition electronic board used to acquire and manipulate the data). Reproduced from ref. 357. Copyright (2022) John Wiley and Sons. Creativecommons.org/licenses/by/4.0/.* ***(e)*** *AFM image of a carbon nanotube decorated with porphyrin-POM, $(H_4TPP)_2[SV_2W_{10}O_{40}]$ nanoparticles and the typical two-level current fluctuations and low-frequency noise (PSD:power spectral density) $1/f^n$ (n = 1.8) recorded for this CNT/porphyrin-POM complex connected between electrodes. Reproduced from ref. 357. Copyright (2022) John Wiley and Sons under Creative commons.org/licenses/by/4.0/.*

## 6. CONCLUSIONS AND PERSPECTIVES.

The electronic transport properties of several types of POMs have been studied in various conditions: from single molecule junctions and monolayer-based junctions to hybrid films and structures where POMs have been embedded with other materials and/or nanostructures. Table 3 summarizes the list of studied POMs and the evaluated device applications reviewed in this work.

| | POM | Basic electron transport studies | | | | Applications in devices | | | | Refs. |
|---|---|---|---|---|---|---|---|---|---|---|
| | | thin film or bulk | mono-layers | single molecule | hybrid networks | capacitive memory | resistive switching memory | spin-tronics | neuro-morphic | |
| **Keggin** | $[PMo_{12}O_{40}]^{3-}$ | | X | | X | | | X | X | 104, 192, 222, 243, 320, 355, 356 |
| | $[PMo_{12}O_{40}]^{3-}$ $[Pt_4]^{5+}$ or $[Pt_2Pd]^{3+}$ | X | | | | | | | | 152 |
| | $[PMo^{VI}_{11}Mo^{V}O_{40}]^{4-}$ | X | X | | | | | | | 156, 222 |
| | $[PMo_{12}O_{40}(VO)_2]^{n-}$ | | | | | | | X | | 326, 329 |
| | $[PW_{12}O_{40}]^{3-}$ | X | X | | | X | X | | X | 74, 92, 104, 137-143, 172, 173, 192, 228, 236, 257, 258, 262, 268, 269, 276 |
| | $[SiW_{12}O_{40}]^{4-}$ | X | | | | | X | | X | 143, 271 |
| | $[GeW_{12}O_{40}]^{4-}$ | | | | | | X | | | 270 |
| | $[XM_{12}O_{40}]^{n-}$ (M= Mo, W; X= P, Si, B, Co) | | X | | | | | | | 174, 177, 178 |
| | $(H_4TPP)_2[SW_{10}V_2O_{40}]$ | | | | | | | | X | 356, 357 |
| | $[PW_{11}O_{39}\{O(SiC_3H_6SH)_2\}]^{3-}$ | | | | X | | | | X | 229 |
| | $[XW_{11}O_{39}\{O(SiC_3H_6SH)_2\}]^{n-}$ (X= P, Si, Al) | | | | X | | | | | 250 |
| | $[PM_{11}O_{39}\{Sn(C_6H_4)C\equiv C(C_6H_4)N_2\}]^{3-}$ M=W,Mo | | X | | | | | | | 104 |
| | $[PW_{11}O_{39}\{Sn(C_6H_4)C\equiv C(C_6H_4)COOH_{0.6}\}]^{4,4-}$ $[PW_{11}O_{39}\{O(SiC_2H_4COOH_{0.8})_2\}]^{3,4-}$ | | X | | | | | | | 198 |
| | $[Co_4(H_2O)_2(B\text{-}\alpha\text{-}PW_9O_{34})_2]^{10-}$ | | | | | | X | | | 272 |
| | $[Fe_4(H_2O)_2(FeW_9O_{34})_2]^{10-}$ | | | | | | | X | | 327, 328 |
| | $[Sm(PW_{11}O_{39})_2]^{11-}$ | | | | | | | | X | 359 |
| | $[PMo_{11}VO_{40}]^{4-}$ | | | | | | | | X | 360 |
| **Wells-Dawson** | $[P_2Mo_{18}O_{62}]^{6-}$ | | X | | | | | | | 207 |
| | $[P_2Mo_{18}O_{62}]^{6-}$ $[Pt_4]^{5+}$ | X | | | | | | | | 152 |
| | $[P_2Mo_{18-x}V_xO_{62}]^{n-}$ (x = 1,2,3) | | X | | | | | | | 207 |
| | $[Mo^{V}_2Mo^{VI}_{16}O_{54}(SO_3)_2]^{6-}$ | X | | | | | | | | 155 |
| | $[W_{18}O_{54}(SeO_3)_2]^{4-}$ | | | | | X | | | | 73 |
| | $[W_{18}O_{56}(WO_6)]^{10-}$ | | | | | X | | | | 73 |
| | $[P_2W_{18}O_{62}]^{6-}$ | X | | | | | | | | 150 |
| | $[P_2W_{17}O_{61}\{O(SiC_6H_4C\equiv CC_6H_4)N_2)_2\}]^{8-}$ | | | | | | X | | | 275 |
| | $[P_2W_{15}Nb_3O_{62}]^{9-}$ | | X | | | | | | | 207 |
| | $[HP_2W_{15}V_3O_{59}\{(OCH_2)_3CCH_2SCH_3\}]^{5-}$ | | X | | | | | | | 113 |
| | $[Co_9(H_2O)_6(OH)_3(p\text{-}RC_6H_4AsO_3)_2(\alpha\text{-}P_2W_{15}O_{56})_3]^{25-}$ | | | | | | | X | | 325 |
| **Preyssler** | $[NaP_5W_{30}O_{110}]^{14-}$ | X | X | | | | | | | 135, 151 |
| | $[DyP_5W_{30}O_{110}]^{12-}$ | | | X | | | | | | 237 |
| | $[Gd(H_2O)P_5W_{30}O_{110}]^{12-}$ | | | | | | | X | | 308 |
| **Anderson-Evans** | $[MnMo_6O_{18}\{(OCH_2)_3CNH_2\}_2]^{3-}$ | | | | | | X | | | 75 |
| | $[MMo_6O_{18}\{(OCH_2)_3CNH_2\}_2]^{3-/4-}$ (M=Fe,Co,Ni, Zn) | | | X | | | | | | 175 |
| | $[MnMo_6O_{18}\{(OCH_2)_3CC_5H_4N\}_2]^{3-}$ | | | X | | | X | | | 108, 274 |
| | $[MnMo_6O_{18}\{(OCH_2)_3CNH(CO)C_5H_4N\}_2]^{3-}$ | | | | | | | X | | 324 |
| **Lindqvist** | $[V_6O_{13}\{(OCH_2)_3CCH_2SR\}_2]^{2-}$ (R=$CH_3$ or $C_6H_5$) | | | X | | | | | | 71 |
| | $[V_6O_{13}\{(OCH_2)_3CCH_2OCO(CH_2)_4C_3H_5S_2\}_2]^{2-}$ $[MnMo_6O_{18}\{(OCH_2)_3CNHCO(CH_2)_4C_3H_5S_2\}_2]^{3-}$ | | X | | | | | | | 242 |
| | $[V_6O_{13}\{(OCH_2)_3CCH_2N_3CH(Ar)C(H)AuL\}_2]^{2-}$ | | | X | | | | | | 217 |
| | $[V_6O_{13}\{(OCH_2)_3CR\}_2]^{2-}$ (R=$CH_3$, $CH_2OH$, $NHCOCH_2Cl$ and $NHCOCH_2\text{-}OOCC_{10}H_{15}$) | | | X | | | | | | 118, 119 |
| | $[HV_{12}O_{32}Cl(LnPc)]^{3-}$ and $[HV_{12}O_{32}Cl(LnPc)_2]^{2-}$ (Ln=Sm, Eu, Gd, Dy, Ho, Er, Yb, Y, Lu) | | X | X | | | | | | 199, 219 |
| | $[Ho(W_5O_{18})_2]^{9-}$ | | | | | | | X | | 335 |
| **Other** | $[\gamma\text{-}Mo_8O_{26}]^{4-}$ | | | | | | X | | | 273 |
| | $[H_7P_8W_{48}O_{184}]^{33-}$ | | X | | | | | | | 97 |

***Table 3.*** *List of the experimentally studied POMs for assessing their basic electron transport properties and device applications, which are reviewed in this paper. The POMs are listed by family type (see also Fig. 1). Note: "monolayers" means one to a few monolayers of POMs, "hybrid networks" refers to works where the POMs are complexed with nanostructures (e.g., carbon nanotubes, metal nanoparticles ).*

At the level of basic electron transport properties, the main acquired knowledge, remaining open questions and perspectives can be summarized as follows:

- The POM thin film conductivity is intrinsically low (< $10^{-8}$ S/cm) but can be significantly increased (a factor $10^3$-$10^5$) by (photo)-reduction or by improving the POM-to-POM connectivity with mixed-valence metal complexes. The highest reported conductivity reached ca. $10^{-4}$ S/cm.
- The conductance and the energy scheme (position of the LUMO with respect to the electrode Fermi energy) of molecular junctions (SAM and single molecule) clearly depend on the nature of the metal atoms and of the heteroatom (in a series of Keggin-type POMs). The energy position of the LUMO in the molecular junction is correlated with the physicochemical properties of the POMs (redox potential), demonstrating that the signature of the POMs is retained in the solid-state devices and unraveling that the POMs play the key role in the electron transport properties of these molecular-scale devices.
- The electronic properties of the POM-based molecular junctions are sensitive to the POM-electrode linkers, as well as to the POM functionalization by ligands at their periphery, as also known for the purely organic molecular junctions.
- The role of the substrate has not been directly assessed. More effort should be devoted to a fundamental understanding of the POM/substrate interactions and the identification of possible interfacial effects, inspired by the thoroughly investigated substrate effect in heterogeneous catalysis.[78]
- The unavoidable counterions play a key role in the electron transport properties of the POM-MJs, far beyond the charge neutrality point of view. For instance, the conductance of the MJ can be modulated by factor 100 just by changing the counterions. The nature of the counterions largely modifies the energy scheme of the molecular junctions (LUMO energy position and delocalization, as well as electronic energy coupling between the POM and the electrode) in a subtile and interdependent way, the exact mechanisms remaining to be clearly understood.
- The multi-level redox properties of POMs are preserved in POM-based molecular junctions. The conductance of the molecular junction increases upon POM reduction, and several reduced states (3 -5) were observed within a low-voltage range (< 2 V) for a Lindqvist-type POMs. The UV photo-reduction of Keggin type was also demonstrated in a solid-state molecular junction. The reduced (open-shell state) gives a 10-fold increase in the MJ conductance, it is stable in the air and at room temperature, the spontaneous return to the oxidized state required several hours to days. The role of ligands and counterions on the multiple resistance switching properties has not been extensively studied and needs further experimental studies and theoretical calculations on the complete molecular devices.

- POMs were successfully combined with other nanostructures (*e.g.,* carbon nanotubes, metallic nanoparticles) to enlarge the device functionalities (*e.g.,* rectifying diodes, neuromorphic devices).

At the nanoelectronic devices level, the take-home messages are:

- The proof-of-principle of memory based on the redox properties of POMs has been demonstrated for both capacitive and resistance switching memory. However, especially for RS memories, a better understanding of the switching and conduction mechanisms remains to be achieved. Several proposed mechanisms seem unrealistic or incomplete and do not consistently explain the complete behaviors of these memories (e.g., conduction filaments based on oxygen vacancies, space-charge effects, interface energetics). Moreover, data on the statistical reproducibility/variability from one device to another are generally missing, which together with the detailed figures of merits are decisive to increase the technological readiness level. The global performances are modest (see "benchmarking" in Tables 1 and 2 and related discussions at the end of sections 3.1 and 3.2), even if a few metrics of POM-based memories are on par with other technologies, and further optimizations are required. For example, the influence of the counterions on memory performances has not been investigated. However, albeit the multi-level RS was observed at the molecular-scale level, no multi-bit memory has been demonstrated to date.
- Albeit the magnetic properties of POMs are widely studied in solution, powders or crystals, the experimental demonstrations of POM-based solid-state spintronic devices (spin valves, qubits) remain scarce or limited to preliminary studies on the anchoring of magnetic POMs on surfaces and electrodes.
- Several synapse-like behaviors (short-term plasticity) were observed with a few POM-based memristor-like devices. When combined with carbon nanotubes or metal nanoparticles, a more complex neuromorphic system known as reservoir computing was implemented, their basic dynamic properties were shown to be prone to neuromorphic computing and simple learned tasks (e.g., recognition and classification) were demonstrated. More realistic applications, beyond these "toy systems », remain a stimulating challenge for POM-based device research community.
- To accelerate the technological deployment, it would be beneficial to identify the most reliable methodologies for POM patterning, enabling both robust devices and their fabrication at a larger scale, with maximal reproducibility. The answer is not straightforward, since the specifications for spintronic or neuromorphic devices, for example, are quite different. Neuromorphic systems are tolerant to default and variability (which even are useful for implementing reservoir computing), while structural disorders may be detrimental to memory and spintronic devices (minoring data retention or increasing spin decoherence through interactions between too close adjacent POMs). Non-covalent routes are easy to implement, however, possible formation of POM aggregates driven by electrostatic interactions between the POMs and their counter-cations is to take into consideration.[74, 236] Covalent routes allow a higher control of the POM/electrode interface at the monolayer scale, hence a higher reproducibility, but are restricted to POM organic-inorganic hybrids, requiring multistep synthesis. POM-polymer composites have been largely used as active materials in molecular junctions (e.g., refs. 75, 143, 268, 277, 359). Soft matter brings flexibility and processability for large-scale production. However, the challenge is still to control the POM distribution and density,[361] which will determine the POM-to-POM communication. The latter will enhance electron charge transport but could be detrimental to spin qubit systems, with loss of information due to quantum decoherence. Besides these bottom-up self-assembly processes, top-down electron beam nano-lithographic techniques reaching the sub-10 nm resolution size are appealing, reminiscent of the earlier purpose of using POM/resist hybrids to develop POM-based devices without the need for additional lithographic steps.[137] In a recent example, nanoscale vanadium oxide structures have been prepared from polyoxovanadate-based negative-tone resists.[362] However, in this example, the molecular nature of the precursors is not maintained. An effort of rationalization of device performance based on benchmark parameters and iterative discussion between chemists, physicists, theoreticians and engineers is needed to devise relevant deposition methodologies, device fabrication, characterization and simulation to push POM based systems to real-world nanoelectronics (e.g., as exemplified for resistive switching memories in ref. 290).

## AUTHOR INFORMATION.

**Corresponding authors.**

**Dominique VUILLAUME** - *Institute for Electronics Microelectronics and Nanotechnology (IEMN), CNRS, Av. Poincaré, F-59652 Villeneuve d'Ascq, France.* Orcid: orcid.org/0000-0002-3362-1669; E-mail: dominique.vuillaume@iemn.fr

**Anna Proust** - *Sorbonne Université, CNRS, Institut Parisien de Chimie Moléculaire, IPCM, 4 Place Jussieu, F-75005 Paris, France*; orcid.org/0000-0002-0903-6507 ; Email: anna.proust@sorbonne-universite.fr

***Notes.***

The authors declare no competing financial interest.

**Biographies.**

Dominique Vuillaume is Emeritus Research Director at the Centre National de la Recherche Scientifique (CNRS). He received the Electronics Engineer degree from the Institut Supérieur d'Electronique du Nord, Lille, France, 1981 and the PhD degree and Habilitation diploma in solid-state physics, from the University of Lille, France in 1984 and 1992, respectively. Since 1992, he has worked at the Institute for Electronics, Microelectronics and Nanotechnology (IEMN), a CNRS laboratory at the University of Lille. In 2000, he created and headed (until 2019) the « Nanostructures, nanoComponents & Devices » (NCM) research group at IEMN. He was head of the department "Physics of materials and nanostructures" at IEMN (2015-2019). His research interests (1982-1992) covered physics and characterization of point defects in semiconductors and MIS devices, physics and reliability of thin insulating films, hot-carrier effects in MOSFET's. Since 1992, he has been engaged in the field of Molecular and Organic Electronics. His current research concerns: (i) design and characterization of molecular and nanoscale electronic devices, (ii) elucidation of fundamental electronic properties of these molecular and nanoscale devices, (iii) study of functional molecular devices and integrated molecular systems, (iv) exploration of new computing paradigms using molecules and nanostructures. He was a scientific advisor for industrial companies (Bull R&D center) on advanced CMOS technology reliability (1988-1990) and for the CEA (Commissariat à l'énergie atomique et aux énergies alternatives) for the "Chimtronique" (Chemistry for nanoelectronics) research program (2006-2013).

Since 2000, Anna Proust has been full professor of inorganic chemistry at Université Pierre et Marie-Curie (UPMC-Paris VI-France) and then Sorbonne Université. She is a graduate of the Ecole Normale Supérieure and of the Université Pierre et Marie Curie. She received her Ph.D. degree in 1992 under the supervision of Professor P. Gouzerh. After a post-doctoral stay at the University of Bielefeld (Germany) with Professor Dr. A. Müller, she returned to UPMC as Assistant Professor, then Associate Professor after her habilitation in 1999. She was junior member of the Institut

Universitaire de France (IUF, 2007-2011) and director of the doctoral school of molecular chemistry. She has been vice-chair then chair of the Coordination Chemistry Division of the French Chemical Society, deputy-director and then director of the Parisian Institute of Molecular Chemistry-IPCM. In 2020, she received the State Prize of the French Academy of Sciences and in 2025 she was elected to the European Academy of Sciences EurASc. Anna Proust is an expert in the chemistry of polyoxometalates (POMs). Her research has covered organometallic oxides, transition metal-substituted POMs, in relation to molecular magnetism or catalysis, and the covalent functionalization and post-functionalization of POMs to use them as modular building blocks for functional molecular materials. Her current endeavors make use of POMs as charge storage nodes, for example in devices for molecular electronics, or as electronic relays in photo- or electro-assisted reduction processes involving small molecules. A common theme is the understanding of the parameters that govern electron transfer/transport and the shaping of POMs, such as their organization on electrodes.

## ACKNOWLEDGEMENTS.

No specific funding to acknowledge.

## REFERENCES.

**ToC graphic**

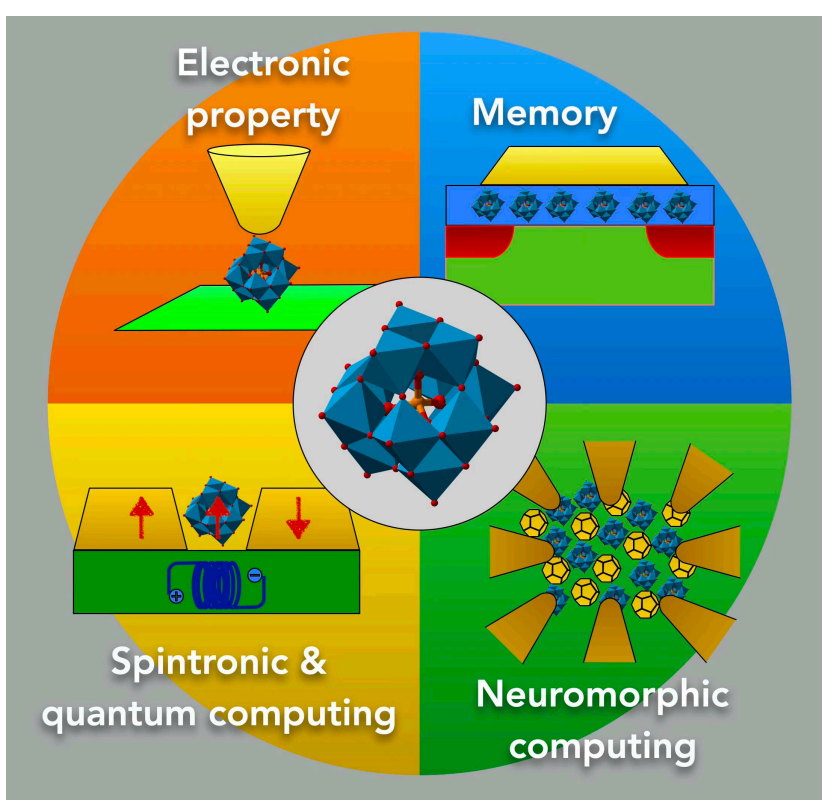